%% file: article.tex
\documentclass[longauth]{aa}  

\usepackage{hyperref}
\usepackage{graphicx}
\usepackage{txfonts}

\def\gaia{\textit{Gaia}\xspace}
\def\gdr3{\textit{Gaia}~DR3\xspace}
\def\egdr3{\textit{Gaia}~EDR3\xspace}
\def\g{$G$\xspace}
\def\bp{$G_{\rm BP}$\xspace}
\def\rp{$G_{\rm RP}$\xspace}
\def\bprp{\mbox{$G_{\rm BP}-G_{\rm RP}$}\xspace}

\newcommand\referee[1]{#1}
\newcommand\OK[1]{#1}

\newcommand{\orcit}[1]{\protect\href{https://orcid.org/#1}{\protect\includegraphics[width=8pt]{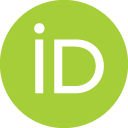}}}

\titlerunning{\gaia~DR3: Performance verification}

\makeatletter
\renewcommand*\maketitle{%
  \thispagestyle{firstpage}
\begingroup
    \if@wideboxfn
    \setlength\bibindent{1.4\parindent}
    \else
    \setlength\bibindent{\parindent}
    \fi
    \renewcommand*\thefootnote{\@fnsymbol\c@footnote}%
    \renewcommand\@makefntext[1]{%
    \ifaa@longfn\hsize\textwidth\fi
    \noindent
    \hb@xt@\bibindent{\hss\@makefnmark\enspace}##1}
  \ifaa@twocolumn
  \begingroup
    \begin{aa@strip}
          \aa@maketitle
    \end{aa@strip}
    \@thanks            
  \endgroup
  \else
    \begingroup
      \let\thanks\footnote
      \aa@maketitle
    \endgroup
  \fi
\endgroup
  \setcounter{footnote}{0}%
}
\makeatother

\usepackage[normalem]{ulem}

\begin{document} 

\title{Gaia Data Release 3}
\subtitle{Pulsations in main sequence OBAF-type stars}

\input{authors.tex}

\date{Received March 15, 2022; accepted April 16, 2022}

\abstract
{
  The third \textit{Gaia} data release provides photometric time series covering 34 months for about 
  10 million stars. For many of those stars, a characterisation in Fourier space and their
  variability classification are also provided. 
  This paper focuses on intermediate- to high-mass (IHM) main sequence pulsators $(M\geq\,1.3\,$M$_\odot$) 
  of spectral types O, B, A, or F, known as $\beta$\,Cep, slowly pulsating B (SPB), $\delta$\,Sct, and $\gamma$\,Dor stars.
  These stars are often multi-periodic and display low amplitudes, making them challenging targets to analyse with
  sparse time series. 
}
{
  We investigate the extent to which the sparse \gdr3 data can be used to detect OBAF-type pulsators
  and discriminate them from other types of variables. We aim to probe the empirical instability strips and
  compare them with theoretical \OK{predictions.}
  The most \OK{populated} variability class \OK{is that of} the $\delta$\,Sct variables.
  For these stars, we aim to confirm their empirical period-luminosity (PL) relation, 
  and verify \OK{the} relation between their oscillation amplitude and rotation.
}
{
  All datasets used in this analysis are part of the \gdr3 data release. The photometric time series 
  were used to perform a Fourier analysis, 
  \OK{while} the global astrophysical parameters necessary for the empirical
  instability strips were taken from the \gdr3 \texttt{gspphot} tables, and the $v\sin i$ data 
  were taken from the \gdr3 \texttt{esphs} tables. 
  The \OK{$\delta\,$Sct} PL relation was derived using the 
  \OK{same} photometric parallax method
 as the one recently
  used to establish the PL relation for classical Cepheids using \textit{Gaia} data.
}
{ We show that for nearby OBAF-type pulsators, the \gdr3 data are precise and accurate enough to pinpoint 
 them in the Hertzsprung-Russell (HR) diagram.
  \OK{We find empirical instability strips covering broader regions than theoretically predicted. 
  In particular, our study reveals}
the presence of fast rotating gravity-mode pulsators outside the strips,
\OK{as well as the co-existence of rotationally
  modulated variables inside the
  strips as reported before in the literature.}
  We 
  \OK{derive an}
extensive period--luminosity relation for 
$\delta$\,Sct stars
  and provide evidence that the relation 
  features 
  different regimes
  depending on the oscillation period. We demonstrate 
  \OK{how} \OK{stellar rotation attenuates the amplitude of the dominant oscillation mode of}
$\delta$\,Sct stars.
}
{\OK{The \gdr3 time-series photometry already allows for the detection of the dominant (non-)radial oscillation mode in 
\OK{about 100 000} 
intermediate- and high-mass dwarfs
\OK{across the entire sky}. This detection capability will increase as the time series becomes longer, allowing the additional delivery of frequencies and amplitudes of secondary pulsation modes.}
}


\titlerunning{Gaia's view on main sequence OBAF-type pulsators}
\authorrunning{De Ridder et al.}

\maketitle

\section{Introduction}
\label{sec:introduction}

Intermediate- and high-mass (IHM) main sequence stars ($M\geq 1.3\,$M$_\odot$,
spectral types O, B, A, and F) have convective cores whose physical conditions cannot 
be extrapolated from dwarf stars like our Sun. The transport processes caused by 
internal convection, rotation, and magnetism in IHM dwarfs are far less understood 
than those in Sun-like stars, yet these processes have a large impact on their evolution 
and their age determination. A large fraction of the IHM stars are oscillators located 
in the $\beta$\,Cep, slowly pulsating B (SPB), $\delta$\,Sct, or $\gamma$\,Dor 
instability strips in the Hertzsprung-Russell (HR) diagram.
\referee{$\beta$\,Cep, SPB, and $\delta$\,Sct stars all exhibit self-excited oscillation 
modes caused by the conversion of thermal energy into mechanical energy by the so-called 
$\kappa$-mechanism \citep{Pamyatnykh1999}. $\gamma$\,Dor oscillations are driven by a
combination of the $\kappa$-mechanism on one hand, and radiative flux modulation caused by 
convection at the bottom of the convective envelope on the other hand. The first mechanism
plays a more important role for the hotter $\gamma$ Dor stars, while the second one is
predominant for the cooler $\gamma$ Dor stars \citep{Guzik2000,Dupret2005,Xiong2016}}. 
For \referee{all four types of these pulsators}, asteroseismology is a particularly promising tool 
to investigate their internal physics \citep[e.g.][]{Aerts2021}.

Space telescopes like CoRoT \citep{Auvergne2009}, {\it Kepler\/} \citep{Koch2010}, and 
TESS \citep{Ricker2016} 
\OK{were}
dedicated to the gathering of long uninterrupted high-cadence photometry with 
$\mu$mag precision. \OK{This type of photometric light curve assembled from space}
brought unprecedented capabilities to test and calibrate the 
theory of stellar structure for stars of various masses and evolutionary stages 
\citep[see][for extensive observational reviews]{HekkerJCD2017,Garcia2019,Kurtz2022}. 
\OK{Thanks to
these high-quality data, the high-dimensional problem of 
asteroseismic modelling of IHM stars becomes achievable}
\citep{Aerts2018}. 
\OK{Asteroseismic modelling requires}
identification of the pulsation
modes in terms of their 
\OK{spherical wave numbers $(\ell,m)$ and radial order $n$.}
\OK{Compared to the simple case of Sun-like stars,}
the Coriolis force as well as the 
non-linear interplay between rotation, convection, and magnetism adds a high level of complexity 
\OK{to the modelling of $\beta$\,Cep, SPB, $\gamma$\,Dor, 
and $\delta$\,Sct stars.}
This is why detailed asteroseismic modelling has so far only been carried out for a few tens 
of such IHM dwarfs, and with inhomogeneous approaches and levels of detail
\citep{Degroote2010,Kurtz2014,Saio2015,Murphy2016,Schmid2016,Deng2018,Szewczuk2018,
Mombarg2019,Wu2020,Mombarg2020,Mombarg2021,Pedersen2021,Sekaran2021}.

In this paper, we present a performance verification of Gaia DR3 in terms of
the mission's capacity to detect radial and non-radial oscillations of oscillating
IHM stars on the main sequence. The Gaia space mission was primarily designed as 
an astrometric mission to
reach $\mu$as precision. To reach this goal, the CCD detections of the sources do
not need to have a very high signal-to-noise ratio (S/N) per field of view transit, nor does
the sampling need to be dense.

As a result, Gaia photometric time series of an
individual star were not meant to be particularly attractive for asteroseismology.
The median number of measurements per star is only around 50 in \gdr3 and their time
sampling is sparse. The photometric precision in the \g passband is of the order
of a mmag. This compares excellently with ground-based surveys, but is 
not (and never was intended to be) able to compete with the photometric precision of
dedicated space missions like {\it Kepler\/} or TESS 
\OK{(cf.\,Appendix\,A)}. 
However, the main advantage
of the Gaia mission for asteroseismology is the sheer number of sources 
for which it gathers data. Unlike the 
\OK{{\it Kepler\/} or TESS} 
space missions focusing on a relatively small part of the sky or only on brighter stars, 
\OK{respectively,}
Gaia is an all-sky survey
covering stars spanning more than 12 orders of magnitude in brightness.
In addition, its high angular resolution allows us to also probe the denser parts of
the sky.  
Therefore, Gaia enables the detection of oscillations in a far larger 
number of variable stars. 
Gaia already proved to be an excellent instrument to detect radially oscillating
high-amplitude variables \citep[e.g.][]{Clementini2019,Mowlavi2018,Ripepi2022,Eyer2019}, 
but 
\OK{its capacity}
has not yet been established for the low-amplitude 
main sequence IHM stars.
While the $\mu$mag-precision space missions so far resulted in asteroseismic 
modelling of thousands of red giants \citep{Hon2018,Yu2018,Yu2020,Yu2021,Stello2021}
and of hundreds of Sun-like dwarfs with masses $M\leq\,1.3\,$M$_\odot$ \citep{Garcia2019},
the samples of asteroseismically modelled IHM dwarfs are much smaller and do
not yet fully cover the parameter space in terms of rotation, binarity, and metallicity. 
It is in this area that Gaia's time series 
\OK{is}
helpful 
\OK{and will allow for the discovery of thousands of}
oscillating IHM
candidates, 
\OK{with the aim being to achieve unbiased and complete samples. These will then become 
suitable for ensemble asteroseismology in the future, once dedicated high-cadence 
high-precision monitoring for their members has become available.}

In terms of impact, Gaia DR3 (and later data releases)
may lead to the discovery of tens of thousands of new IHM pulsating dwarfs,
delivering the frequencies and amplitudes of their dominant (non)radial
oscillation mode(s), aside from a high-precision parallax.  Here, we assess
Gaia's performance on this front, offering 
\OK{as well}
an observational census of the
instability strips along the upper main sequence for samples of new
pulsating IHM dwarfs with $M\geq 1.3\,$M$_\odot$. The borders of those
instability strips depend on the physics of internal rotation, gravitational
settling, radiative levitation, shear mixing, and so on, as well as on the bulk metallicity
and possible binarity at birth. Given the complex (often non-linear)
interplay of these astrophysical phenomena during the evolution and the initial
conditions at birth, samples of thousands of (non-)radial pulsators are required
to interpret the borders of the instability strips in terms of the
excitation physics 
\OK{\citep{Townsend2005,Szewczuk2017} and transport processes \citep{Aerts2019}}.

This paper is one among several publications led by the Gaia Data Processing
Analysis Consortium (DPAC) accompanying the Gaia DR3. 
We present the Gaia photometric dataset of variable IHM dwarfs in
Sect.\,2 and develop our strategies to detect \referee{IHM} main sequence
pulsators in Sect.\,3, distinguishing non-radial gravity-mode (g mode) pulsators
from (non-)radial pressure-mode (p mode) pulsators. We compare the samples of IHM
dwarf pulsators with those in the literature in Sect.\,4 and move on to Gaia's
performance in Sect.\,5, highlighting some dedicated applications. We conclude in
Sect.\,6 with an explicit encouragement and invitation to the worldwide
astrophysics community to exhaustively exploit the Gaia time-series data in its full glory.

\section{The dataset}
\label{sec:dataset}

Within the \gaia Data Processing and Analysis Consortium (DPAC), the CU7 coordination unit responsible for analysing variable
stars processed a grand total of 1\,840\,651\,642 sources, of which 25\% of the most variable sources were selected for further investigation. 
This variability was tested in the time domain rather than the Fourier domain for computational reasons, 
as described in \citet{Eyer2022}. These variable sources were then further characterised in the Fourier domain and 
classified by multiple classifiers each with their own setup \citet{Rimoldini2022a}. We also refer to the documentation provided by \citet{Rimoldini2022b} 
for further processing details.

For this paper, we started with those 450\,605 variable sources from the \gdr3 table \texttt{vari\_classifier\_result} 
with at least 40 photometric measurements in the \gaia \g-band,
and which the CU7 classification team classified as either $\beta$\,Cep, $\delta$\,Sct, SX Phe, SPB, or $\gamma$\,Dor \OK{star}. 
No constraints were put on the apparent \g magnitude. 
To this sample we added the 54\,476 \referee{IHM main sequence 
pulsator} candidates from the \gdr3 table \texttt{vari\_ms\_oscillator} in 
\gdr3, which used a different set of classification criteria.
We filtered out any duplicates common to both samples
\OK{to end up with a sample of 460\,519 candidate main sequence variables with at least 40 \gaia \g data points. }
Both classifications mentioned here are based on a list of attributes
that includes the main frequency $\nu_1$ found in the generalised Lomb-Scargle \OK{periodogram} \citep{Zechmeister2009}.
\referee{This main frequency optimises the fit of the following model to the data:}
\begin{equation}
    G_n = C + A \sin(2\pi\nu_1 t_n + \varphi)
,\end{equation}
\referee{where $G_n$ is the observed \gaia \g magnitude at time $t_n$, $C$ is a constant term, and $A$ and $\varphi$ are 
the amplitude and phase of the sine wave, respectively.}

\OK{The former classification}
did not automatically exclude those variables for which $\nu_1$ is not statistically significant.
\OK{Hence, for this sample} the classification may be partly based on  \OK{detected variability} 
that has no astrophysical origin. For this reason we imposed a first threshold on the false
alarm probability \citep[FAP;][]{Baluev2008} of $\nu_1$, which we limit to be no greater than $0.01$. 

\begin{figure}
    \resizebox{\hsize}{!}{\includegraphics{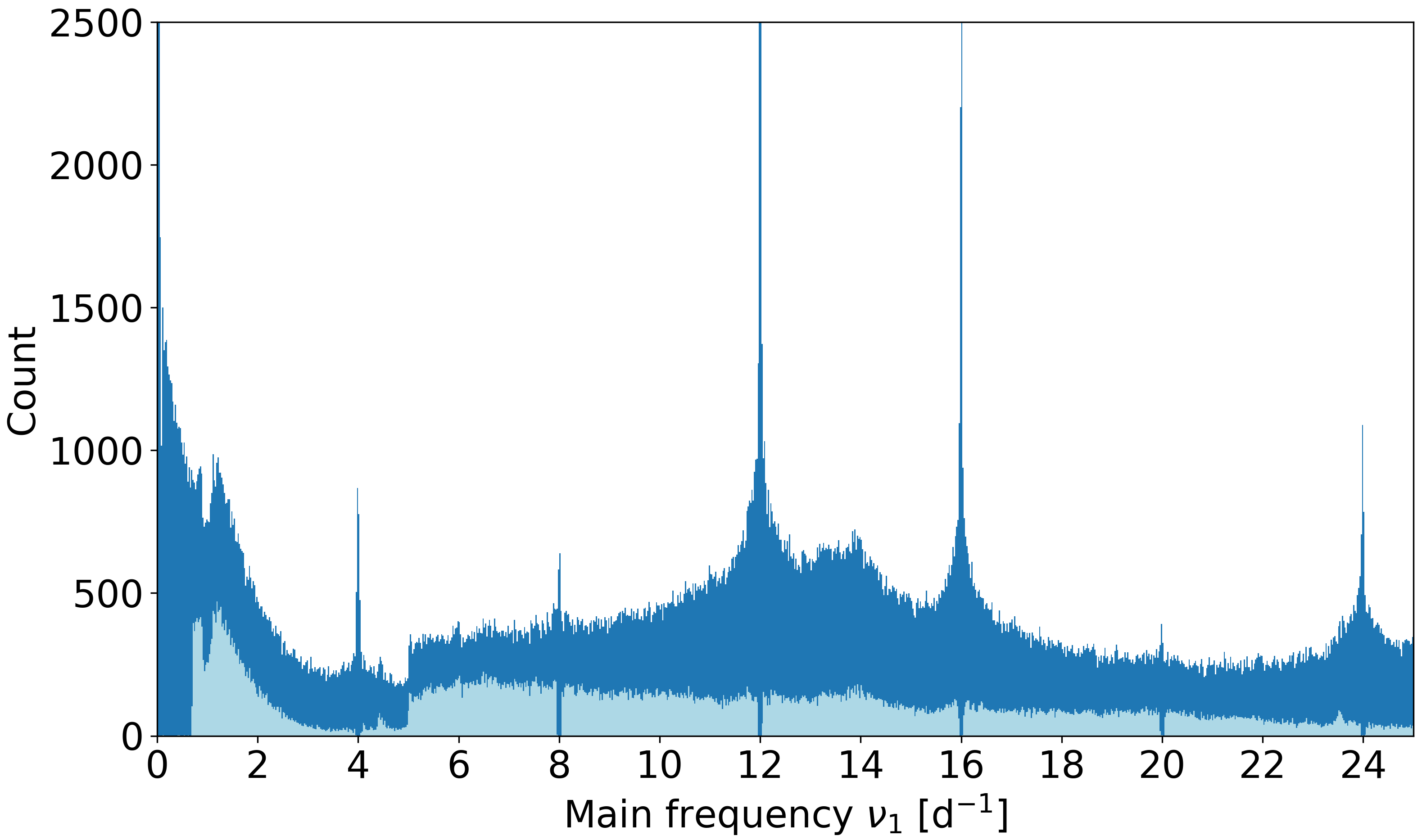}}
    \caption{A histogram of the first frequency $\nu_1$.  
             Dark blue: histogram of the first frequency $\nu_1$ of the 460\,519 candidate main sequence variables. 
             The highest peaks go beyond the maximum value of the y-axis. Light blue: Same histogram for the 108\,663 
             for which we filtered on the primary frequency $\nu_1$ and its corresponding FAP value to avoid 
             instrumental artefacts. \referee{Because of the 34-month time-span of the DR3 time series, the typical 
             uncertainty on the frequency is about $10^{-3}$ $d^{-1}$. }}
    \label{fig:f1_largesample}
\end{figure}

Examination of the time series revealed that a stricter FAP
threshold is required for some cases. The histogram of the first 
frequency $\nu_1$ for our sample of stars is shown in Fig.\,\ref{fig:f1_largesample} in dark blue. This histogram shows that, 
although the quality of the photometric data reduction is steadily and significantly improving with every data release, 
the photometric data are not yet completely free of instrumental artefacts. We see high peaks around harmonics of 4\,d$^{-1}$, 
which is the spin frequency of the spacecraft \citep[cf.\,][]{Prusti2016}. The strong wings around these peaks suggest that 
sources in a fairly broad frequency range are affected. For these stars, the Fourier analysis therefore extracted the signature 
of an instrumental variation rather than an astrophysical variation.
\citet{Holl2022} and \citet{Distefano2022} also observed instrumental periodic variations in the
photometric data of \gdr3, be it while focusing on a different frequency range ($\nu < 0.5$ d$^{-1}$) from that relevant for
this article. Although these authors were not able to clean the affected time series from instrumental variations, 
they were able to efficiently filter out the impacted sources. Their filtering method proved not as efficient for our
sources, casting doubt on whether the instrumental frequencies observed at very low frequencies are of the same origin
as those we observe at high frequencies. There are multiple possible reasons for this; in addition to genuine instrumental variations, 
some of the peaks observed in Fig. \ref{fig:f1_largesample} might be caused by frequency aliasing of very low frequencies 
into high frequencies.
For the final release (Gaia DR5), we expect the instrumental variations 
to be mostly calibrated out. 

To mitigate these variations for the benefit of the current 
\OK{paper,} we experimented with many different observables 
to filter out the affected sources. The only effective one proved to be a stricter FAP threshold, 
which we limit to be no greater than $10^{-3}$ for sources with $\nu_1 \in [4, 25]$\,d$^{-1}$.  
This strongly reduces the wings of the high peaks seen in 
Fig.\,\ref{fig:f1_largesample}. It does not completely eliminate the 
peaks themselves, and so we also removed all sources with a frequency $\nu_1$ near 4, 8, 12, 16, 20, and 24\,d$^{-1}$ using 
a proximity threshold of 0.05\,d$^{-1}$. Lastly, we also discarded those sources with $\nu_1 < 0.7\,$d$^{-1}$ 
\OK{to avoid dominant long-term variability of unknown origin, because 
\OK{most} 
of the dominant g modes found in \gaia DR3 data of the best characterised 63 bona fide {\it Kepler\/} $\gamma\,$Dor and SPB stars have frequencies above this value (cf.\,Appendix\,A).}
The histogram of $\nu_1$ of the resulting sample of 108\,663 sources is shown in light blue in Fig.\,\ref{fig:f1_largesample}.

Looking 
\OK{further at} 
the main frequency $\nu_1$, we find that the CU7 classifiers 
\OK{assign}
some of the stars to the
$\beta$\,Cep, SPB, $\gamma$\,Dor, or $\delta$\,Sct 
\OK{classes} even if 
$\nu_1$ is far from the 
\OK{typical}
frequency ranges associated with 
\OK{the dominant modes of such stars in this}
particular type of variables, even 
\OK{when taking into account} frequency shifts due to rapid
rotation. 
\OK{An} underlying reason is that the classifiers did not 
\OK{assign sufficient} 
weight to the main frequency
\OK{of the variability}.
In a next step we therefore further filtered the main frequency $\nu_1$ and 
only retained stars classified as $\beta$ Cep stars with $\nu_1 \in [3, 8]$\,d$^{-1}$
\citep{Stankov2005}, 
stars classified as $\delta$\,Sct stars or SX Phe stars with $\nu_1 \in [5, 25]$\,d$^{-1}$
\citep{Rodriguez2000a,Rodriguez2000b}, 
stars classified as $\gamma$\,Dor stars with $\nu_1 \in [0.7, 3.2]$\,d$^{-1}$
\citep{GangLi2020}, 
and stars classified as SPB stars with $\nu_1 \in [0.7, 5]$\,d$^{-1}$ \citep{Pedersen2021}, which
\OK{ leaves a sample of 106\,207 variables.}

\begin{figure}
    \resizebox{\hsize}{!}{\includegraphics{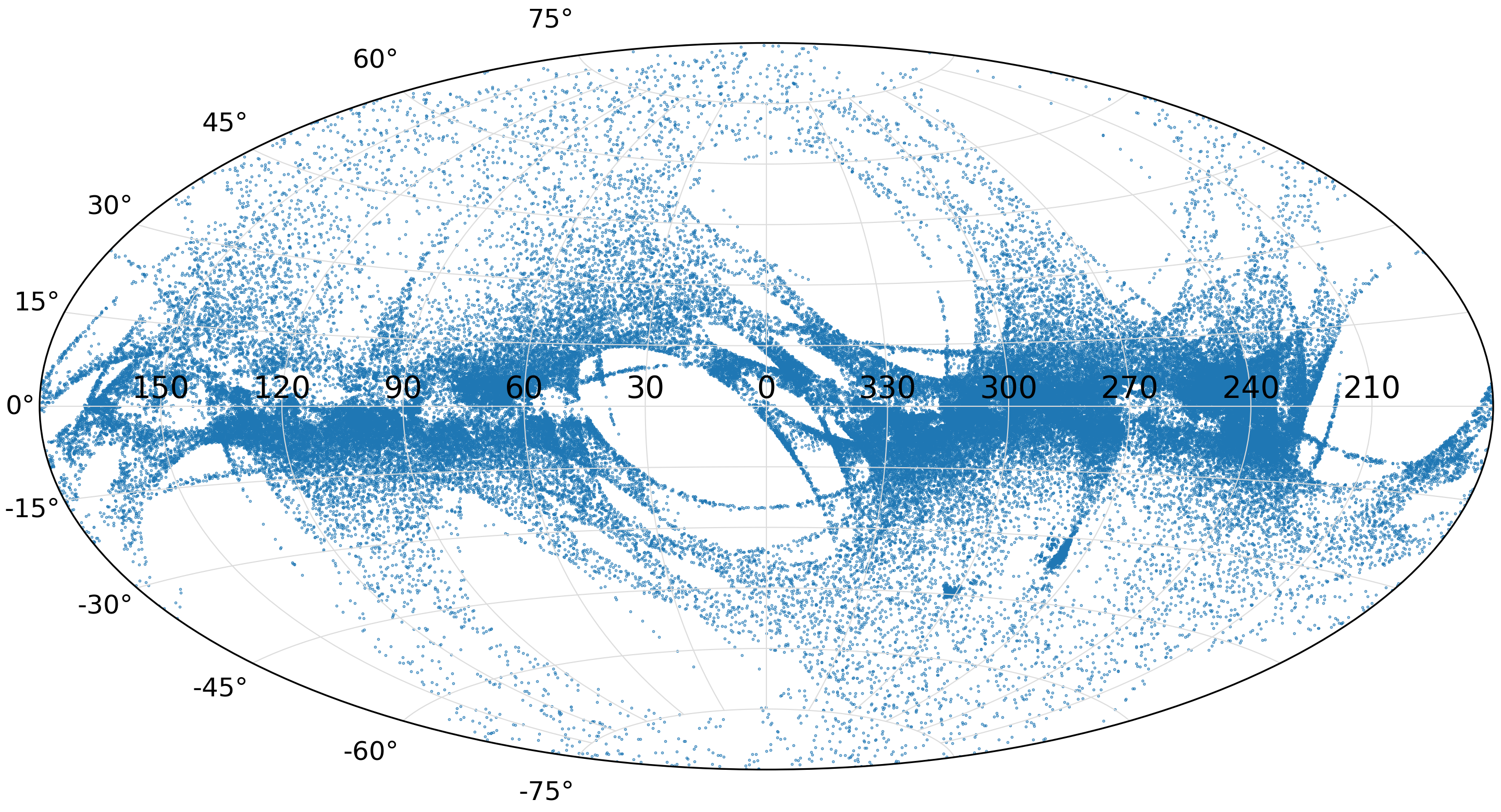}}
    \caption{Sky distribution in Galactic coordinates of the 106\,207 sources in the data set \OK{resulting from various steps of filtering as} described in
             Section \ref{sec:dataset}.}
    \label{fig:mso_skydistrib}
\end{figure}

\begin{figure}
    \resizebox{\hsize}{!}{\includegraphics{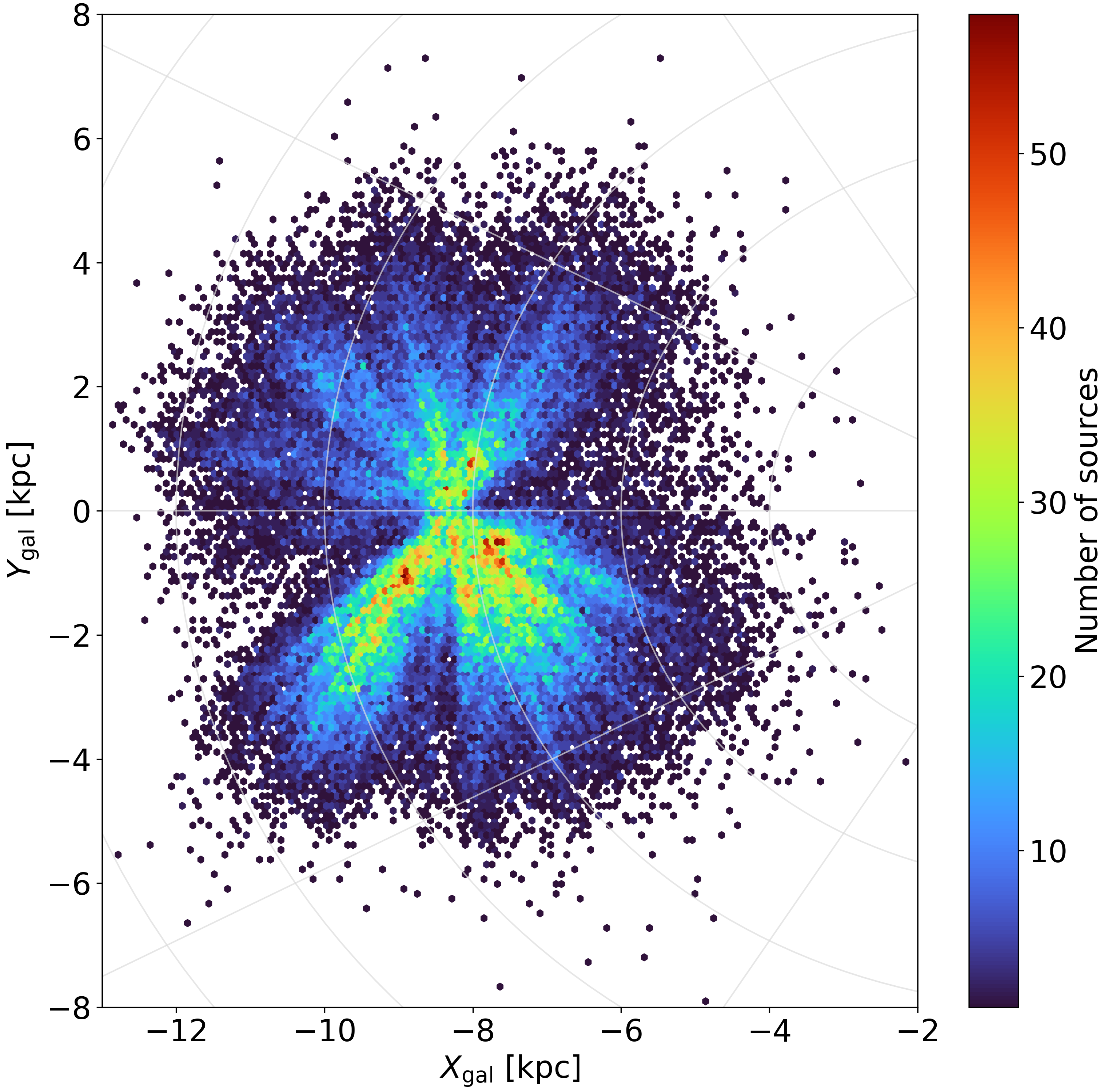}}
    \caption{Distribution of the Galactic sources of the dataset 
    \OK{shown in Fig.\,\ref{fig:mso_skydistrib}, limited to those 71\,490 stars with}
             $0 < \sigma_{\varpi}/\varpi \le 0.25$, projected on the equatorial plane of the 
             Galactocentric reference frame.}
    \label{fig:mso_3Ddistrib}
\end{figure}

Figure\,\ref{fig:mso_skydistrib} shows the sky distribution of our sample 
in galactic coordinates, 
and Fig.\,\ref{fig:mso_3Ddistrib} shows the projection on the $XY$ galactocentric plane of the subset of 
71\,490 sources with 
a relative parallax uncertainty $0 < \sigma_{\varpi}/\varpi < 0.25$. 
The galactocentric coordinates $(X_{\rm gal}, Y_{\rm gal})$ were computed using the median photogeometric distances computed 
from Gaia EDR3 data by \citet{BailerJones2021}.

\begin{figure*}
  \includegraphics[width=\textwidth]{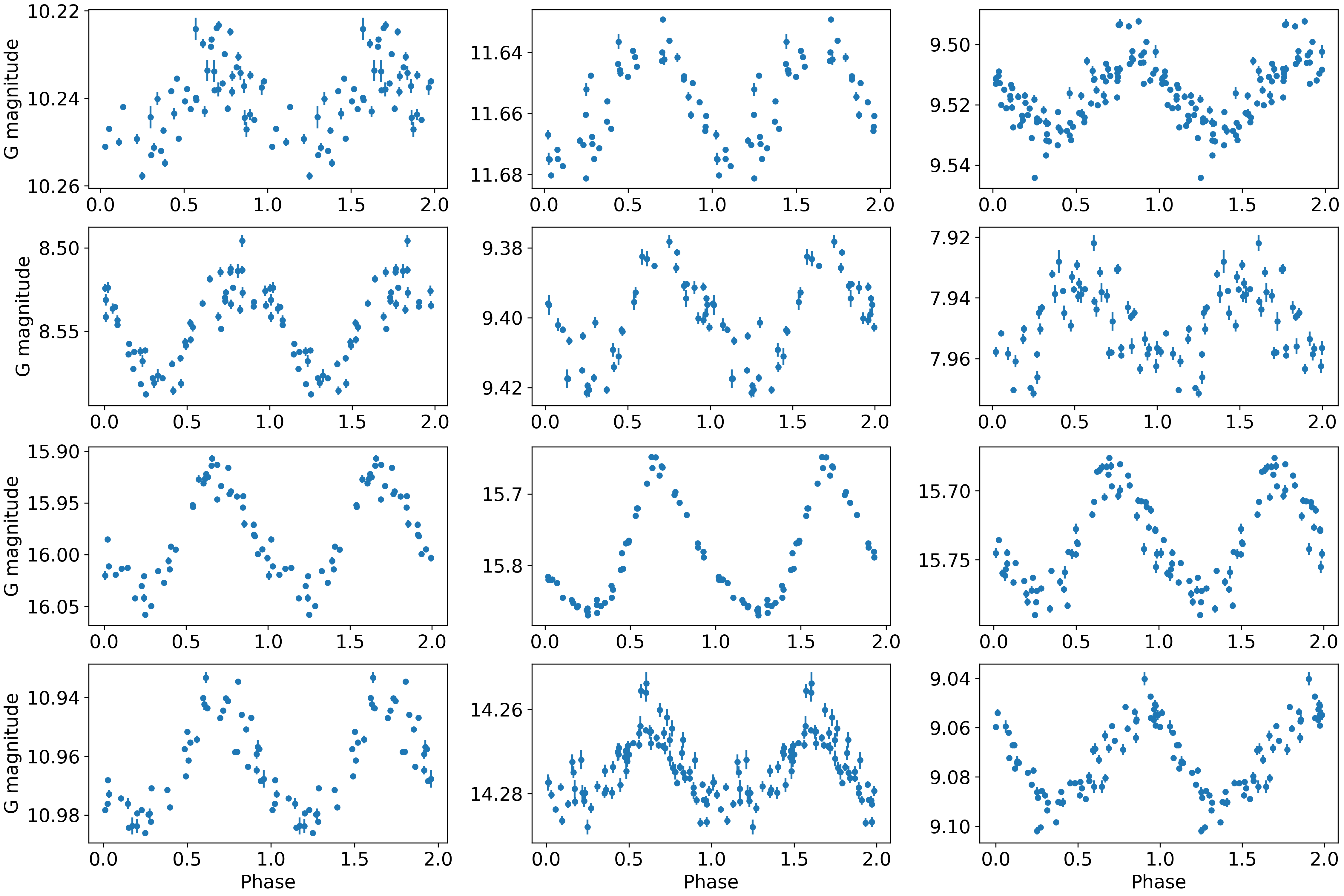}
  \caption{Example phase diagrams of sources classified as main sequence OBAF-type pulsators, 
           folded with the primary frequency $\nu_1$ found in Gaia \g-band photometry. 
           From top to bottom the rows show three examples of stars classified as
           respectively $\beta$\,Cep, SPB, $\delta$\,Sct, and $\gamma$\,Dor variables.}
  \label{fig:example_phasediagrams}
\end{figure*}

The gaps in the spatial distribution are caused by \gaia's scanning law and the 
lower constraint on the number of time points ($N \ge 40$) for our sources. 
Finally, Fig.\,\ref{fig:example_phasediagrams} shows examples of phase diagrams of high-S/N light curves for 
each of the OBAF-type pulsators.

\section{The IHM instability strips as observed by Gaia}
\label{sec:instabstrips}

\OK{In a global sense,} pulsation theory successfully predicts the occurrence of 
\OK{modes in} 
OBAF-type main sequence stars in the HR diagram 
\citep[see e.g.][and references therein]{Dupret2005, Townsend2005,Miglio2007, Daszynska-Daszkiewicz2013,  Xiong2016, Szewczuk2017,Burssens2020}.
\OK{However, discrepancies between mode excitation predictions 
and observations occur, both at the level of the observed samples as a whole and for the various detected modes in individual pulsators. As an}
example \OK{of the former}, \citet{Murphy2019} found that the observational instability 
strip of about 2000 $\delta$\,Sct stars detected in \textit{Kepler} data covered
a broader temperature regime than predicted by state-of-the-art pulsation theory
based on time-dependent convection theory 
\OK{
\citep{Dupret2005}. 
\OK{For individual hotter stars,} two examples of IHM pulsators with 
detected modes predicted not to be excited were discussed in great detail by \cite{Daszynska2017} 
and \citet{Moravveji2015} for a $\beta\,$Cep and SPB star, respectively.}
\OK{These two B-type pulsators are representative of numerous additional cases, pointing to overly low opacities for iron peak elements used as input physics for the computation of stellar models.}
Moreover, even in state-of-the-art stellar models, very approximative treatments of internal rotation 
and core convection are used, leading to unreliable predictions about their non-radial oscillations 
\citep{Aerts2021}. 
On the other hand, from the observational side, the empirical location of 
\OK{IHM} stars in the
HR-
\OK{or Kiel diagrams is often plagued by large systematic uncertainties on the observed stellar distance, 
interstellar reddening, and bolometric corrections \citep{Pedersen2020}, as well as on the star's gravity, 
helium abundance, metallicity 
\OK{($z$)}, and, to a lesser extent,}
effective temperature \OK{\citep{Burssens2020,Gebruers2021}.}

For the present Gaia sample, we face the additional challenge that the classification 
was 
based on one frequency only, which makes the presence of contaminants in our sample
likely. In particular, rotationally modulated stars are prevalent at spectral 
types A and B 
\citep[e.g.~Bp and Ap stars,][]{Kurtz2000, Briquet2007, 
Balona2011b, Bowman2018a}. 
\OK{Bp and Ap stars tend to be
slow to moderate} rotators showing 
surface spots, leading to variability in the same frequency domain as g-mode pulsations. 
On top of that, although the window function of Gaia time series is significantly better
than those of typical ground-based surveys, the highest peak in the normalised 
\OK{spectral window}
\OK{of the \gaia DR3 time series}
can still reach 60\% or more for certain regions in the sky. 
\OK{
We cannot therefore exclude
that a fraction of the low-frequency oscillators are actually $\delta$\,Sct or 
$\beta$ Cep stars for which an alias frequency was found instead of the true frequency.
Likewise,} high-frequency oscillators might have been 
classified as SPB or $\gamma$\,Dor star
because of frequency aliasing (cf.\,Appendix\,A).

For our sample of OBAF-type IHM stars, we 
\OK{extract} the effective temperature $T_{\rm eff}$ 
from the field \texttt{teff\_gspphot} in the table \texttt{astrophysical\_parameters\_supp} 
in \gdr3. The luminosities were computed using the absolute \g magnitude 
\texttt{mg\_gspphot} field in the same table, and a bolometric correction BC (depending on 
$T_{\rm eff}$) from \citet{Pedersen2020} for stars hotter than 10000 K and from 
\citet{Eker2020} for the cooler stars. In both cases, we used the metallicity-independent
BC prescription, as we do not have a reliable estimate of $[M/H]$ for most of our stars.
\referee{We propagated the errors on $T_{\rm eff}$ and $M_G$ to obtain the errors on the 
bolometric corrections and the luminosities. The typical $T_{\rm eff}$ uncertainty of
the stars in our sample as reported in table \texttt{astrophysical\_parameters\_supp} 
is about 150 K, leading to a typical uncertainty of 0.08 on $\log L$. However, as 
\citet{Andrae2022} discusses in depth, the uncertainties on $T_{\rm eff}$ and $M_G$
in the \gdr3 release are likely underestimated, in some cases by a factor of three. 
Given the experience of ground-based spectroscopy, 
we can expect a minimum uncertainty of about 400 K on $T_{\rm eff}$,  especially for the hotter
B-type pulsators. 
}

\begin{figure}
  \resizebox*{\hsize}{!}{\includegraphics{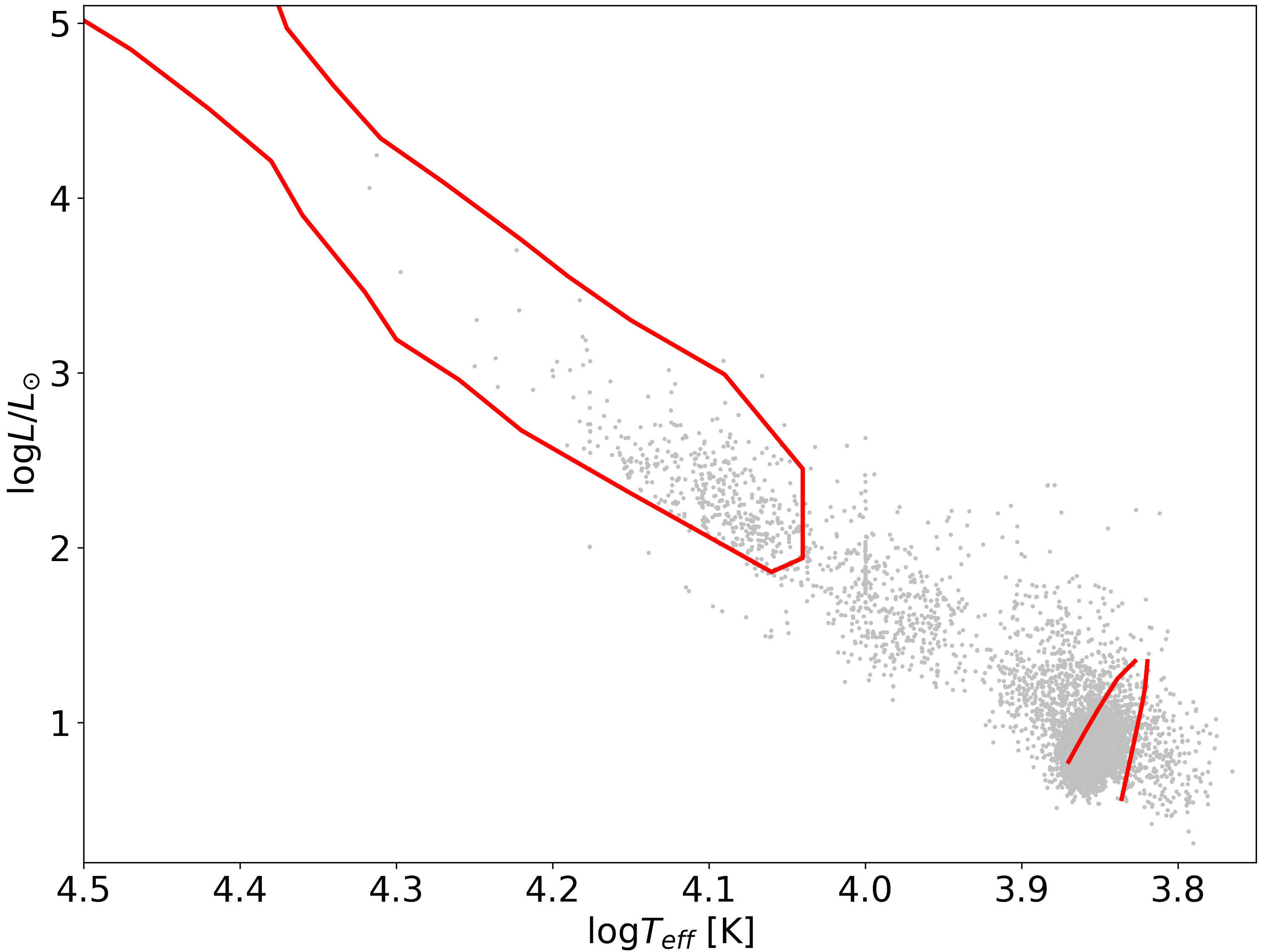}}
  \caption{HR diagram of 10\,402 nearby \OK{g-mode pulsators} 
  classified as SPB or $\gamma$\,Dor star with a relative
           parallax uncertainty of better than 3\%. The red line is the SPB instability
           strip from \OK{\citet[][at hot temperatures]{Burssens2020}}
and the $\gamma$\,Dor instability strip
from \citet[][at low temperatures]{Dupret2005}, both for solar metallicity.}
\label{fig:hrdiagram_spbgdor_p3}
\end{figure}

As our sample includes stars from magnitudes \g=6 to \g=18, and the uncertainty on the 
\texttt{gspphot} stellar parameters is often too large to 
\OK{make them useful for placement} in the HR diagram, 
we limit ourselves to the nearby IHM pulsators with a relative \gdr3 
parallax uncertainty 
$0 < \sigma_{\varpi}/\varpi < 0.03$. 
Figure~\ref{fig:hrdiagram_spbgdor_p3} shows 
a HR diagram with the nearby \gaia sources classified as $\gamma$\,Dor 
or SPB as described in Section \ref{sec:dataset}. In the same figure, the SPB-star instability 
strip from 
\OK{\citet{Burssens2020}}
is plotted ($\ell \in \{0,1,2\}$, $n\in\{-50,\cdots,-1\}$, $Z=0.02$) along with the $\gamma$\,Dor 
instability strip ($\ell=1$, $Z=0.02$)  
from \citet{Dupret2005} \OK{for a value of the mixing length, $\alpha_{\rm MLT}=2$.}

\begin{figure}
  \resizebox*{\hsize}{!}{\includegraphics{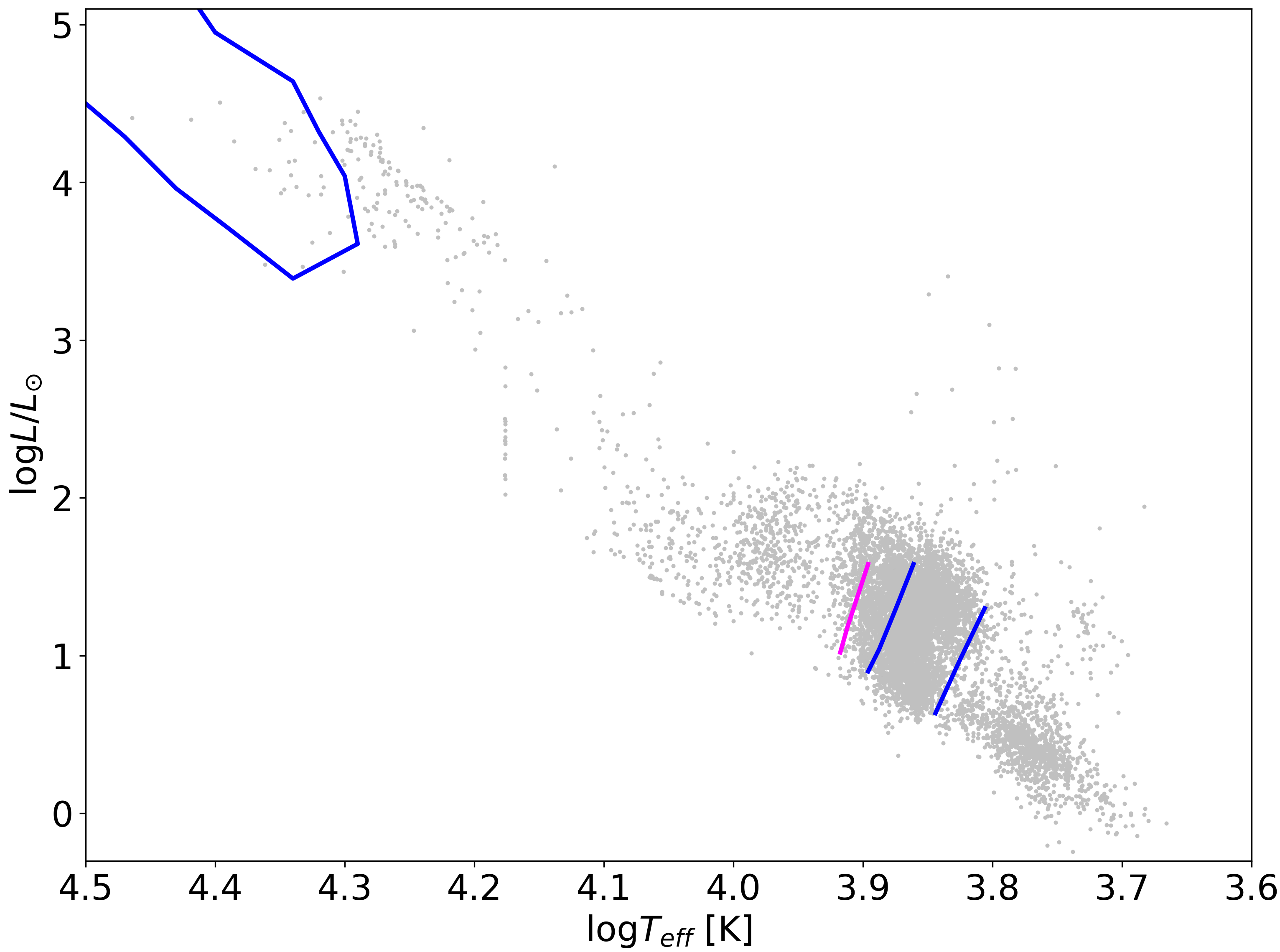}}
  \caption{HR diagram of 13\,974 nearby \OK{p-mode pulsators} 
  classified as $\beta$\,Cep or $\delta$\,Sct stars.
  For the $\beta$\,Cep stars, we imposed $0<\sigma_{\varpi}/\varpi<0.2$, \OK{while we demanded} $0<\sigma_{\varpi}/\varpi<0.03$ 
  for the $\delta$\,Sct stars. 
  The blue line at hot temperatures is the $\beta$ Cep star instability strip \OK{from \citet{Burssens2020} 
  for spherical degree $\ell \in \{0,1,2\}$ and radial order $n\in\{0,\cdots,6\}$ . 
  The blue lines represent the blue and red edge of the $\delta$\,Sct instability strip for $n=1$ modes, and the fuchsia 
  line the blue edge for $n=4$ modes. All instability strips are for solar metallicity.}} 
  \label{fig:hr_diagram_highfreq_p3}
\end{figure}

The highest concentration of \OK{g-mode}
pulsators is, as expected, inside the
$\gamma$\,Dor instability strip. 
The slight offset towards lower luminosities is easily explained as
due to systematic uncertainties in the reddening or bolometric correction BC.
The low-frequency variables 
\OK{blueward of the $\gamma$\,Dor instability strip fall
inside 
\OK{or above} the $\delta$\,Sct instability strip 
(cf.\,Fig.\,\ref{fig:hr_diagram_highfreq_p3}).}
Aside from the possibility that they might be 
\OK{hot $\gamma$\,Dor stars \citep{GangLi2020}
or $\gamma\,$Dor stars} with
an uncertain location in the HR diagram, they could also represent 
\OK{hybrid p- and g-mode $\gamma\,$Dor/$\delta\,$Sct pulsators \citep{Grigahcene2010,Bowman2016},} 
aliased $\delta$\,Sct stars, 
Bp/Ap stars \citep{Balona2019}, 
\OK{fast-rotating SPB stars \citep{Townsend2005,Szewczuk2017},}
or rotationally modulated g-mode pulsators following the recent 
pulsation excitation theory for fast rotating 
early-type stars \citep{Ouazzani2020,Lee2020,Lee2021}. This latter
theory predicts g-mode pulsators to occur in rapid rotators between the SPB and $\gamma\,$Dor instability strips, 
that is,~at effective temperatures representative of spectral types A0-A3. 
This concerns fast rotating spotted stars with gravito-inertial 
envelope modes coupled to inertial core modes as observed in 
{\it Kepler\/} observations \citep{Saio2018}.

\OK{Occurrences of rapidly rotating g-mode pulsators between the SPB and $\delta\,$Sct instability strips are 
well established from modern \OK{space} photometry. Indeed,
after initial discoveries from CoRoT data of SPB 
stars \citep{Degroote2009,Degroote2011}, numerous
detections were achieved from ground-based
cluster studies as well \citep{Saesen2010,Saesen2013,Mowlavi2013,Mozdzierski2014,
Mowlavi2016,Saio2017-PL,Mozdzierski2019}. 
\OK{These detections are interpreted as fast-rotating SPB stars from instability computations based on the so-called traditional approximation of rotation applied to g modes \citep{Salmon2014}, following the interpretation by \citet{AertsKolenberg2005} based on spectroscopic time-series data.} 
Many fast-rotating SPB and Be pulsators were also found near the red edge or below the SPB instability strip 
from {\it Kepler\/} and 
TESS data \citep[]{Balona2011a,Pedersen2019,Burssens2020,Sharma2022}.
Our Gaia results in Fig.\,\ref{fig:hrdiagram_spbgdor_p3} reveal
plenty of g-mode pulsators between the SPB and 
$\delta$\,Sct instability strips. Within this group of g-mode pulsators,} 
we observe an {under-}density between the hot A stars 
and the blue edge of the $\delta$\,Sct instability strip.
At this point, we cannot exclude that this is 
an artefact rather than having an astrophysical origin.

\OK{The distribution of 
stars within the SPB instability strip 
in Fig.\,\ref{fig:hrdiagram_spbgdor_p3}
is concentrated in or near the cooler bottom half, 
in close agreement with mode density excitation predictions \citep[cf.\,Fig.\,1 in ][]{Papics2017}. 
These g-mode pulsators are in the same position as those observed in high-cadence space photometry 
of slow- and fast-rotating B and Be stars \citep{Neiner2009,Huat2009,Diago2009,Neiner2012,Baade2016,
White2017,Baade2018a,Baade2018b,Pedersen2019,Burssens2020,Pedersen2021,Szewczuk2021,Sharma2022}. 
It is noteworthy that the \gaia CU7 classification scheme also has a class of $\gamma\,$Cas variables 
representing classical Be stars \citep{Rivinius2013}. It has long since been known from line-profile 
variability that many Be stars are rapidly rotating non-radial p- and/or g-mode pulsators 
\citep{Rivinius2003}. High-cadence space photometry indeed reveals most of 
the $\gamma\,$Cas variables to have non-radial pulsations aside from rotational 
modulation \citep{Balona2011a,Balona2016}, including the prototype $\gamma\,$Cas itself 
\citep{Labadie2021,Smith2021}. We did not include the $\gamma\,$Cas variables in our study, 
because this class definition within CU7 does not rely on them being non-radial pulsators. 
We therefore anticipate that many more rapidly rotating non-radial B-type pulsators occur in \gaia DR3, 
but were assigned to the $\gamma\,$Cas class.}

In Fig.\,\ref{fig:hr_diagram_highfreq_p3} we show the HR diagram dedicated to
the stars that were classified as either $\beta$\,Cep or $\delta$\,Sct
\OK{star}s. As before,
we imposed a relative parallax of better than 3\% for the $\delta$\,Sct stars. For the
$\beta$ Cep stars, which are much more rare, we relaxed this constraint to 
$0<\sigma_{\varpi}/\varpi<0.2$.   
The vast majority of $\delta$\,Sct stars are well concentrated in the expected instability
strip \OK{of p modes}, 
demonstrating both the good performance of the variability classifier as well as the
precision of the astrophysical parameter determination by the CU8 processing pipeline, at
least for the nearby stars. 
\OK{Higher up the main sequence,} the 
sample of nearby very luminous $\beta$\,Cep stars is small.
Most of them have an observed location in the HR diagram 
outside the expected instability 
strip, which is difficult to explain 
\OK{in terms of metallicity.} 
The most likely explanation is a systematic bias in the astrophysical parameters of hot stars as derived 
from \gdr3 \texttt{gspphot} data.
\OK{Indeed, 
the observed shifts to cooler temperatures can easily occur due to poorly estimated reddening. 
However, the effects of fast rotation may also play a role, as it is not or only incompletely treated in 
published instability strips for p modes, while its impact on the properties of the modes is large 
\citep{Daszynska2002}. Many fast rotators with p modes have indeed been found in high-cadence space 
photometry of early-type B and Be stars \citep{Balona2011a,Burssens2020,Balona2020}.}

The classifier also picked up a sample of stars with high
frequencies $\nu_1$ right below the SPB instability strip, with $T_{\rm eff}$ around 
10000 K or slightly cooler. 
\OK{These variables are in line with CoRoT and TESS discoveries of what appears to be p-mode pulsators 
in that position of the HR diagram \citep{Degroote2009,Balona2020}. The 
variability of such pulsators is not yet well understood. We note that the frequencies of 
gravito-inertial modes in moderate to fast rotators can easily get shifted into the regime of p modes 
\citep[e.g.][]{AertsKolenberg2005,Salmon2014}. Such frequency shifts may be
particularly large for g modes with 
$|m|>1$, which complicates the classification of the variability based on the frequencies from photometry 
alone. This is illustrated nicely
by the two cases of the slowly rotating variable star Maia \citep{Struve1955} and the 
magnetic moderately rotating SPB pulsator HD\,43317 \citep{Papics2012}. While Maia 
was originally classified as a pulsator by \citet{Struve1955} and even led Struve to introduce a 
seemingly new class of pulsators (the so-called Maia variables), \citet{White2017} showed it to be 
a rotationally modulated star rather than a pulsator. On the other hand, 
 the high-frequency mode of HD\,43317  at 4.3\,d$^{-1}$ was originally misinterpreted as a p mode, while 
it is a rotationally shifted quadrupole g mode \citep{Buysschaert2018}. These two examples reveal 
that misclassifications of B-type variables based on photometric data alone  are easily made and 
that correct interpretation of the   frequencies of such stars in the regime of a few to several d$^{-1}$ 
is only possible from adding independent data ---such as high-precision spectroscopy--- to photometric light curves.
Given the recent {\it Kepler-}guided understanding of fast rotating BAF-type pulsators with mode 
frequencies between pure g and pure p~modes, such variables were}
not yet included as a separate variability class
in the CU7 classifiers. Therefore, \OK{their strongly shifted non-radial mode frequencies due to fast 
rotation in the inertial frame of the observer may imply that these variables get 
spuriously 
classified as p-mode pulsators below the SPB instability strip in Fig.\,\ref{fig:hr_diagram_highfreq_p3}.}

\begin{figure}
  \resizebox*{\hsize}{!}{\includegraphics{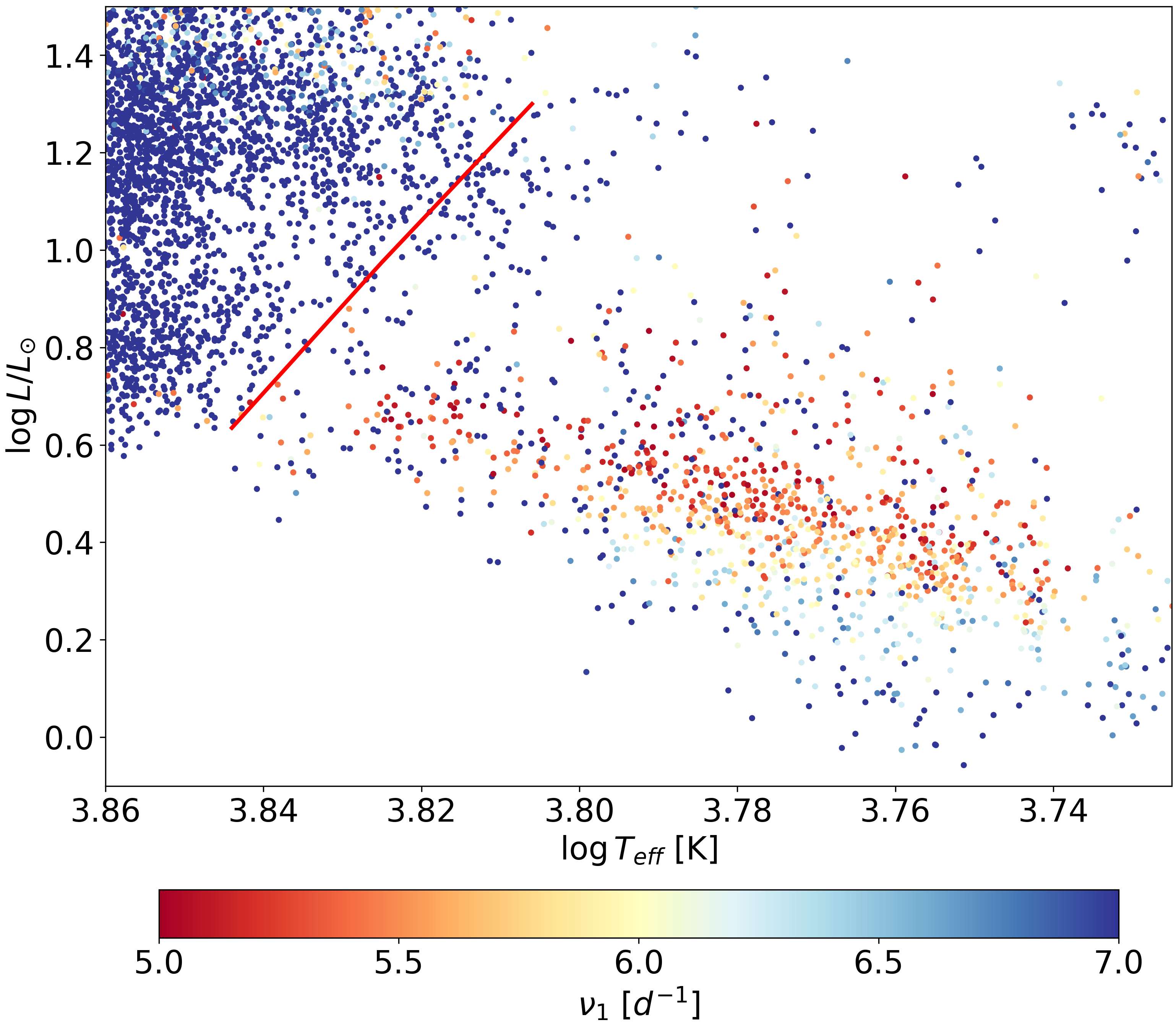}}
  \caption{Zoom into the overdensity below the $\delta$\,Sct instability strip at lower
effective
           temperatures in the HR diagram shown in Fig.~\ref{fig:hr_diagram_highfreq_p3}. The red line is the red edge of the $\delta$\,Sct instability strip for $n=1$ modes.}
  \label{fig:hrdiagram_weirdos_p3}
\end{figure}

An even larger overdensity of high-frequency variables \OK{also} pops up at cooler temperatures well 
\OK{below} the $\delta$\,Sct instability strip. A closer look at this part of the
HR diagram is shown in Fig.\,\ref{fig:hrdiagram_weirdos_p3}.
This concerns mainly stars with \OK{dominant} frequency below 7\,d$^{-1}$, 
\OK{obeying} a period--luminosity relation (albeit a noisy one) as can be seen from the colour gradient in the figure. 
\OK{We postulate that this may be a cooler-temperature, lower mass relation analogous to the period--luminosity (PL) 
relations found for rapidly rotating B-type pulsators with g modes occurring below the SPB instability strip.
Such PL relations were initially deduced by \citet{Saio2017-PL} based on the B-type cluster pulsators 
found by \citet{Mowlavi2013,Mowlavi2016}. \citet{Sharma2022} found similar PL relations from TESS photometry of 
rapidly rotating g-mode pulsators occurring between the SPB and $\delta\,$Sct strips
(cf.\,Fig.\,\ref{fig:hrdiagram_spbgdor_p3}). 
\citet{Sharma2022} interpreted these PL relations in terms of dipole ($l=|m|=1$) and quadrupole ($l=|m|=2$) g modes. 
It may well be that the PL relation revealed in Fig.\,\ref{fig:hrdiagram_weirdos_p3} can be explained similarly, that is, 
as the result of a mixture of dipole or quadrupole gravito-inertial, Rossby, or Yanai modes in very fast rotators 
of spectral type F, as such types of modes have been detected in {\it Kepler\/} photometry of cool $\gamma\,$Dor 
stars \citep{VanReeth2016,VanReeth2018,Saio2018}. 
}


\section{Comparison with the literature}
\label{sec:litcomparison}

Starting from our 106K sample described in Section \ref{sec:dataset}, we removed the distinct
population of stars far beyond the red edge of the $\delta$\,Sct instability strip 
(cf.~Fig \ref{fig:hrdiagram_weirdos_p3} for a nearby subsample of them), as we cannot rule out the possibility that these are misclassifications
at this point. This leaves us with a sample
of 88\,872 sources, which in the remainder of this section we refer to as the 88K sample.

We compare our sample of 88K candidate \referee{IHM} main sequence oscillators with two different
compilations from literature catalogues. The first compilation consists of 2121 $\delta$\,Sct stars
and 603 $\gamma$\,Dor stars in the 
\OK{
{\it Kepler\/} field as identified by
\citet{Bowman2016, Murphy2019} and 
\citet{VanReeth2015, GangLi2020}, respectively. }
This sample has the advantage that the classification
was carried out manually based on dense {\it Kepler\/} time series, and is therefore very reliable. The
drawback is that the sample covers only the fairly small magnitude range observed by {\it Kepler\/} and is very 
localised in the sky. This sample therefore does not cover the large magnitude range of
our \gdr3 sample, nor does it cover all the window functions caused by the sky-coordinate-dependent 
{\it Gaia} sampling.

Focusing first on the $\delta$\,Sct stars, we find that 746 of them (35\%) are identified
as an OBAF-type pulsator by the CU7 classifiers. The majority were therefore either not found to be
variable in the time domain or ended up as another type of variable.
This illustrates the challenges 
that come with identifying these small-amplitude multi-periodic variables with the fairly small number 
of {\it Gaia} measurements. Of these 746 pulsators, 73 (10\%)
have a FAP of smaller than our threshold of $10^{-3}$ (cf.~Section \ref{sec:dataset}). 
Our measures to mitigate the presence of instrumental effects have a significant
impact on the selection of bona fide pulsators as well. The vast majority (82\%) of the $\delta$\,Sct 
stars that did pass the FAP criterion were correctly identified as $\delta$\,Sct stars by the 
CU7 classifiers.

The results for the $\gamma$\,Dor stars are similar. We find that 233 of them (39\%) passed the 
CU7 classifier as an OBAF-type pulsator, and 26 of these stars have a FAP value smaller than our
threshold of $10^{-2}$ for g-mode pulsators. For most of these stars, their small amplitudes
\OK{(often below 5 mmag, cf.\,Appendix\,A)}
make the peaks in the Fourier spectrum barely statistically significant and distinguishable
from peaks purely caused by noise fluctuations. 

The second literature compilation with which we compare our sample is that of \citet{Gavras2022}, who 
compiled a large database of variable stars using existing catalogues published in the literature 
(before the \gdr3 release), including 
those of \citet{Stankov2005}, \citet{Pigulski2009}, \citet{Palaversa2013}, \citet{Drake2017}, 
and \citet{Chen2020} among others. The compilation ensured an all-sky coverage and a large
range of \g magnitude (and thus S/N levels). It consists of catalogues compiled using a manual \OK{variability}
classification, as well as catalogues \OK{based on} a probabilistic machine learning approach. 
We refer to \citet{Gavras2022} for more details. Although these latter authors adopted a stringent classification
score threshold, we should expect a small fraction of these literature classifications to be inaccurate. 

\begin{figure}
    \resizebox{\hsize}{!}{\includegraphics{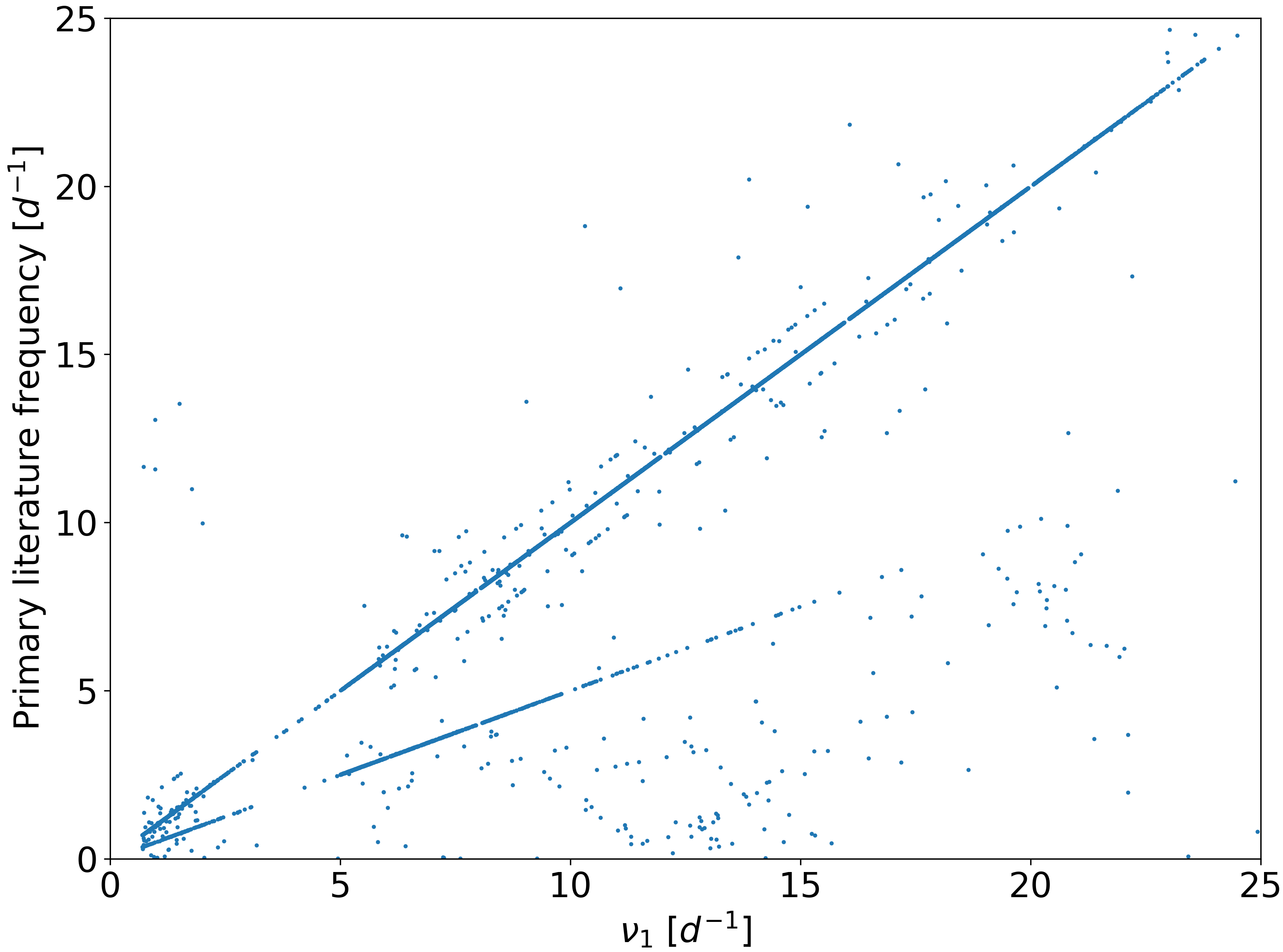}}
    \caption{Comparison with the main frequency $\nu_1$ found with the CU7 pipeline using 
             \gdr3 data and the main frequency $f_1$ reported in the literature.}
    \label{fig:freq_comp_lit.png}
\end{figure}

For 12\,042 stars in our sample, the cross-match catalogue of \citet{Gavras2022} includes 
a primary frequency found in the literature. The comparison of these literature frequencies 
with the ones derived by CU7 using \gdr3 photometric time series in the \g passband is 
shown in Figure \ref{fig:freq_comp_lit.png}.
Overall there is an excellent match between the \gdr3 and the literature frequencies. 
A small sample of sources forms a parallel line above and below the main diagonal. 
These are sources for which ground-based observations found a one-day alias of the true
frequency, while the oscillation frequency found by Gaia is not affected by Earth's 
day--night rhythm. For a smaller sample, the main frequency derived from \gdr3 data is 
twice the frequency found in the literature. We find that most of them are stars catalogued 
as a binary star in the literature, but erroneously classified as a g-mode pulsator by the 
CU7 classifiers. 

The literature cross-match compilation of \citet{Gavras2022} contains 108 $\beta$ Cep stars,
98 SPB stars, 421 $\gamma$\,Dor stars, and 14350 $\delta$\,Sct stars (+ SX Phe) for which
the corresponding source in \gdr3 has at least 40 data points, and which could therefore in principle
have been retained after the data selection criteria explained in Section \ref{sec:dataset}.
Of these stars, 41 $\beta$ Cep stars (38\%), 49 SPB stars (50\%), 69 $\gamma$\,Dor stars (16\%), 
and 9780 $\delta$\,Sct + SX Phe star (68\%) also appear in our \gdr3 sample of 88K variables. 
Of this subselection the CU7 classification pipeline obtained the 
same classification as the literature for 41 $\beta$ Cep stars (100\%), 42 SPB stars (86\%), 
64 $\gamma$\,Dor stars (93\%), and 9765 $\delta$\,Sct stars (99.8\%). The miss rate
(i.e.~the false negative rate FN/(FN+TP)) is therefore fairly good for $\delta$\,Sct stars (32\%) 
but high ($\ge 57\%$) for the other OBAF-type pulsators. This is not surprising, given that the stars in our 88K sample
have a median number of time points equal to 56. 
Nevertheless, the OBAF-type pulsators of the literature that do end up in our 88K sample are 
mostly correctly classified. We remind the reader that the CU7 classification pipeline
used XGBoost and Random Forest classifiers to classify 12 million variable stars into 25 
variability classes. More details can be found in \citet{Rimoldini2022a}.

Starting from our 88K sample of \gdr3 OBAF-type pulsators, it turns out that 66 $\beta$ Cep 
stars, 119 SPB stars, 793 $\gamma$\,Dor stars, and 11064 $\delta$\,Sct (+ SX Phe) stars also appear 
in the literature compilation of \citet{Gavras2022}, although not necessarily classified as 
the same type. 
For the $\beta$ Cep stars, the main contaminants are the $\gamma$\,Cas
variables (20\%) and the $\delta$\,Sct variables (6\%). For SPB \OK{stars,} the main contaminants are
$\alpha^2$ Canum Venaticorum (ACV) variables (24\%), rotationally modulated stars (21\%),
and $\gamma$\,Cas variables (9\%). 
\OK{The $\gamma$\,Dor stars} are confused most often with RS Canum 
Venaticorum variables (33\%), rotationally modulated stars (23\%) and, as expected,
with binary stars (12\%). Finally, the $\delta$\,Sct (+ SX Phe) stars are sometimes
confused with binaries (6\%), RR Lyr stars (2\%), or RS Canum Venaticorum variables.
Assuming again that the literature class label is always reliable, we can conclude that 
the false discovery rate (i.e.~FP/(FP+TP)) is fairly good to excellent for the high-frequency
variables, as low as 12\% for the $\delta$\,Sct stars, but is high for the low-frequency
variables ($\ge 65\%$). However, we note that we used our entire 88K sample for this
assessment, including the faint stars. As can be seen from the previous section, 
the comparison gives much better results for the nearby and brighter stars.

To investigate why the classification of $g$-mode pulsators is so difficult with \gdr3 data in more detail, we analyse in Appendix\,\ref{appendix:kepleranalysis} the
\gdr3 time series in the Gaia \g-band for 63 well-known bona fide g-mode pulsators
listed in \citet{Aerts2021-IGW}. For these \referee{stars,} we know the true oscillation frequency
from densely sampled {\it Kepler} photometric time series. Our analysis confirms
the comparison with the literature above. For about 15\% of these stars, the main
frequency identified from {\it Gaia\/} photometry coincides with the main frequency
found from {\it Kepler} photometry. For two of these stars, the second
frequency also coincides. The main culprit for mismatches is either an aliased frequency
making the g-mode pulsator look like a $\delta$\,Sct star, or an instrumental
frequency as described in Section \ref{sec:dataset}.

\section{The period--luminosity relation of $\delta$\,Sct stars}
\label{sec:plrelation}

$\delta$\,Sct stars pulsating in the radial fundamental (F) and/or in the first overtone (1O) mode show a 
period--luminosity relation, albeit with a larger scatter than for Cepheids (see e.g.~\citet{McNamara2007}, 
\citet{McNamara2011}, \citet{Ziaali2019}, \citet{Jayasinghe2020}, \citet{Poro2021}). For these modes, the pulsation 
constant $Q \equiv P\sqrt{\bar{\varrho}/\bar{\varrho}_{\odot}}$ is more or less constant over the instability strip, 
and the dominant stellar property that correlates with the stellar mean density is the luminosity $L$,
hence the PL relation \citep{Breger2000}.

\begin{figure}
  \resizebox{\hsize}{!}{\includegraphics{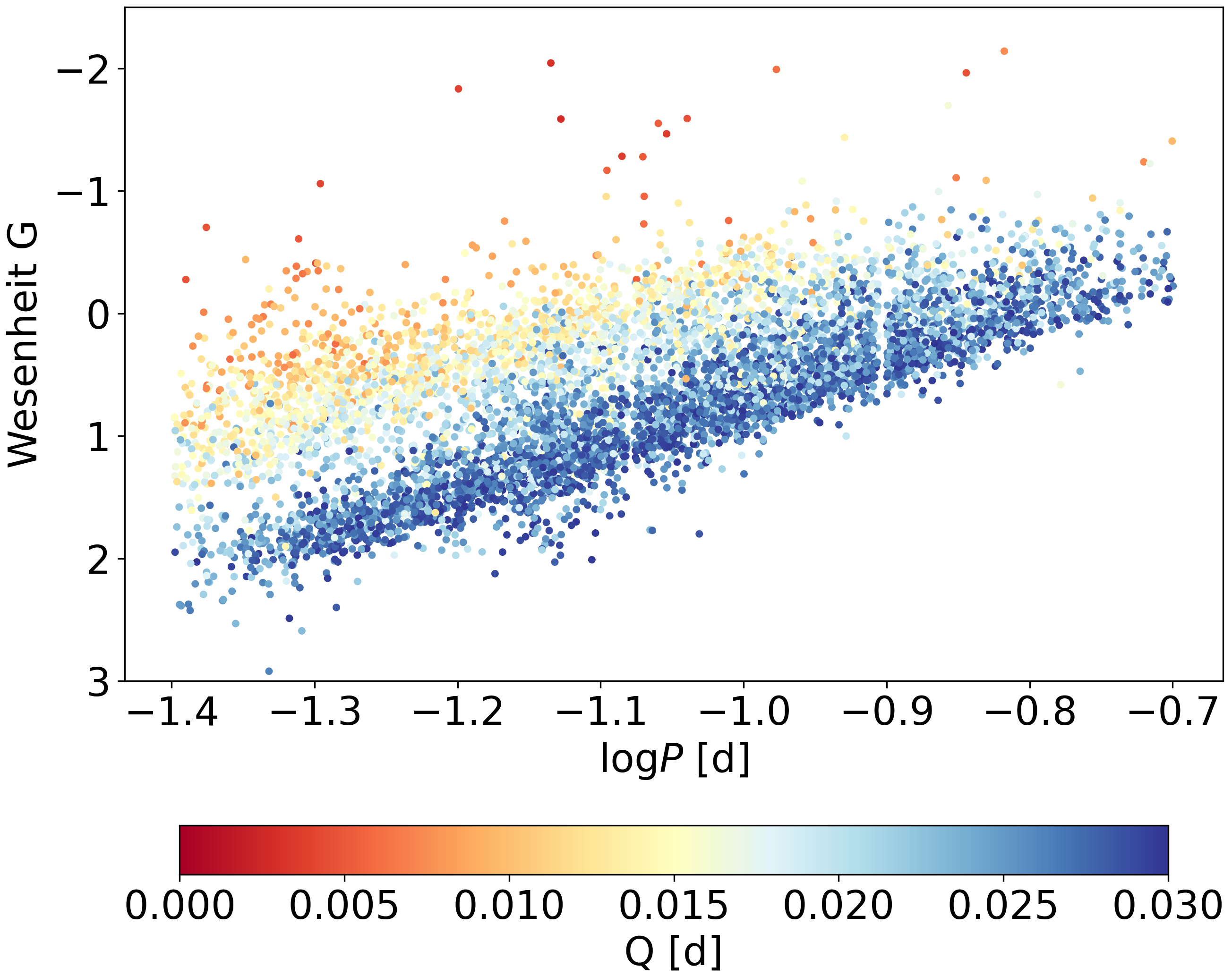}}
  \caption{Period--Wesenheit relation for 6511 $\delta$ Sct stars with $\sigma_{\varpi} / \varpi < 0.05$, 
           $T_{\rm eff} \in [6400, 8700]$ K, and Q < 0.03. \referee{The uncertainty on $\log P$ is smaller than the symbol size.
           The typical uncertainty of the Wesenheit G index is between 2 and 4 mmag.} }
  \label{fig:PWrelation_F_1O}
\end{figure}

To establish the relation as seen using the \gdr3 dataset, we first computed the reddening-insensitive 
period--Wesenheit (PW) relation. The absolute Wesenheit index $W$ was computed using
\begin{equation}
  W = M_G - 1.9 \ (G_{\rm BP} - G_{\rm RP}),
\end{equation} 
where $M_G$ is the absolute magnitude in the \g-band taken from the \gdr3 \texttt{astrophysical\_parameters}
table (\texttt{gspphot}). 
Inspection of the PW relation obtained with this sample revealed some contamination 
from the well-defined group of low-luminosity stars with a frequency $f<6.6$\,d$^{-1}$ (P$>$0.15 days) described
in the last paragraph of Section \ref{sec:instabstrips}.  Because of their compactness, it proved easy to select 
and remove them with the criteria $f<6.6$\,d$^{-1}$ and $W<1.0$ mag. In this way, we discarded 233 objects.
Figure~\ref{fig:PWrelation_F_1O} shows the PW relation  for nearby $\delta$ Sct stars with
$\sigma_{\varpi} / \varpi < 0.05$ and  
$T_{\rm eff} \in [6400, 8700]$ K colour coded with the empirical $Q$ value (in days). 
The latter was computed using the relation derived by \citet{Breger1990}:
\begin{equation}
\log_{10}\left(\frac{Q}{P}\right) = -0.25 \log_{10}\left(\frac{L}{L_{\odot}}\right) + 0.5 \log_{10}\left(\frac{g}{g_{\odot}}\right)
                 + \log_{10}\left(\frac{T_{\rm eff}}{T_{\rm eff,\odot}}\right)
.\end{equation}
As before, the quantities $L/L_{\odot}$, $g$ and $T_{\rm eff}$ were taken from the \gdr3 
\texttt{astrophysical\_parameters} table (\texttt{gspphot}).
Comparing the empirical pulsation constant $Q$ with the typical values given by \citet{Breger1979}, we
see that the \gdr3 data not only show the ridge of the fundamental mode (dark blue), but also the ridges of the
first and the second overtone.

\begin{figure}
  \resizebox{\hsize}{!}{\includegraphics{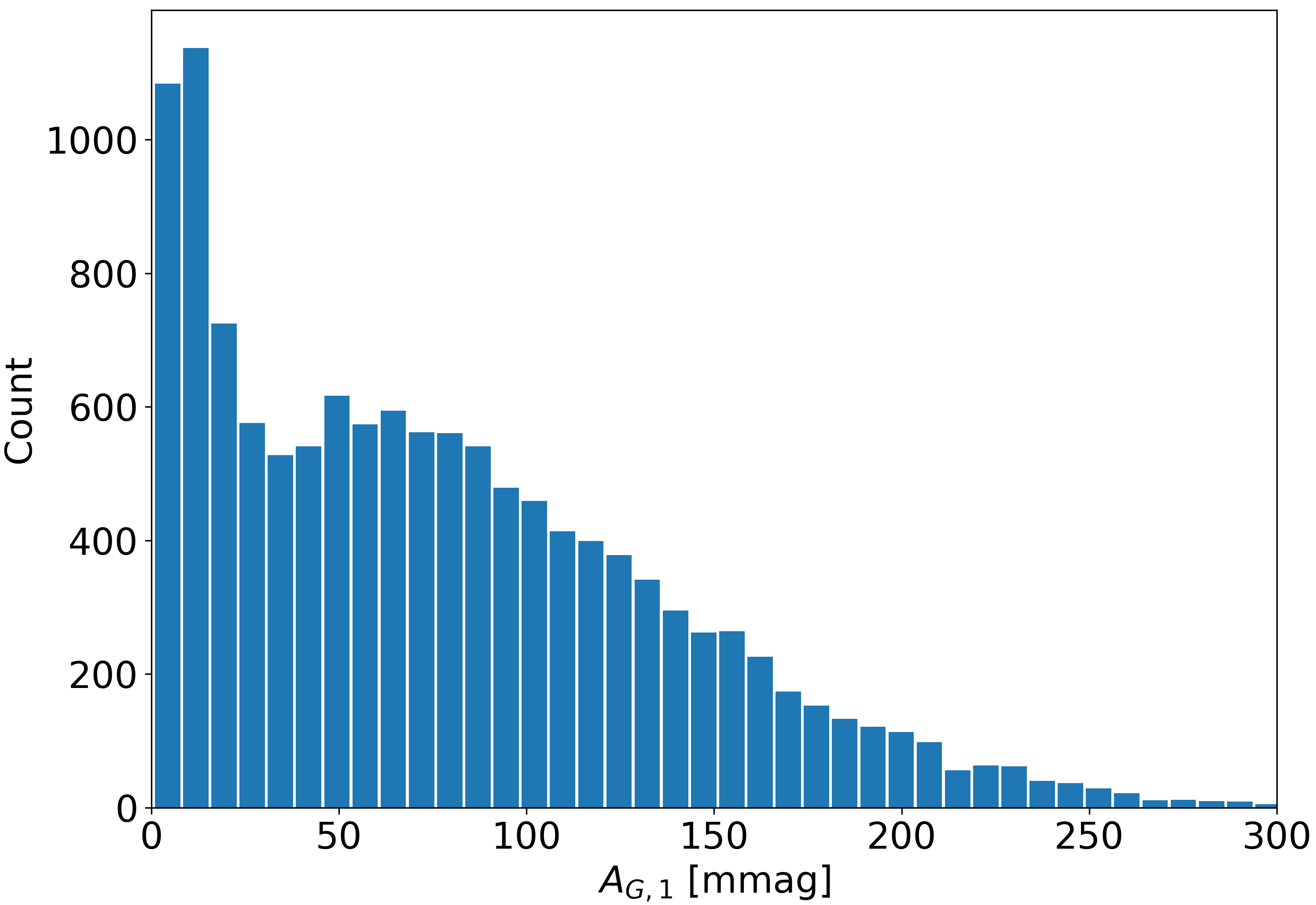}}
  \caption{Histogram of the amplitude $A_{G,1}$ of the main frequency $\nu_1$ of our sample of 
           $\delta$\,Sct stars.}
  \label{fig:dsct_AG1_histogram}
\end{figure}

\begin{figure}
  \resizebox{\hsize}{!}{\includegraphics{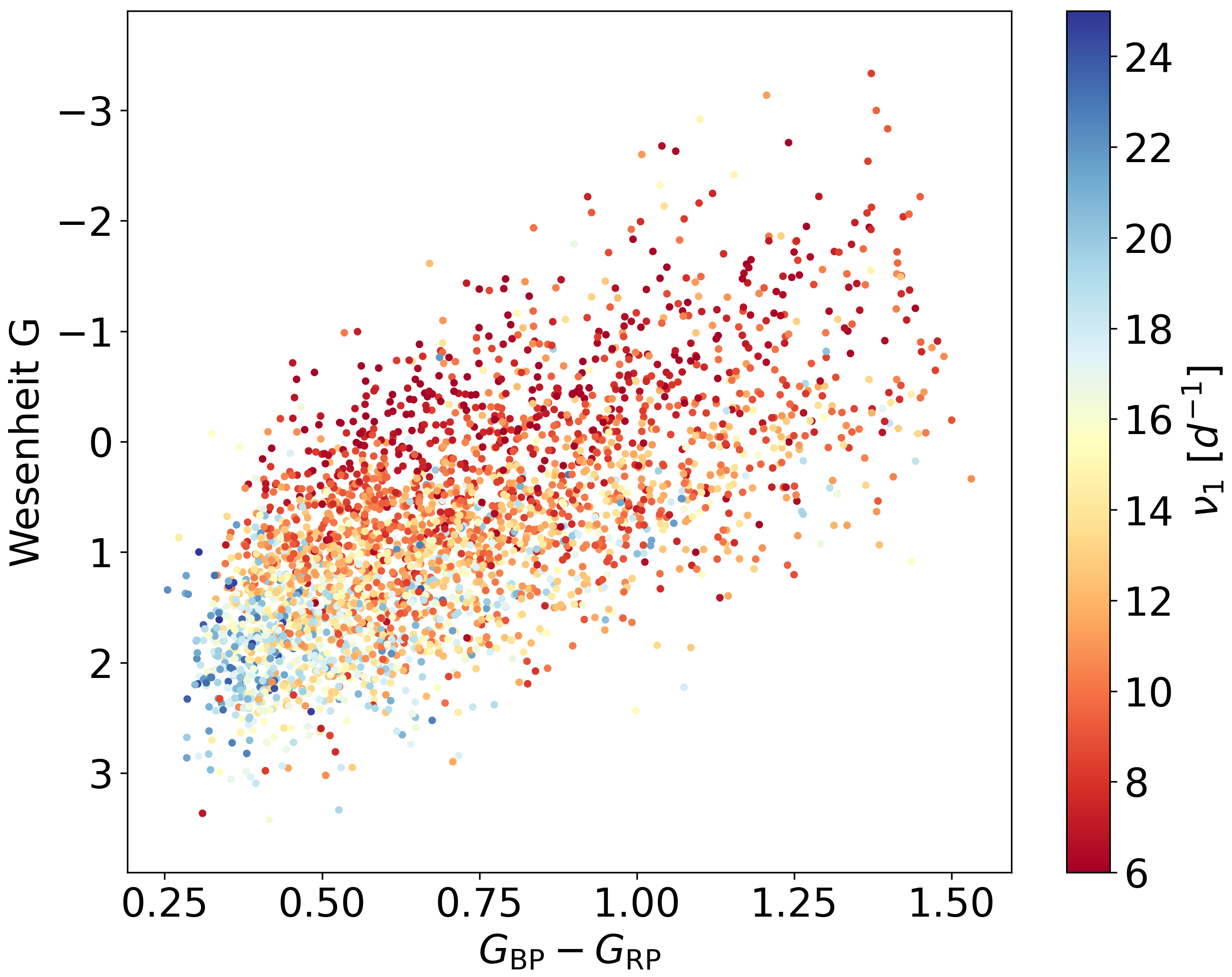}}
    \caption{Location of 3122 high-amplitude $\delta$\,Sct stars with $0<\sigma_{\varpi}/\varpi < 0.2$ 
             in a colour--Wesenheit \g diagram. The points are colour coded according to frequency $\nu_1$.
             \referee{Typical uncertainties of the mean \bprp are between 1 and 2 mmag and between 2 and 4 mmag for
             the Wesenheit G index.}
    }
    \label{fig:dsct_hads_HRdiagram}
  \end{figure}

In the remainder of this section, we focus on the most populous ridge in Fig.~\ref{fig:PWrelation_F_1O}
which is that of the fundamental mode pulsators, for which we follow a more precise procedure. 
The high-amplitude $\delta$\,Sct (HADS) variables are known to pulsate in the F mode, the 1O mode, or both \citep{McNamara2007}. 
Figure~\ref{fig:dsct_AG1_histogram} shows a histogram of the amplitude $A_{G,1}$ of the main frequency
$\nu_1$ for our sample of $\delta$\,Sct stars, in which the HADS population is clearly visible. To avoid
too much contamination of the non-radially pulsating lower amplitude population, we only retained 
$\delta$\,Sct stars with $A_{G,1}\geq50$ mmag. As before, we constrained the effective 
temperature of our sample stars to the range of the $\delta$\,Sct instability strip, i.e.~6400 
K $\leq T_{\rm eff} \leq$ 8700 K. This left us with 8894 stars. 
In addition we also filtered on the quality of the astrometric parallax as this is a
key ingredient in the derivation of the PW relation. To do so, we followed the approach of \citet{Rybizki2022}, 
who used a neural network based on 17 proper {\it Gaia} catalogue entries to discriminate objects with poor 
astrometry by means of a single parameter which they called `astrometric fidelity'. 
We retained only objects with their \texttt{fidelity\_v2} $> 0.5$. This last selection defined our final sample
of 8760 $\delta$\,Sct stars useful for the PW relation determination. 
We plotted the $\delta$ Sct stars in this sample with a relative parallax uncertainty of better than 20\% in 
Fig.~\ref{fig:dsct_hads_HRdiagram}. The points are colour coded with the frequency $\nu_1$ so that 
the frequency gradient due to the period--luminosity relation is clearly visible.

To calculate the PW relation, we adopted the so-called photometric parallax 
\citep[see][and references therein]{Riess2021}, which was recently used to derive the period--Wesenheit--metallicity 
relation for the classical Cepheids in the {\it Gaia} bands by \citet{Ripepi2022}. This approach allows us to use 
the parallax linearly, retaining the Gaussian property of its uncertainty. Moreover, as we do not make any 
selection in parallax (even negative parallaxes can be used) we are not introducing biases in the PW relation.
The photometric parallax (in mas) is defined as follows: 
\begin{equation}
\varpi_{phot}=10^{-0.2\ (w-W-10)}\, ,
\end{equation}
\noindent where $W$ is the absolute Wesenheit magnitude in the \g passband, 
and $w$ is the corresponding apparent Wesenheit magnitude:
\begin{equation}
w = G - 1.9 \ (G_{\rm BP} - G_{\rm RP}).
\end{equation} 
The absolute Wesenheit index $W$ can be written as
\begin{equation}
W=\alpha + \beta\,(\log_{10} P-\log_{10} P_0)\, , 
\label{eq:PWZ}
\end{equation}
where P is the period and $P_0$ a `pivoting' period (in days) chosen approximately as the mean 
of the period distribution (i.e.~$\log_{10}P_0=-1.1$).
This serves to reduce the correlation between the zero point $\alpha$ and the slope 
$\beta$ of the PW relation.  

Denoting with $\varpi_{\rm EDR3}$ the parallaxes of the pulsators corrected for the zero-point offset (ZPO) 
\citep[see][]{Lindegren2021}, we seek to minimise the following quantity:
\begin{equation}
\chi^2=\sum \frac{(\varpi_{EDR3}-\varpi_{phot})^2}{\sigma^2}\, ,   
\label{eq:chi}
\end{equation}
where $\sigma^2=\sigma_{\varpi_{EDR3}}^2+\sigma_{\varpi_{phot}}^2$. In our case, $\sigma_{\varpi_{\rm EDR3}}$
is composed of the standard error of the parallax as reported in the \egdr3 catalogue, here conservatively increased 
by 10\%, and the uncertainty on the individual ZPO corrections, that is 13\,$\mu$as \citep{Lindegren2021}. 
To calculate the uncertainty on the photometric parallax we followed the procedure detailed by \citet{Ripepi2022}: 
$\sigma_{\varpi_{\rm phot}} = 0.46 \cdot \sigma_\mu \cdot \varpi_{\rm phot}$,  where $\mu=w-W$ is the distance modulus 
and  $\sigma_\mu^2 = \sigma_{w}^2+\sigma_{W}^2$ its variance. We calculated $\sigma_{w}$ from error propagation assuming 
a constant error of 0.01 mag in each of the three \gaia\ bands (\g, \bp, \rp), where $\sigma_{W}$ is the 
intrinsic dispersion of the PW relation. Similarly to the case of classical Cepheids, we assumed a value of 0.1 mag 
for this quantity.  

We minimised Eq.~(\ref{eq:chi}) using the {\tt Python} minimisation routine {\tt optimize.minimize} 
\citep[included in the {\tt Scipy} package][]{Virtanen2020}. To provide robust uncertainties on the coefficients 
$\alpha$ and $\beta$ of the PW relation, we used a bootstrap procedure in which we repeated the fit to 
the data of Eq.~(\ref{eq:chi}) 1000 times, each time with a randomised bootstrap sample, and obtained a different $\alpha$
and $\beta$ value for each repetition. Our best estimate for these two parameters is then the median of the resulting 
distributions, while for the uncertainty we used 1.4826 $\cdot$ MAD (median absolute dispersion). 

\begin{figure}
  \resizebox{\hsize}{!}{\includegraphics{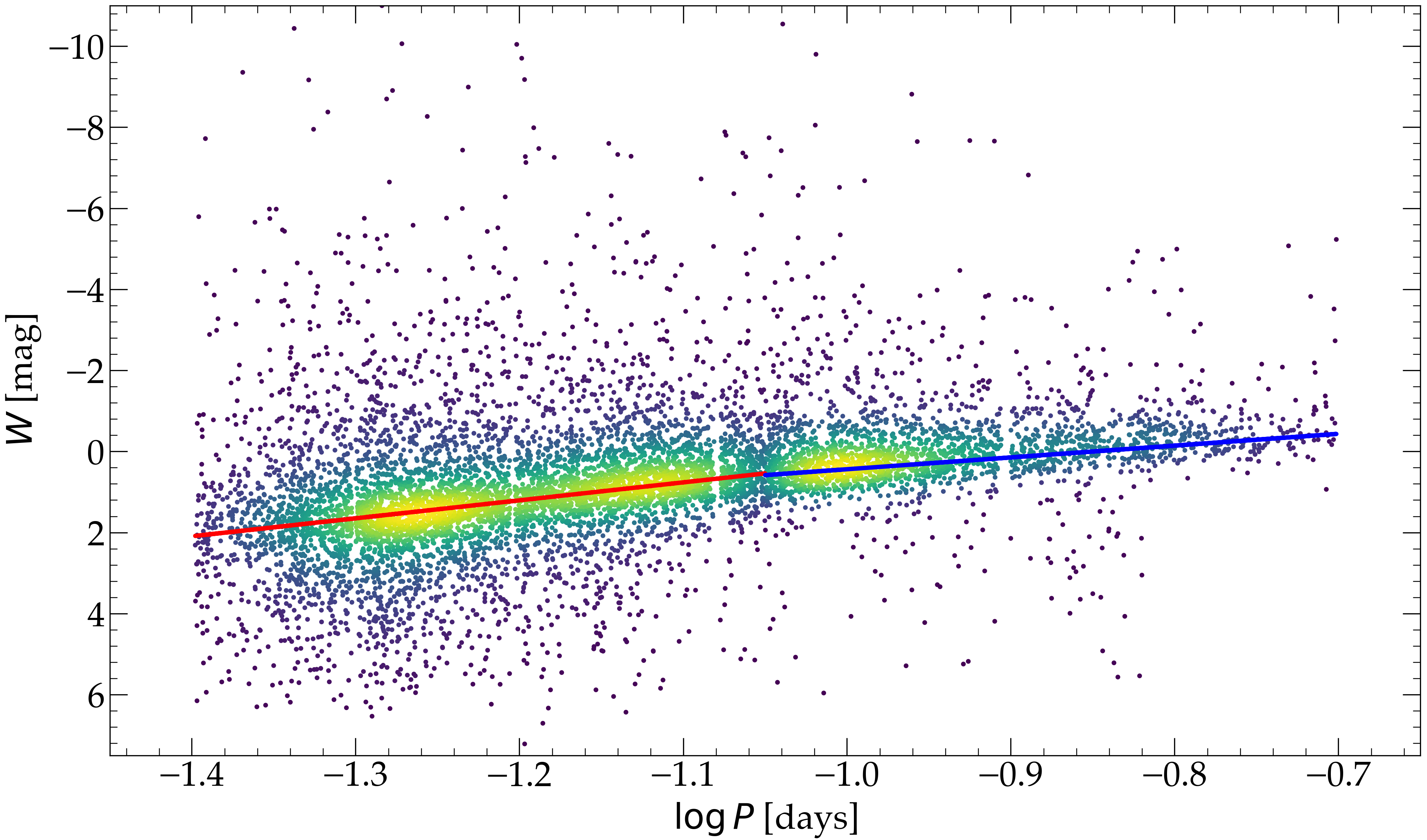}}
  \resizebox{\hsize}{!}{\includegraphics{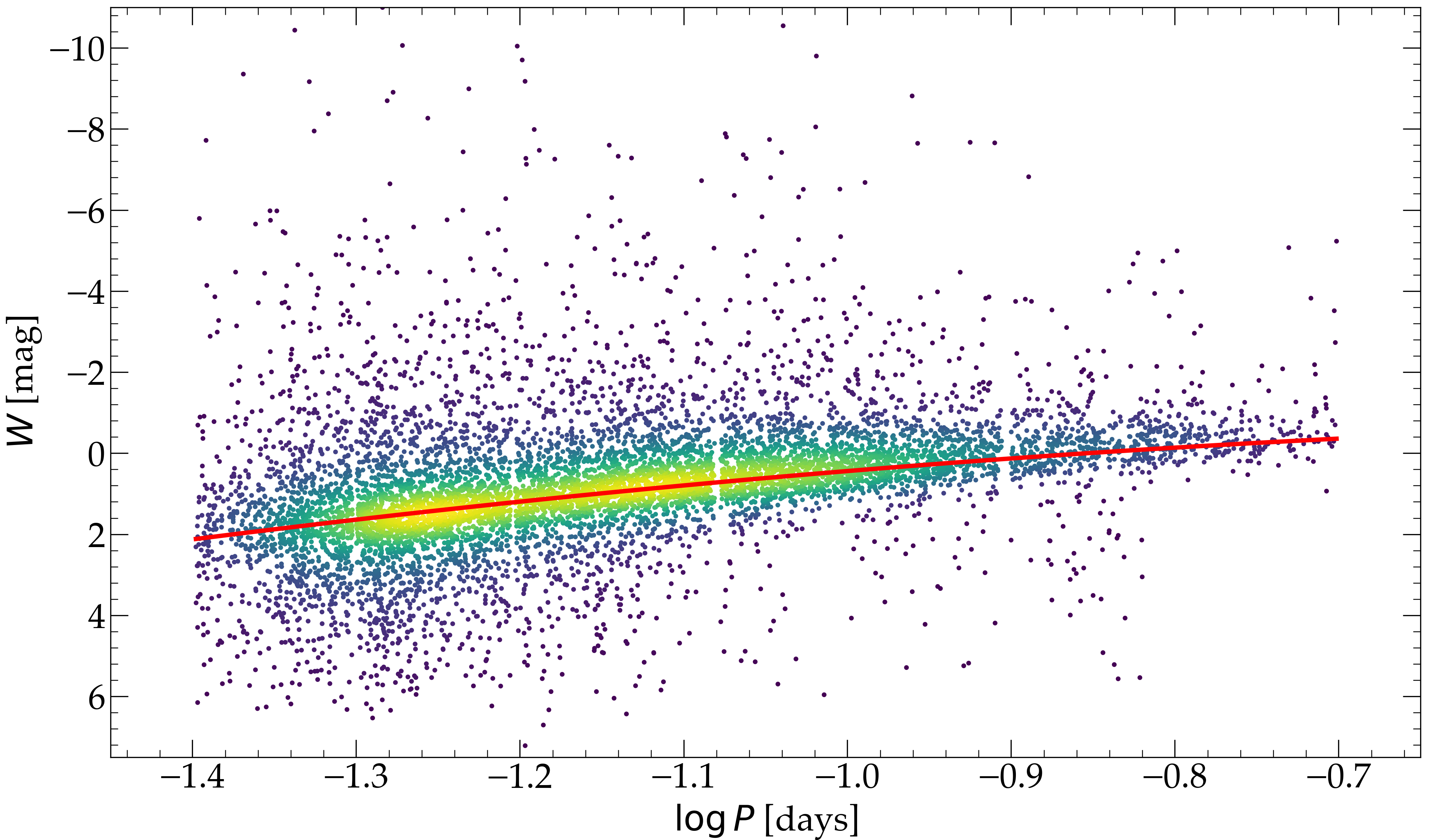}} 
  \caption{PW fit to the data. The top panel shows the two-line fit (see first two lines of Table~\ref{tab:PWfit})
  and the bottom panel shows the quadratic fit to the data (see first two lines of Table~\ref{tab:PWfit}).}
  \label{fig:PWs}
\end{figure}

In a first attempt, we fitted the PW relation over the whole period range of 0.04-0.20 days, but it became
clear that a single \OK{linear fit}
is not able to represent the empirical PW relation as the data show a break or a 
non-linearity (see Fig.~\ref{fig:PWs}). We therefore applied two approaches, first fitting the PW with a piecewise
linear function, and then with a quadratic PW relation of the form 
\begin{equation}
W = \alpha + \beta\,(\log P-\log_{10} P_0) + \gamma\,(\log P-\log_{10} P_0)^2.
\end{equation}
For the piecewise linear model, we find that the breaking point \OK{occurs} approximately at 
$\log P = -1.05 \pm 0.05$. The best piecewise linear fit to the data 
is shown in the top panel of Fig.~\ref{fig:PWs} and the corresponding fitted parameter values are
listed in the first two \OK{rows} of Table~\ref{tab:PWfit}. 
The quadratic PW relation fit to the data is shown in the bottom panel of Fig.~\ref{fig:PWs}, 
while the corresponding fit parameters are listed in the 
\OK{bottom row}
of Table~\ref{tab:PWfit}.  

As for the piecewise linear fit, the slopes obtained 
\OK{for}
the two intervals \OK{of the dominant pulsation period} 
are different with a significance of about 
10$\sigma$, indicating that the PW relation of the $\delta$\,Sct stars with a 
\OK{pulsation} period P shorter than $\sim$0.09 days 
has a much steeper slope than 
\OK{the one valid}
for the longer-period 
\OK{pulsators.}
Also, the quadratic term has a high significance, 
indicating that the non-linearity or the two-regime nature of the $\delta$\,Sct PW relation is a genuine physical 
feature. 

\begin{figure}
  \resizebox{\hsize}{!}{\includegraphics{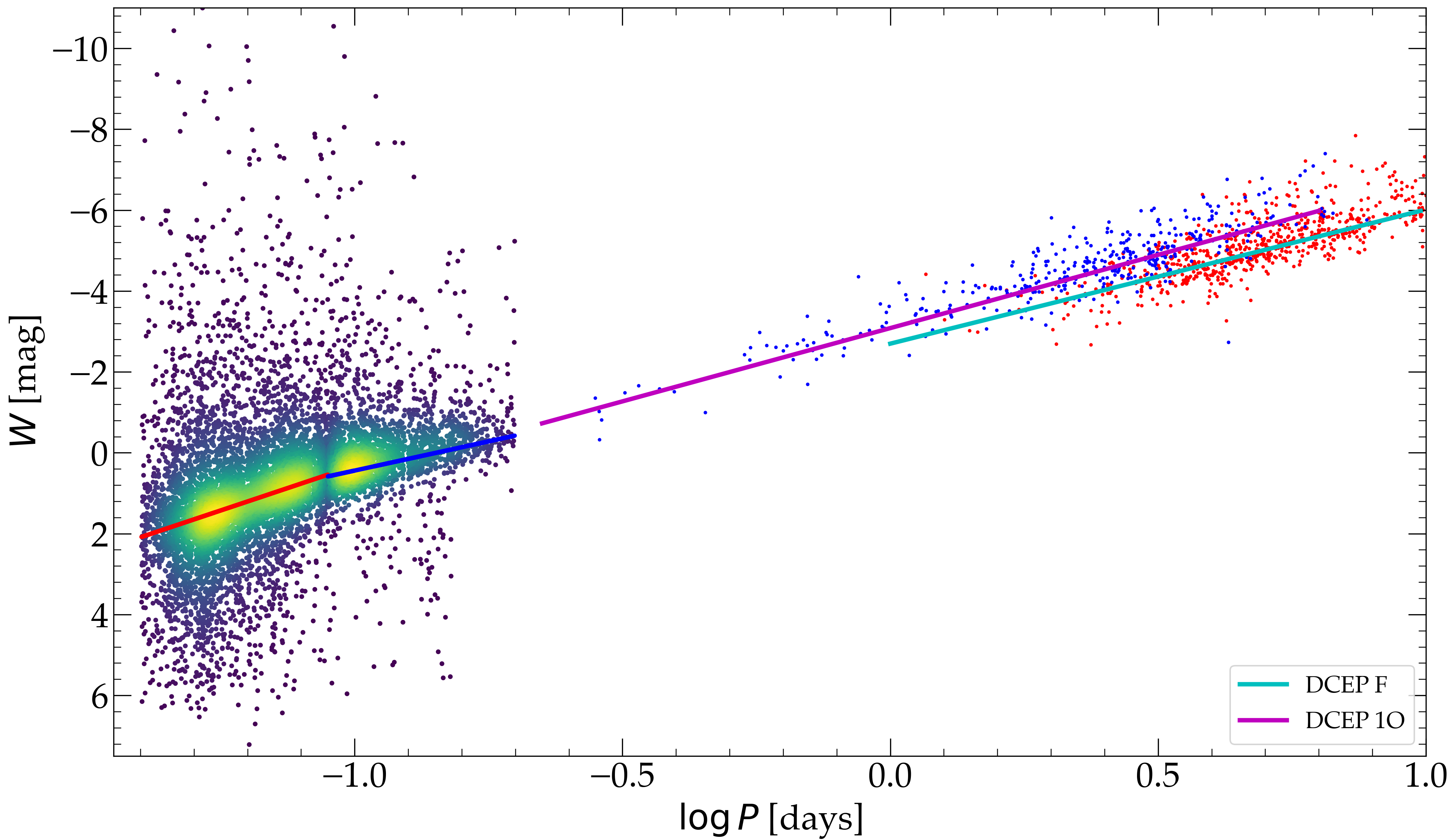}}
  \caption{Comparison between the PW relation(s) calculated in this paper for the fundamental F-mode 
           $\delta$\,Sct (as in Fig.~\ref{fig:PWs}) and those of F- (red dots) and the first overtone 
           1O-mode (blue dots) classical Cepheids. For the latter pulsators, the data and the PWs 
           were taken from \citet{Poggio2021}. For clarity, only objects with relative error on 
           the photometric parallax better than 20\% are plotted.}
  \label{fig:PW_comparisonWithDCEPs}
\end{figure}

It is instructive to compare the $\delta$\,Sct PW relation with that of classical Cepheids as both classes \OK{of pulsators}
pulsate due to the well-known $\kappa$ mechanism. 
Therefore, the instability strip of \OK{the} $\delta$\,Sct \OK{stars} can be 
considered the low-luminosity extension of the classical Cepheid instability strip. To make the comparison, 
we considered the classical Cepheids' PW relations obtained in the {\it Gaia} bands by \citet{Poggio2021}, 
which 
\OK{were}
derived in the same way as for the $\delta$\,Sct stars in this paper. Figure~\ref{fig:PW_comparisonWithDCEPs} 
shows the same $\delta$\,Sct stars and PW relations as in the top panel of Fig.~\ref{fig:PWs} with, in addition, 
the location of the F- and 1O-mode classical Cepheids and the relative PW relations from \citet{Poggio2021}. 
For clarity, we only show objects with relative errors on the \egdr3 parallax of better than 20\%.
The distribution of \OK{the} classical Cepheids \OK{pulsating in the first overtone} 
extends up to that of \OK{the} $\delta$\,Sct F-mode \OK{pulsators.}
\footnote{The instability strip of 
\OK{the} F-mode classical Cepheids \OK{becomes} 
too narrow at low luminosity to allow \OK{for} the 
\OK{occurrence} 
of pulsators with period smaller than about one day.} 
It is clear that the PWs of both variability 
classes predict similar luminosities 
\OK{for modes}
in the period range $\sim$0.20--0.25 days, where 
\OK{the}
$\delta$\,Sct and classical 
Cepheid \OK{period} distributions 
\OK{overlap.}
The slopes of the F- and 1O-mode classical Cepheids PW relations as calculated 
by \citet{Poggio2021} are $-3.317\pm0.028$ and $3.624\pm0.017$, respectively. Comparing these values with those 
listed in the first two \OK{rows} of Table~\ref{tab:PWfit}, we find that the 
\OK{slopes for the relations found for the} classical Cepheids are 
intermediate between those calculated for the low- and high-period $\delta$\,Sct samples. If we were to ignore the
observed curvature in the $\delta$\,Sct PW data and fit a single straight line, this would produce an intermediate
slope similar to that of the classical Cepheids.

\begin{table*}
\centering
\caption{Results for the PW fitting for $\delta$\,Sct variables. The fitted equations are 
$W=\alpha+\beta(\log_{10}P+\log_{10}P_0)$ or $W=\alpha+\beta(\log_{10}P+\log_{10}P_0)$+$\gamma(\log_{10}P+\log_{10}P_0)^2$,
where we took $\log_{10}P_0 = 1.1$. 
The Notes column lists the period interval over which the different PW relations were calculated.}
\label{tab:PWfit}
\setlength{\tabcolsep}{3.5pt}
\begin{tabular}{cccccc}
\hline
\hline            
\noalign{\smallskip}
         $\alpha$  &        $\beta$ & $\gamma$     &  $\chi^2$ & n.Obj & Notes \\
 \hline            
\noalign{\smallskip}
 0.764 $\pm$ 0.016  &  -4.444 $\pm$ 0.128 &               & 1.6$\pm$0.1 & 6403  & $-1.40 \leq \log_{10} P < -1.05$     \\
 0.730 $\pm$ 0.027  &  -2.921 $\pm$ 0.130 &               & 3.8$\pm$0.6 & 2357  & $-1.05 \leq \log_{10} P \leq -0.70$  \\
 0.796 $\pm$ 0.012  &  -3.773 $\pm$ 0.063 & 2.20$\pm$0.21 & 2.3$\pm$0.2&  8760  & $-1.40 \leq \log_{10} P \leq -0.70$  \\
\hline                                      
\end{tabular}
\end{table*}

\section{Pulsation amplitude attenuation due to rotation}

There are 
\OK{at least}
two ways 
\OK{in which}
stellar rotation can affect the observed photometric oscillation 
amplitudes.\footnote{We deliberately ignore the possible presence of a magnetic field,
which may make the interaction between rotation and pulsations even more complex.}
The first is through a change in the photometric visibility, that is,~how the pulsation manifests 
itself on the stellar surface, and the second is through a change in the intrinsic oscillation 
amplitude. \OK{Theoretical predictions of} photometric visibilities 
\OK{for modes in rotating stars}
have been analysed in depth in the literature (see e.g.~
\citet{Daszynska-Daszkiewicz2007} and \citet{Reese2013,Reese2021}
\OK{for p~modes and \citet{Henneco2021,Dhouib2021a,Dhouib2021b} for g~modes).}
For non-radial oscillations \OK{in fast rotators}, the
surface pattern 
\OK{of the modes}
can no longer be described by a single spherical harmonic $Y_{\ell}^m$.
\OK{Depending on the type of mode, its frequency,} and the angular rotation speed, 
\OK{the pulsation patterns}
can have a 
complex structure making
them appreciable only at certain latitudes, which affects the geometrical disc-integration factor.

The effect of rotation on the intrinsic amplitude of \OK{$\kappa$-driven pulsators is poorly} understood.
The vast majority of available pulsation codes \OK{rely on linear pulsation theory,}
which does not allow  the 
\OK{mode} amplitudes to be derived. Analysing the effect of rotation on the pulsation amplitudes 
\OK{in rapidly rotating IHM pulsators requires non-linear pulsation theory, which takes into account the effects of 
mode coupling in the presence of 
the Coriolis and 
centrifugal forces. While nonlinear mode coupling is detected in many IHM pulsators \citep{Bowman2016,VanBeeck2021}, 
modelling of their observed mode amplitudes has not yet been achieved.} 

\OK{Observations suggest that there is a link between the rotation of a star and its photometric pulsation amplitudes, 
but the relationship between the two is not well understood
\citep[e.g.][in the case of B-type pulsators]{Aerts2014}. 
This is partly due to the limited sample sizes of IHM pulsators with the appropriate information.
Given the sample sizes of new IHM pulsators presented here, and the precision of $v\sin i$ delivered by Gaia for these stars, 
we focus here on the observed relationship between rotation and amplitude of the dominant pulsation mode for the 
$\delta\,$Sct stars. The extreme case of the $\delta\,$Sct star Altair 
rotating at an equatorial speed of $\sim 310\,$km\,s$^{-1}$ \citep{Bouchaud2020}
reveals a high level of complexity in its pulsations \citep{Altair2021} and makes a study of the relationship between mode amplitudes and 
rotation for an entire sample worthwhile.} 

\begin{figure}
\resizebox{\hsize}{!}{\includegraphics{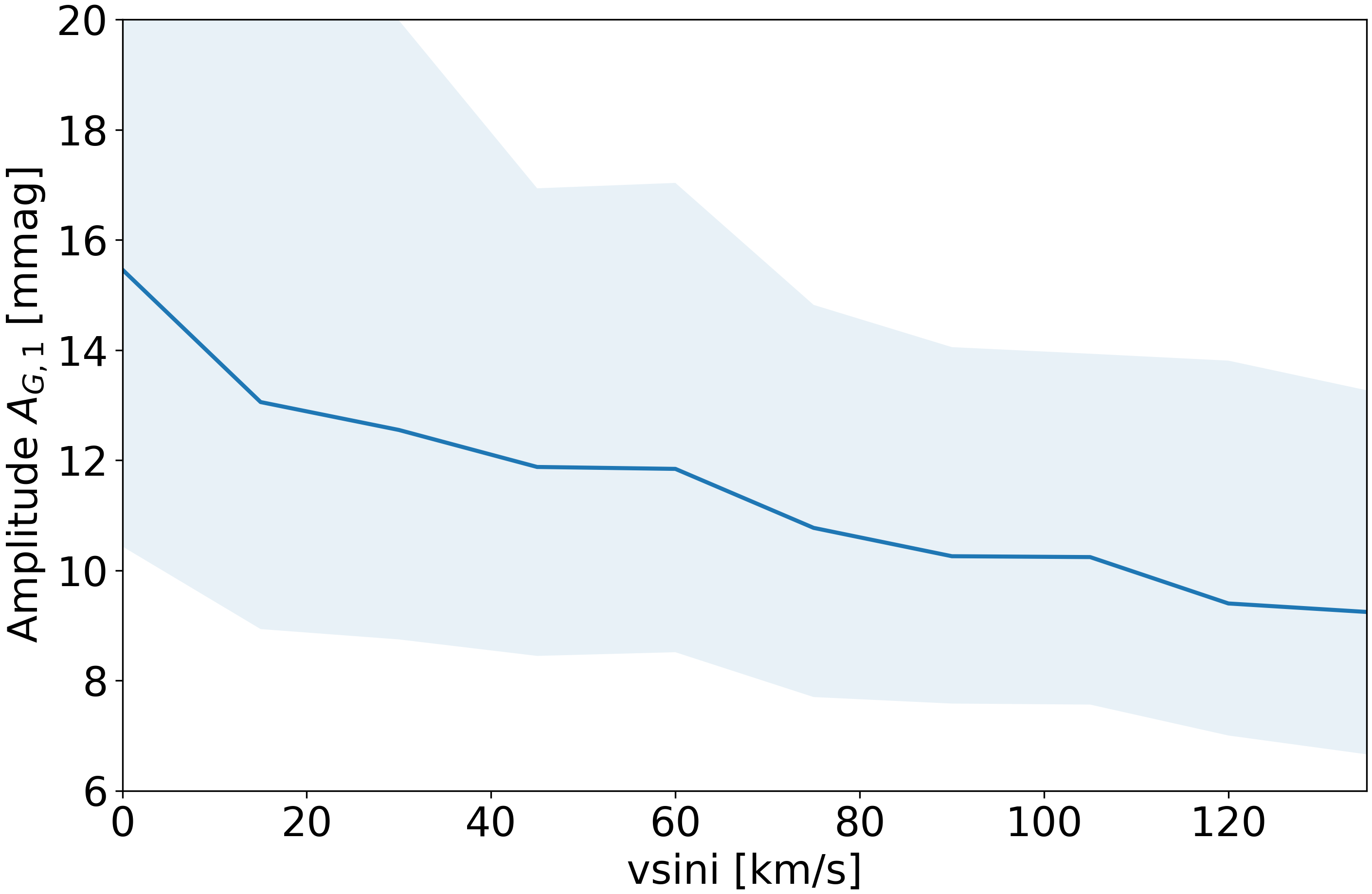}}
    \caption{
    Photometric amplitude $A_1$ in the \gaia \g-band as a function of the $v\sin i$ derived from
             spectroscopic \gaia\  DR3 data for a sample of 3515 $\delta$\,Sct stars. \OK{The dark blue line is 
             the median amplitude for each 
        $v\sin i$ bin of 15\,km\,s$^{-1}$ while the light blue band shows the 25\%-75\% quantile interval.}}
    \label{fig:dsct_vsini_ampl}
\end{figure}

The high-amplitude $\delta$\,Sct stars tend to rotate slower (typically $v\sin i \le 30$ km/s) than 
normal $\delta$\,Sct stars, which tend 
to have $v\sin i \ge 150$ km/s \citep[see e.g.~][]{Breger2000, Antoci2019}. 
\OK{This suggests that different angular momentum transport mechanisms
are at work in these two groups of $\delta\,$Sct stars, or 
different efficiencies of the same mechanism. The origin of this transport is poorly understood
\citep{Aerts2019}.}
To investigate the relation between the oscillation amplitude and the stellar rotation, we cross-matched 
the stars classified as $\delta$\,Sct stars in our sample described in section \ref{sec:dataset} 
with those in the \gaia-DR3 archive for which a $v\sin i$ value is available from \gaia spectroscopy
(\texttt{vsini\_esphs}) with a formal precision better than 50\%. This returned 3515 sources, which we 
partitioned into ten bins in $v\sin i$, each with a width of 15 km\,s$^{-1}$. For each bin, we computed the 
median value of the photometric amplitude $A_{G,1}$. As expected, the bins are not equally populated.
All bins contain at least 220 stars, except for the 
bin with the most rapid rotators for which
the constraint on the $v\sin i$ precision resulted in only 29 stars. The result is shown in  
Fig.~\ref{fig:dsct_vsini_ampl}. 
The dark blue line denotes the median amplitude for each bin, and 
the light blue band shows the 25\%-75\% quantile interval of photometric amplitude $A_{G,1}$ in
each bin. A clear and steady decline in median amplitude can be seen with increasing $v\sin i$ 
at a rate of roughly 46\,$\mu$mag per km\,s$^{-1}$. 
We note that the \OK{sample of
3525 sources contains both HADS and normal $\delta$\,Sct stars.
The vast majority of HADS occur in the lowest $v\sin i$ bin, where they} 
are responsible for the
strong decrease in amplitude in that bin. The corresponding curve for normal $\delta$\,Sct stars
is far less pronounced in the low-$v\sin i$ bins, 
\OK{yet} is clearly present in 
\OK{all} 
$v\sin i$ bins. 
\OK{
We cover $\delta\,$Sct stars with slow to moderate rotation rates of 
$v\sin\,i<150$\,km\,s$^{-1}$. This is well below the $v\sin\,i=270\,$km\,s$^{-1}$ of the fast-rotating $\delta\,$Sct star Altair, 
whose dominant mode has an amplitude of only about 0.6\,mmag in the white-light broad-band filter of the MOST space telescope 
\citep{Altair2021}.}

\section{Summary and Conclusions}

In this paper, we present our investigation of the \gdr3 time series of stars classified as $\beta$ Cep,
SPB, $\gamma$\,Dor, or $\delta$\,Sct by the CU7 classification pipeline. These stars are
non-radial pulsators of intermediate and high mass and have proven to be excellent targets
for asteroseismology \citep{Aerts2021,Kurtz2022}. They are often multi-periodic and have much lower amplitudes than 
RR Lyr stars or 
\OK{long-period variables. This}
makes them more challenging to detect and discriminate
from other types of variables, in particular with the modest size and sparseness of a 
typical \g-band time series in \gdr3. For this reason, we focused on time series with a least
40 data points.

Roughly half a million stars with at least 40 data points were classified 
\OK{in}
one of the \OK{four} 
classes 
by the variability processing pipeline of CU7. Closer inspection
of the results of the automated Fourier analysis of this pipeline showed that  a variation of instrumental origin
\OK{was detected for the majority of these stars.}
This 
\OK{points to}
imperfect 
calibrations of the photometric time series
\OK{at mmag level.}
We expect these instrumental variations to be calibrated out in 
the \textit{Gaia} DR5 release. We succeeded in filtering out most of the stars that show 
instrumental variations, at the cost of significantly reducing the sample size
\OK{of IHM pulsators. Nevertheless, the approximately $106$K pulsators of this kind constitute an unseen sample in terms of size, magnitude range, and sky coverage.}
Our comparison
with a subsample of the remaining stars in Section \ref{sec:litcomparison} shows that the 
main frequency $\nu_1$ agrees very well with that found in the literature.

To assess the quality of our sample of IHM 
\OK{pulsators,}
we compared the position of the stars
in the HR diagram with the theoretical instability strips. This requires the empirical
$T_{\rm eff}$ and luminosity of each star, which we took from the \gdr3 \texttt{gspphot} 
data as derived by DPAC-CU8. We focused on the nearby stars with the most precise location 
in the HR diagram as observed by Gaia. For these stars, the empirical location of 
\OK{SPB,}
$\delta$\,Sct, and 
$\gamma$\,Dor stars matched 
the theoretical instability strips \OK{quite well. We did find a fraction of $\sim$20\% pulsators to occur outside the strips, 
in between the SPB and $\delta\,$Sct strip on the one hand, and below the $\gamma\,$Dor strip on the other. 
These pulsators have a dominant frequency expected for fast rotators pulsating in gravito-inertial modes 
\citep{Salmon2014,Saio2017-PL,Aerts2019}.}
For some of the $\beta$ Cep stars, we noticed 
discrepant positions with respect to their strip.
This may be caused by a
systematic bias of the $T_{\rm eff}$ and/or the luminosity $L$ as derived from 
\texttt{gspphot} data for the hottest stars. 
\OK{However, it may also suggest that the instability strips of OB-type stars are incompletely covered by the current 
physical descriptions of their interiors.}
The HR diagrams also revealed the presence
of rotationally modulated stars.
This is 
\OK{as expected since both ground-based data}
\citep[e.g.][]{Briquet2007} 
\OK{
and space photometry \citep[][among others]{Degroote2011,Balona2016,Bowman2018a,Balona2019}}
already showed that rotationally modulated
and pulsating 
stars co-exist inside the instability strips, and that the two populations
of stars 
\OK{largely} overlap. 

Our analysis of the period--luminosity relation of 
$\delta$\,Sct stars provides a good demonstration of the uniqueness of \textit{Gaia} data, as \gdr3
 is able to bring both ingredients to investigate this relation. To our knowledge,
the empirical PW relation in Fig.~\ref{fig:PWs} is the most extensive
reported in the literature for $\delta$\,Sct stars so far. We find evidence that the relation
has two different regimes depending on the period 
\OK{of the dominant
oscillation mode.}

Finally, we are able to confirm that stellar rotation has a direct impact on the 
\OK{observed}
oscillation amplitude of $\delta$\,Sct stars, in the sense that 
\OK{increasing}
rotation
\OK{decreases the amplitude. The $v\sin i$ data necessary to establish this result}
were 
derived from \gdr3 spectroscopic \texttt{esphs} data. We were able to quantify the order 
of magnitude of the gradient, arriving at roughly 46\,$\mu$mag per km\,s$^{-1}$. 

Despite the not yet perfect photometric calibration 
\OK{of the \gaia DR3 data,}
we were able to demonstrate the great
potential of 
\OK{combined and homogeneously sampled}
photometric, spectroscopic, and astrometric \textit{Gaia} data to investigate
OBAF-type main sequence pulsators, and we look forward to the 
\OK{more extensive}
\textit{Gaia} DR4 and DR5 data
for which we 
expect the time series 
\OK{to be less prone to} high-frequency instrumental variations.


\section*{Acknowledgements\label{sec:acknowl}}

This work presents results from the European Space Agency (ESA) space mission \gaia. \gaia\ data are being processed by the \gaia\ 
Data Processing and Analysis Consortium (DPAC). Funding for the DPAC is provided by national institutions, in particular the 
institutions participating in the \gaia\ MultiLateral Agreement (MLA). The \gaia\ mission website is \url{https://www.cosmos.esa.int/gaia}. 
The \gaia\ archive website is \url{https://archives.esac.esa.int/gaia}. 
Further acknowledgements are given in Appendix \ref{appendix:acknowledgements}.


\bibliographystyle{aa}
\bibliography{bibliography}

\appendix

\section{\label{appendix:acknowledgements}Acknowledgements}

The \gaia\ mission and data processing have financially been supported by, in alphabetical order by country:

\begin{itemize}
\item the Algerian Centre de Recherche en Astronomie, Astrophysique et G\'{e}ophysique of Bouzareah Observatory;
\item the Austrian Fonds zur F\"{o}rderung der wissenschaftlichen Forschung (FWF) Hertha Firnberg Programme through grants T359, P20046, and P23737;
\item the BELgian federal Science Policy Office (BELSPO) through various PROgramme de D\'{e}veloppement d'Exp\'{e}riences scientifiques (PRODEX) 
      grants, and the Research Foundation Flanders (Fonds Wetenschappelijk Onderzoek) through grant VS.091.16N, 
      and the Fonds de la Recherche Scientifique (FNRS), and the Research Council of Katholieke Universiteit (KU) Leuven through 
      grant C16/18/005 (Pushing AsteRoseismology to the next level with TESS, GaiA, and the Sloan DIgital Sky SurvEy -- PARADISE);  
\item the Brazil-France exchange programmes Funda\c{c}\~{a}o de Amparo \`{a} Pesquisa do Estado de S\~{a}o Paulo (FAPESP) and Coordena\c{c}\~{a}o de Aperfeicoamento de Pessoal de N\'{\i}vel Superior (CAPES) - Comit\'{e} Fran\c{c}ais d'Evaluation de la Coop\'{e}ration Universitaire et Scientifique avec le Br\'{e}sil (COFECUB);
\item the Chilean Agencia Nacional de Investigaci\'{o}n y Desarrollo (ANID) through Fondo Nacional de Desarrollo Cient\'{\i}fico y Tecnol\'{o}gico (FONDECYT) Regular Project 1210992 (L.~Chemin);
\item the National Natural Science Foundation of China (NSFC) through grants 11573054, 11703065, and 12173069, the China Scholarship Council through grant 201806040200, and the Natural Science Foundation of Shanghai through grant 21ZR1474100;  
\item the Tenure Track Pilot Programme of the Croatian Science Foundation and the \'{E}cole Polytechnique F\'{e}d\'{e}rale de Lausanne and the project TTP-2018-07-1171 `Mining the Variable Sky', with the funds of the Croatian-Swiss Research Programme;
\item the Czech-Republic Ministry of Education, Youth, and Sports through grant LG 15010 and INTER-EXCELLENCE grant LTAUSA18093, and the Czech Space Office through ESA PECS contract 98058;
\item the Danish Ministry of Science;
\item the Estonian Ministry of Education and Research through grant IUT40-1;
\item the European Commission’s Sixth Framework Programme through the European Leadership in Space Astrometry (\href{https://www.cosmos.esa.int/web/gaia/elsa-rtn-programme}{ELSA}) Marie Curie Research Training Network (MRTN-CT-2006-033481), through Marie Curie project PIOF-GA-2009-255267 (Space AsteroSeismology \& RR Lyrae stars, SAS-RRL), and through a Marie Curie Transfer-of-Knowledge (ToK) fellowship (MTKD-CT-2004-014188); the European Commission's Seventh Framework Programme through grant FP7-606740 (FP7-SPACE-2013-1) for the \gaia\ European Network for Improved data User Services (\href{https://gaia.ub.edu/twiki/do/view/GENIUS/}{GENIUS}) and through grant 264895 for the \gaia\ Research for European Astronomy Training (\href{https://www.cosmos.esa.int/web/gaia/great-programme}{GREAT-ITN}) network;
\item the European Cooperation in Science and Technology (COST) through COST Action CA18104 `Revealing the Milky Way with \gaia (MW-Gaia)';
\item the European Research Council (ERC) through grants 320360, 647208, and 834148 and through the European Union’s Horizon 2020 research and innovation and excellent science programmes through Marie Sk{\l}odowska-Curie grant 745617 (Our Galaxy at full HD -- Gal-HD) and 895174 (The build-up and fate of self-gravitating systems in the Universe) as well as grants 687378 (Small Bodies: Near and Far), 682115 (Using the Magellanic Clouds to Understand the Interaction of Galaxies), 695099 (A sub-percent distance scale from binaries and Cepheids -- CepBin), 716155 (Structured ACCREtion Disks -- SACCRED), 951549 (Sub-percent calibration of the extragalactic distance scale in the era of big surveys -- UniverScale), and 101004214 (Innovative Scientific Data Exploration and Exploitation Applications for Space Sciences -- EXPLORE);
\item the European Science Foundation (ESF), in the framework of the \gaia\ Research for European Astronomy Training Research Network Programme (\href{https://www.cosmos.esa.int/web/gaia/great-programme}{GREAT-ESF});
\item the European Space Agency (ESA) in the framework of the \gaia\ project, through the Plan for European Cooperating States (PECS) programme through contracts C98090 and 4000106398/12/NL/KML for Hungary, through contract 4000115263/15/NL/IB for Germany, and through PROgramme de D\'{e}veloppement d'Exp\'{e}riences scientifiques (PRODEX) grant 4000127986 for Slovenia;  
\item the Academy of Finland through grants 299543, 307157, 325805, 328654, 336546, and 345115 and the Magnus Ehrnrooth Foundation;
\item the French Centre National d’\'{E}tudes Spatiales (CNES), the Agence Nationale de la Recherche (ANR) through grant ANR-10-IDEX-0001-02 for the `Investissements d'avenir' programme, through grant ANR-15-CE31-0007 for project `Modelling the Milky Way in the \gaia era’ (MOD4Gaia), through grant ANR-14-CE33-0014-01 for project `The Milky Way disc formation in the \gaia era’ (ARCHEOGAL), through grant ANR-15-CE31-0012-01 for project `Unlocking the potential of Cepheids as primary distance calibrators’ (UnlockCepheids), through grant ANR-19-CE31-0017 for project `Secular evolution of galxies' (SEGAL), and through grant ANR-18-CE31-0006 for project `Galactic Dark Matter' (GaDaMa), the Centre National de la Recherche Scientifique (CNRS) and its SNO \gaia of the Institut des Sciences de l’Univers (INSU), its Programmes Nationaux: Cosmologie et Galaxies (PNCG), Gravitation R\'{e}f\'{e}rences Astronomie M\'{e}trologie (PNGRAM), Plan\'{e}tologie (PNP), Physique et Chimie du Milieu Interstellaire (PCMI), and Physique Stellaire (PNPS), the `Action F\'{e}d\'{e}ratrice \gaia' of the Observatoire de Paris, the R\'{e}gion de Franche-Comt\'{e}, the Institut National Polytechnique (INP) and the Institut National de Physique nucl\'{e}aire et de Physique des Particules (IN2P3) co-funded by CNES;
\item the German Aerospace Agency (Deutsches Zentrum f\"{u}r Luft- und Raumfahrt e.V., DLR) through grants 50QG0501, 50QG0601, 50QG0602, 50QG0701, 50QG0901, 50QG1001, 50QG1101, 50\-QG1401, 50QG1402, 50QG1403, 50QG1404, 50QG1904, 50QG2101, 50QG2102, and 50QG2202, and the Centre for Information Services and High Performance Computing (ZIH) at the Technische Universit\"{a}t Dresden for generous allocations of computer time;
\item the Hungarian Academy of Sciences through the Lend\"{u}let Programme grants LP2014-17 and LP2018-7 and the Hungarian National Research, Development, and Innovation Office (NKFIH) through grant KKP-137523 (`SeismoLab');
\item the Science Foundation Ireland (SFI) through a Royal Society - SFI University Research Fellowship (M.~Fraser);
\item the Israel Ministry of Science and Technology through grant 3-18143 and the Tel Aviv University Center for Artificial Intelligence and Data Science (TAD) through a grant;
\item the Agenzia Spaziale Italiana (ASI) through contracts I/037/08/0, I/058/10/0, 2014-025-R.0, 2014-025-R.1.2015, and 2018-24-HH.0 to the Italian Istituto Nazionale di Astrofisica (INAF), contract 2014-049-R.0/1/2 to INAF for the Space Science Data Centre (SSDC, formerly known as the ASI Science Data Center, ASDC), contracts I/008/10/0, 2013/030/I.0, 2013-030-I.0.1-2015, and 2016-17-I.0 to the Aerospace Logistics Technology Engineering Company (ALTEC S.p.A.), INAF, and the Italian Ministry of Education, University, and Research (Ministero dell'Istruzione, dell'Universit\`{a} e della Ricerca) through the Premiale project `MIning The Cosmos Big Data and Innovative Italian Technology for Frontier Astrophysics and Cosmology' (MITiC);
\item the Netherlands Organisation for Scientific Research (NWO) through grant NWO-M-614.061.414, through a VICI grant (A.~Helmi), and through a Spinoza prize (A.~Helmi), and the Netherlands Research School for Astronomy (NOVA);
\item the Polish National Science Centre through HARMONIA grant 2018/30/M/ST9/00311 and DAINA grant 2017/27/L/ST9/03221 and the Ministry of Science and Higher Education (MNiSW) through grant DIR/WK/2018/12;
\item the Portuguese Funda\c{c}\~{a}o para a Ci\^{e}ncia e a Tecnologia (FCT) through national funds, grants SFRH/\-BD/128840/2017 and PTDC/FIS-AST/30389/2017, and work contract DL 57/2016/CP1364/CT0006, the Fundo Europeu de Desenvolvimento Regional (FEDER) through grant POCI-01-0145-FEDER-030389 and its Programa Operacional Competitividade e Internacionaliza\c{c}\~{a}o (COMPETE2020) through grants UIDB/04434/2020 and UIDP/04434/2020, and the Strategic Programme UIDB/\-00099/2020 for the Centro de Astrof\'{\i}sica e Gravita\c{c}\~{a}o (CENTRA);  
\item the Slovenian Research Agency through grant P1-0188;
\item the Spanish Ministry of Economy (MINECO/FEDER, UE), the Spanish Ministry of Science and Innovation (MICIN), the Spanish Ministry of Education, Culture, and Sports, and the Spanish Government through grants BES-2016-078499, BES-2017-083126, BES-C-2017-0085, ESP2016-80079-C2-1-R, ESP2016-80079-C2-2-R, FPU16/03827, PDC2021-121059-C22, RTI2018-095076-B-C22, and TIN2015-65316-P (`Computaci\'{o}n de Altas Prestaciones VII'), the Juan de la Cierva Incorporaci\'{o}n Programme (FJCI-2015-2671 and IJC2019-04862-I for F.~Anders), the Severo Ochoa Centre of Excellence Programme (SEV2015-0493), and MICIN/AEI/10.13039/501100011033 (and the European Union through European Regional Development Fund `A way of making Europe') through grant RTI2018-095076-B-C21, the Institute of Cosmos Sciences University of Barcelona (ICCUB, Unidad de Excelencia `Mar\'{\i}a de Maeztu’) through grant CEX2019-000918-M, the University of Barcelona's official doctoral programme for the development of an R+D+i project through an Ajuts de Personal Investigador en Formaci\'{o} (APIF) grant, the Spanish Virtual Observatory through project AyA2017-84089, the Galician Regional Government, Xunta de Galicia, through grants ED431B-2021/36, ED481A-2019/155, and ED481A-2021/296, the Centro de Investigaci\'{o}n en Tecnolog\'{\i}as de la Informaci\'{o}n y las Comunicaciones (CITIC), funded by the Xunta de Galicia and the European Union (European Regional Development Fund -- Galicia 2014-2020 Programme), through grant ED431G-2019/01, the Red Espa\~{n}ola de Supercomputaci\'{o}n (RES) computer resources at MareNostrum, the Barcelona Supercomputing Centre - Centro Nacional de Supercomputaci\'{o}n (BSC-CNS) through activities AECT-2017-2-0002, AECT-2017-3-0006, AECT-2018-1-0017, AECT-2018-2-0013, AECT-2018-3-0011, AECT-2019-1-0010, AECT-2019-2-0014, AECT-2019-3-0003, AECT-2020-1-0004, and DATA-2020-1-0010, the Departament d'Innovaci\'{o}, Universitats i Empresa de la Generalitat de Catalunya through grant 2014-SGR-1051 for project `Models de Programaci\'{o} i Entorns d'Execuci\'{o} Parallels' (MPEXPAR), and Ramon y Cajal Fellowship RYC2018-025968-I funded by MICIN/AEI/10.13039/501100011033 and the European Science Foundation (`Investing in your future');
\item the Swedish National Space Agency (SNSA/Rymdstyrelsen);
\item the Swiss State Secretariat for Education, Research, and Innovation through the Swiss Activit\'{e}s Nationales Compl\'{e}mentaires and the Swiss National Science Foundation through an Eccellenza Professorial Fellowship (award PCEFP2\_194638 for R.~Anderson);
\item the United Kingdom Particle Physics and Astronomy Research Council (PPARC), the United Kingdom Science and Technology Facilities Council (STFC), and the United Kingdom Space Agency (UKSA) through the following grants to the University of Bristol, the University of Cambridge, the University of Edinburgh, the University of Leicester, the Mullard Space Sciences Laboratory of University College London, and the United Kingdom Rutherford Appleton Laboratory (RAL): PP/D006511/1, PP/D006546/1, PP/D006570/1, ST/I000852/1, ST/J005045/1, ST/K00056X/1, ST/\-K000209/1, ST/K000756/1, ST/L006561/1, ST/N000595/1, ST/N000641/1, ST/N000978/1, ST/\-N001117/1, ST/S000089/1, ST/S000976/1, ST/S000984/1, ST/S001123/1, ST/S001948/1, ST/\-S001980/1, ST/S002103/1, ST/V000969/1, ST/W002469/1, ST/W002493/1, ST/W002671/1, ST/W002809/1, and EP/V520342/1.
\end{itemize}

The GBOT programme  uses observations collected at (i) the European Organisation for Astronomical Research in the Southern Hemisphere (ESO) with the VLT Survey Telescope (VST), under ESO programmes
092.B-0165,
093.B-0236,
094.B-0181,
095.B-0046,
096.B-0162,
097.B-0304,
098.B-0030,
099.B-0034,
0100.B-0131,
0101.B-0156,
0102.B-0174, and
0103.B-0165;
and (ii) the Liverpool Telescope, which is operated on the island of La Palma by Liverpool John Moores University in the Spanish Observatorio del Roque de los Muchachos of the Instituto de Astrof\'{\i}sica de Canarias with financial support from the United Kingdom Science and Technology Facilities Council, and (iii) telescopes of the Las Cumbres Observatory Global Telescope Network.

In addition we thank Dr.~Eric Gosset as well as the anonymous referee for useful comments. 

For an up-to-date list of acknowledgements, contributors, and former contributors to the \gaia project, 
we refer to the following URL: 
\url{https://gea.esac.esa.int/archive/documentation/GDR3/Miscellaneous/sec_acknowl/}

This work made use of Python (Python Software Foundation) and the IPython software \citep{IPython2007}, 
as well as of the Python packages Numpy \citep{harris2020}, Pandas \citep{Mckinney2010}, Matplotlib \citep{Hunter2007}, 
Astropy \citep{Astropy2013,Astropy2018}. In addition we made use of the visualisation software \href{}{TOPCAT} \citep{Taylor2005},
and the National Aeronautics and Space Administration (NASA) Astrophysics Data System (ADS).

\section{\label{appendix:kepleranalysis}Gaia DR3 data of asteroseismically modelled {\it Kepler\/} $\gamma\,$Dor and SPB stars}

We extracted the \gdr3 time series in the Gaia \g-band for 63
well-known bona fide g-mode pulsators assembled in \citet{Aerts2021-IGW}. 
This sample consists of 37 $\gamma$\,Dor and 26 SPB pulsators whose internal structure and 
evolutionary stage have been modelled asteroseismically using \textit{Kepler} 
photometric time series that were reduced by \citet{VanReeth2015} and 
\citet{Pedersen2021}.  
The {\it Kepler\/} light curves typically have a total time base of $\sim 1470\,$d and between some
24\,000 and 66\,000 data points, with an even sampling time of $\sim 30$ minutes
\citep{Koch2010}. The \gdr3 epoch photometric data of these g-mode pulsators
is sparsely sampled, containing between 34 and 52 data points spread over a
time base of between about 910 and 920\,d.
We re-analysed these {\it Kepler\/} light curves in the Fourier
domain in the same way as was done for the \gdr3 \g-band epoch photometry,
using the generalised Lomb-Scargle periodogram \citep{Zechmeister2009}, and extracted the 
two dominant frequencies in the interval $[0,25]\,$d$^{-1}$. After derivation of the
dominant periodic signal, the second strongest frequency was extracted after
prewhitening of the dominant frequency and its harmonics up to fourth order.
\begin{figure}
\begin{center} 
\resizebox{\hsize}{!}{\includegraphics{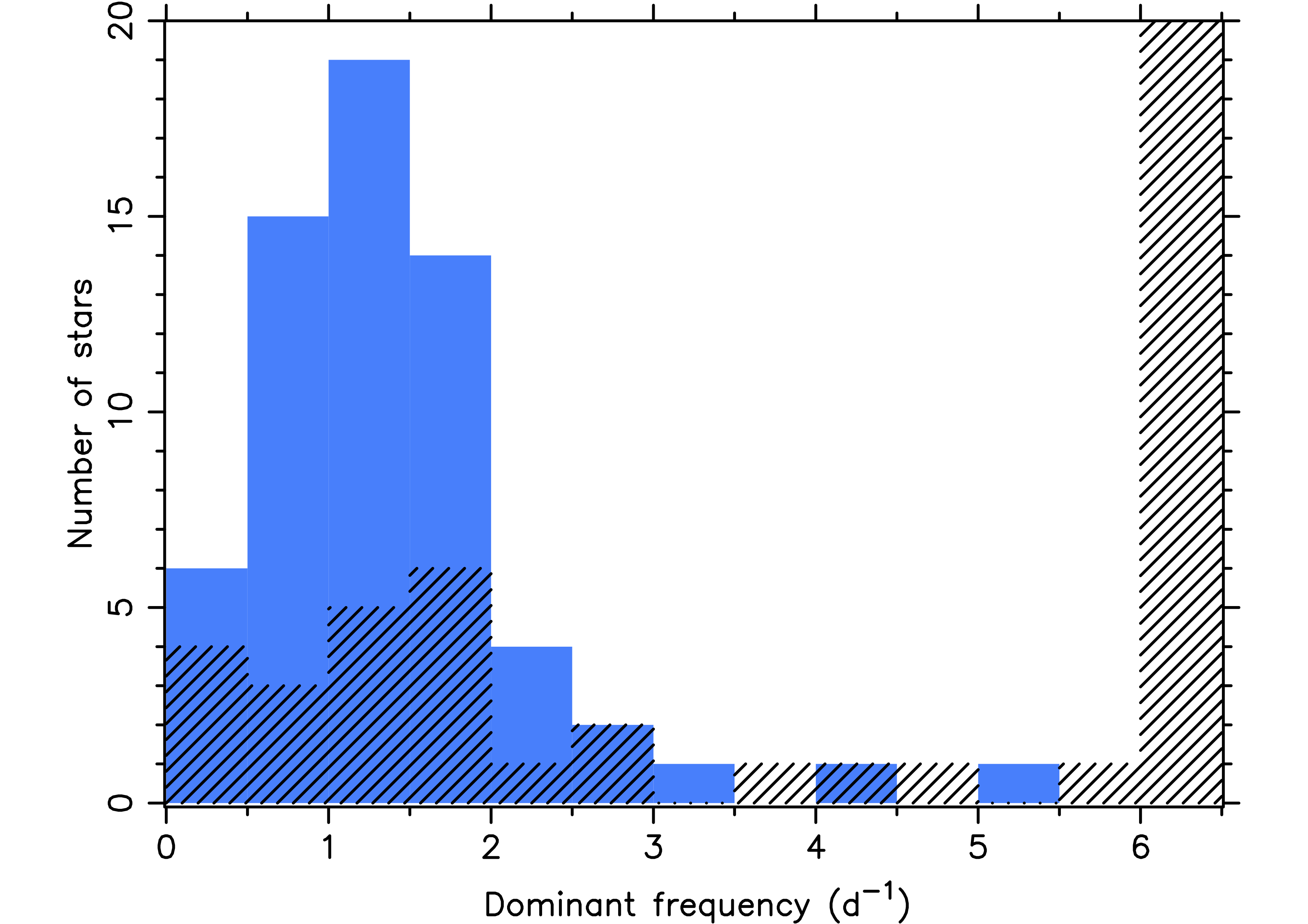}}
\end{center}
\caption{\label{histogram} Histogram of the dominant frequency in the {\it
    Kepler\/} light curves of 63 asteroseismically modelled g-mode
  pulsators (blue) compared with the dominant frequency occurring in their Gaia
  DR3 G-band epoch photometry (black hatched). For reasons of visibility, the
  $x-$axis is cut at 6.5\,d$^{-1}$ and the $y-$axis at 20; 38 of the 63 g-mode
  pulsators have their Gaia data dominated by instrumental effects with dominant
  frequency above 6\,d$^{-1}$ rather than by their g modes.}
\end{figure}
\begin{figure}
\begin{center} 
\resizebox{\hsize}{!}{\includegraphics{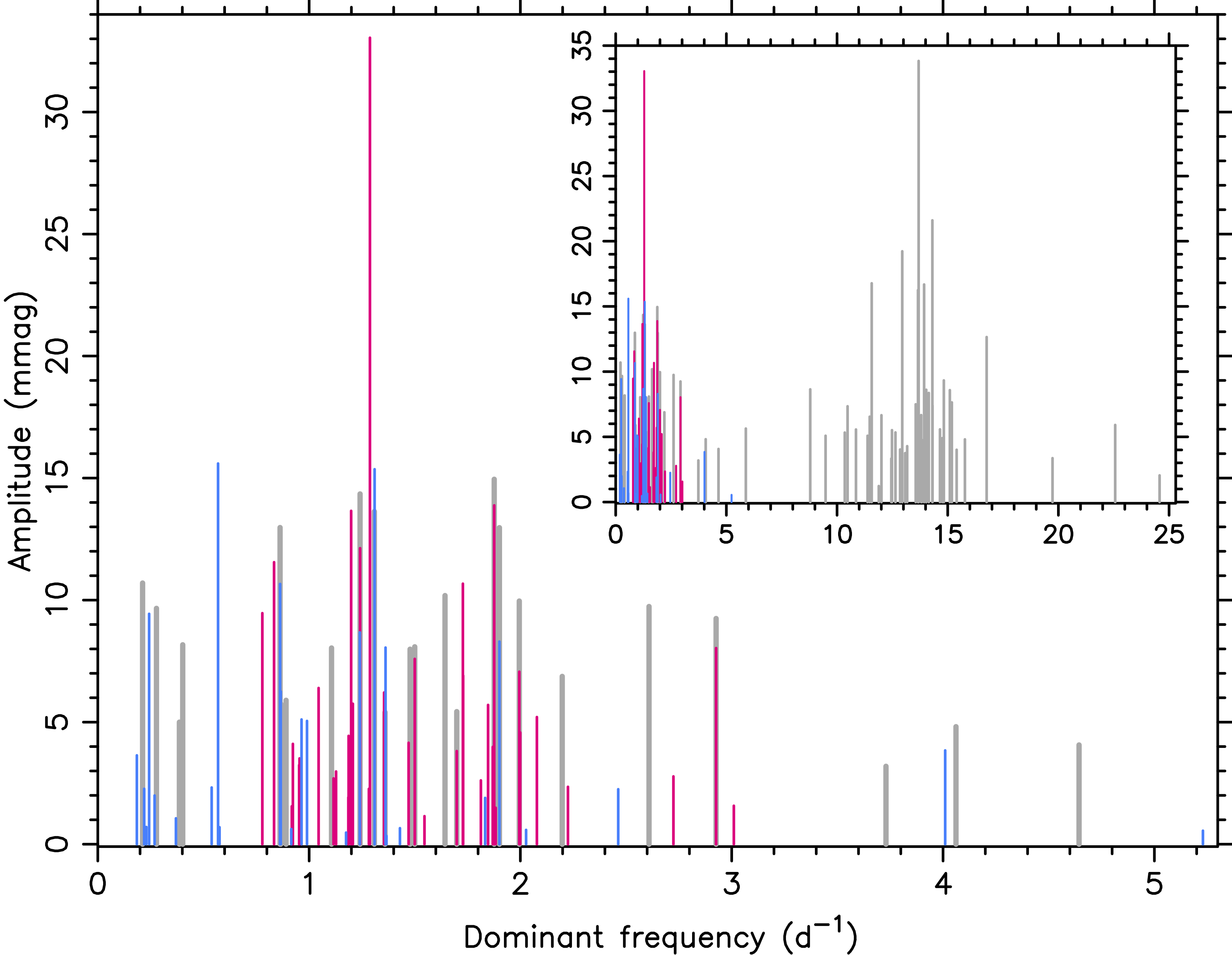}}
\end{center}
\caption{\label{amplitudes} Amplitudes of the dominant frequencies in the Gaia
 G-band DR3 light curves (grey) and in the {\it Kepler\/} light curves of the
 $\gamma\,$Dor (pink) and SPB (blue) pulsators. The inset shows the entire
 frequency range, while the main panel focuses on the range of the 
true dominant g-mode
 frequencies of the 63 stars.} 
\end{figure}
\begin{figure}
\begin{center} 
\resizebox{\hsize}{!}{\includegraphics{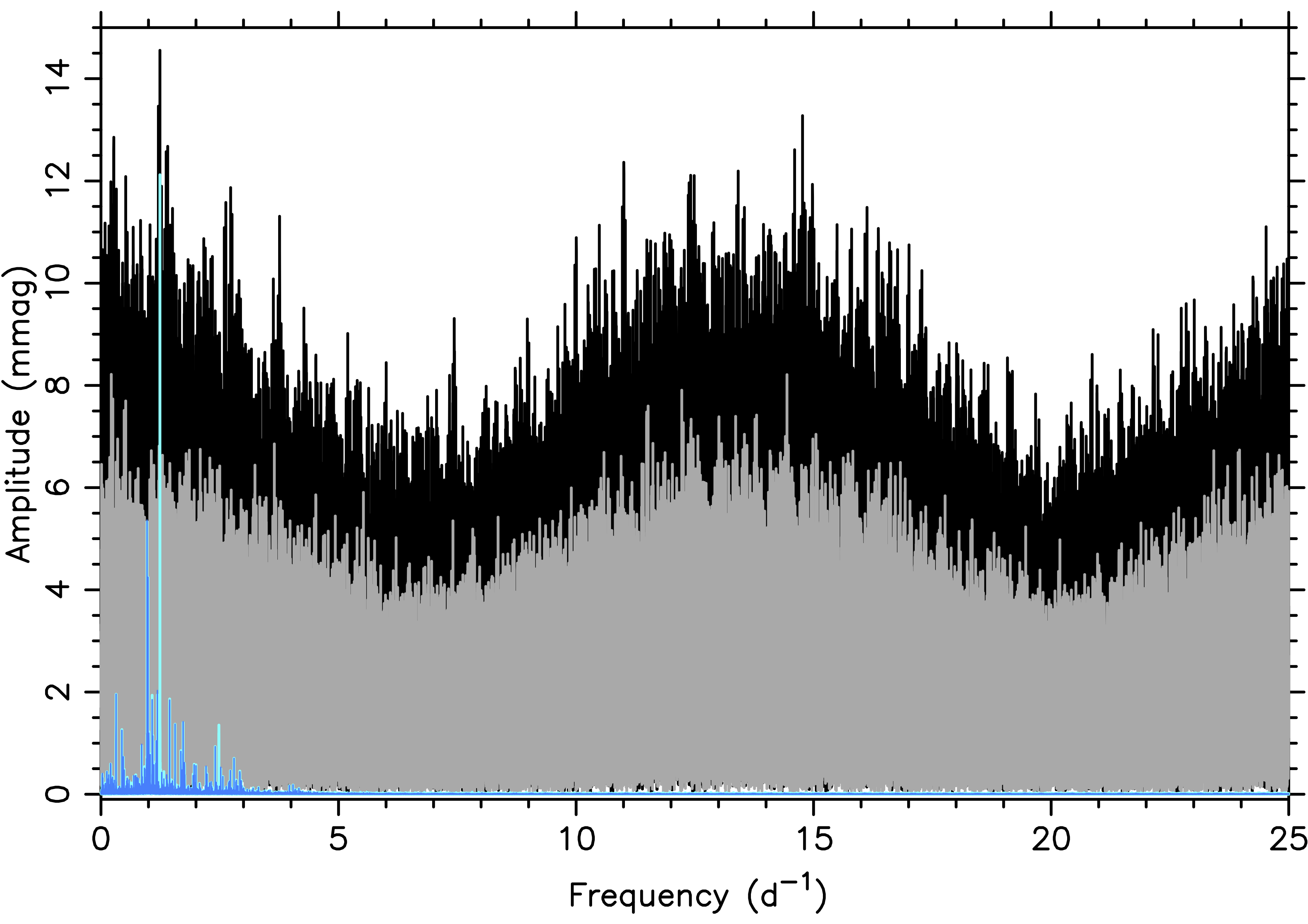}}
\resizebox{\hsize}{!}{\includegraphics{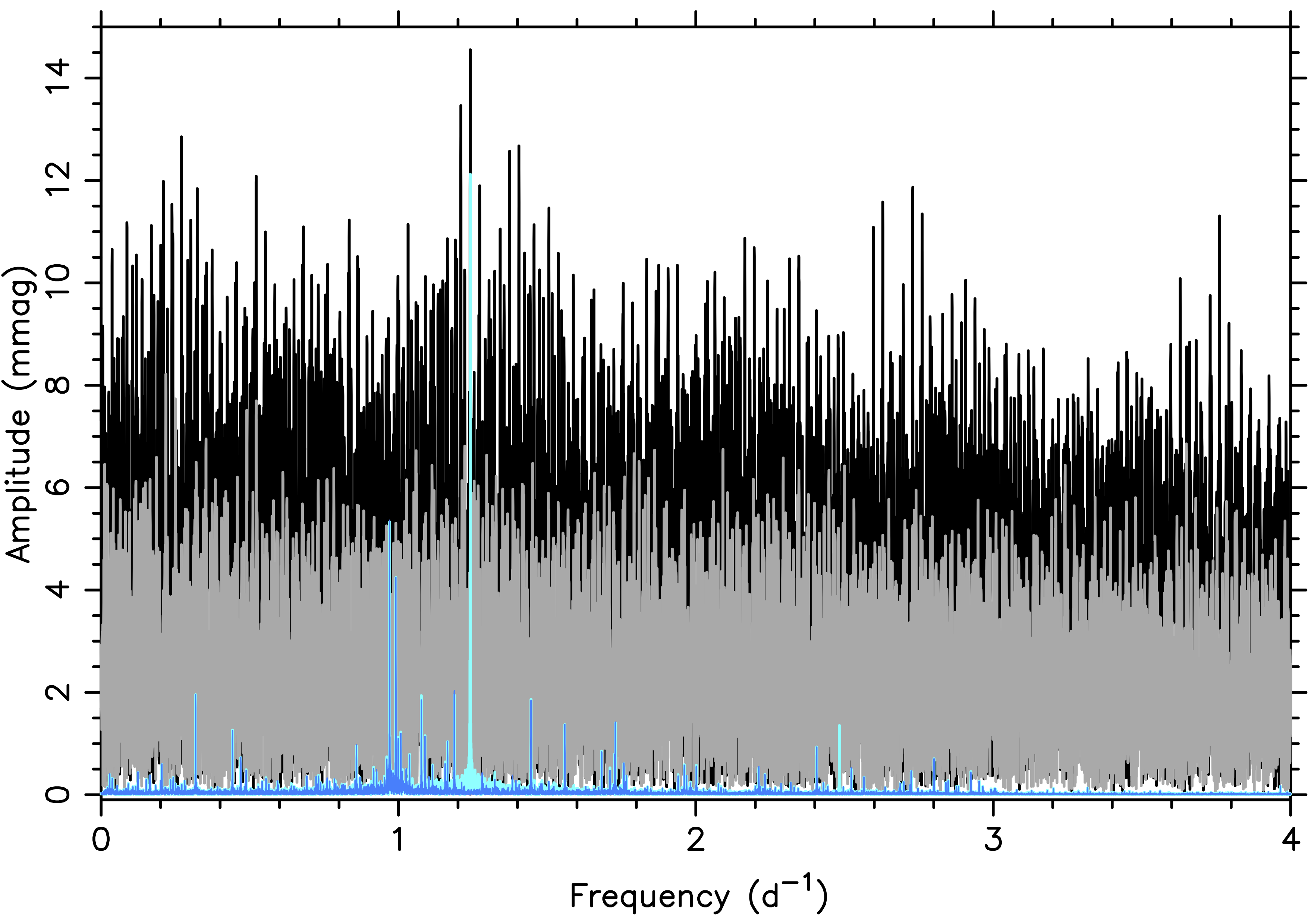}}
\resizebox{\hsize}{!}{\includegraphics{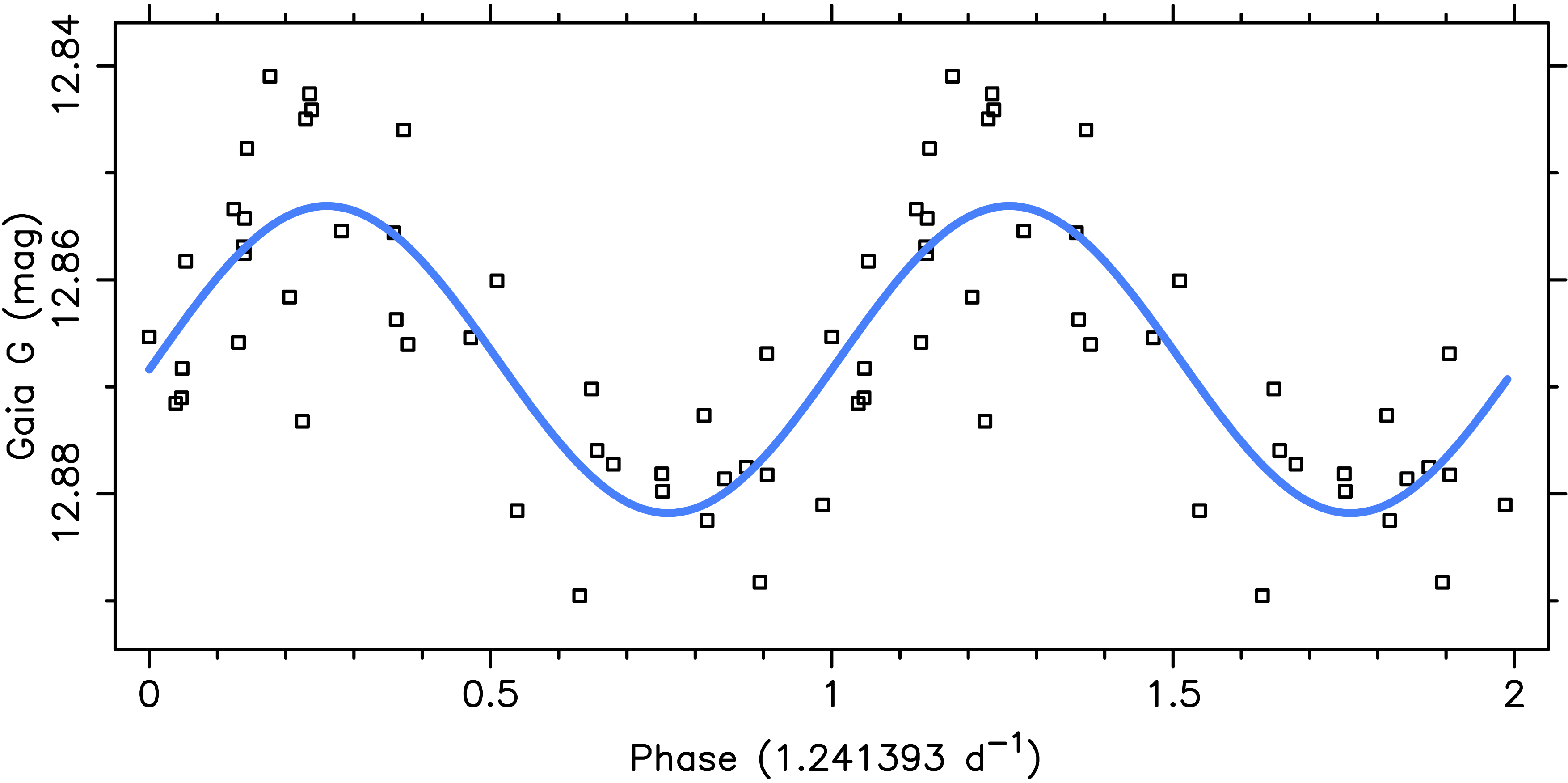}}
\end{center}
\caption{\label{kic11080103} Four Lomb-Scargle periodograms (upper: full
  frequency range; middle; zoomed version) of the $\gamma\,$Dor star
  KIC\,11080103. The black and cyan curves stand for the Gaia DR3 G-band and
  {\it Kepler\/} light curves, respectively. The grey and blue periodograms
  result from prewhitening the Gaia G and {\it Kepler\/} data with the dominant
  frequency. The lower panel shows the Gaia DR3 G-band data (black squares)
  folded with the dominant frequency detected in common in both light curves; a
  harmonic fit with that frequency is overplotted (full blue line). For
  visibility purposes, the phase is shown for two cycles. }
\end{figure}
\begin{figure}
\begin{center} 
\resizebox{\hsize}{!}{\includegraphics{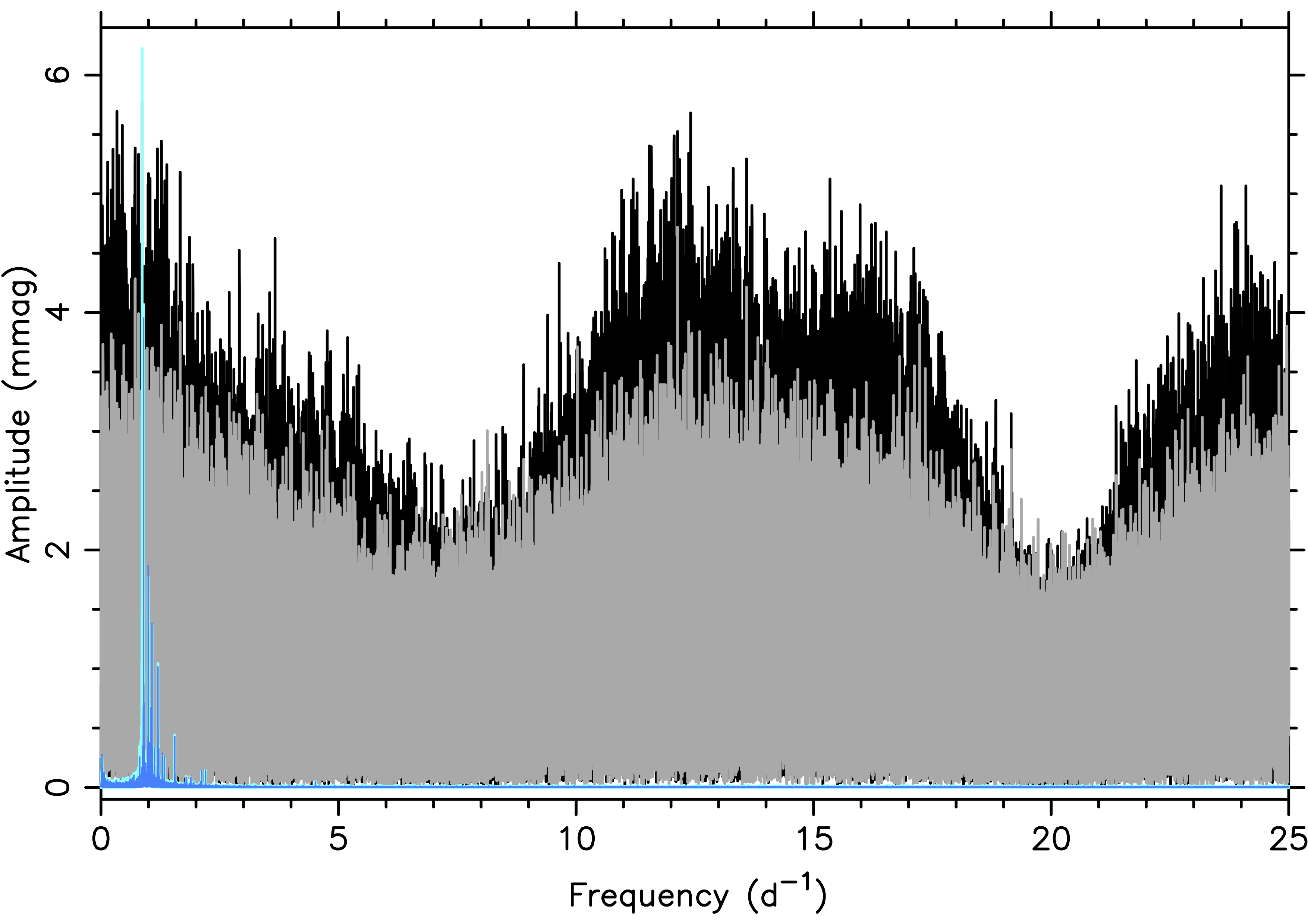}}
\resizebox{\hsize}{!}{\includegraphics{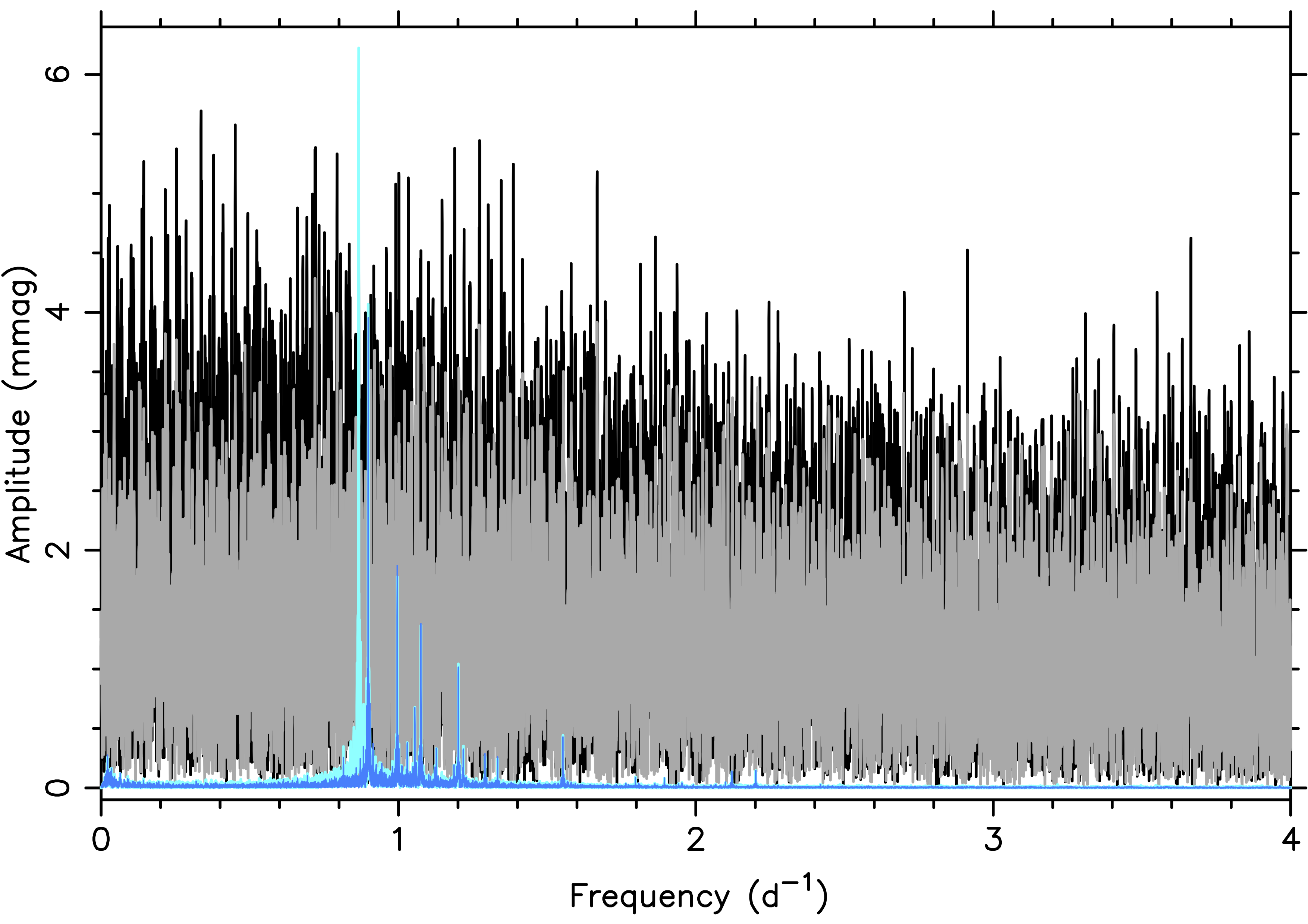}}
\resizebox{\hsize}{!}{\includegraphics{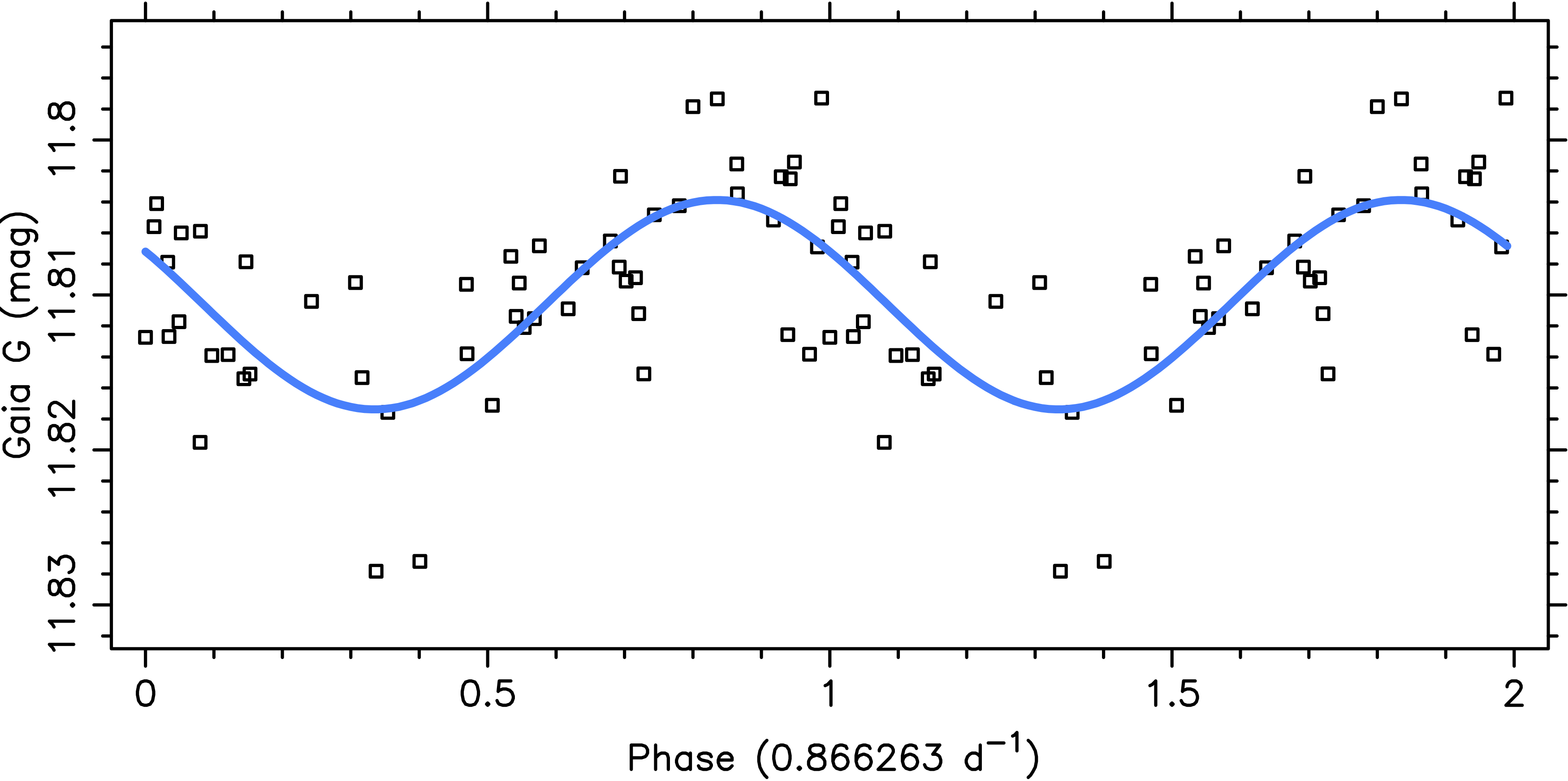}}
\end{center}
\caption{\label{kic4936089} Same as Fig.\,\ref{kic11080103}, but for the 
SPB star KIC\,4936089.}
\end{figure}
\begin{figure*}
\begin{center} 
  \begin{tabular}{cc}
    \includegraphics[width=85mm]{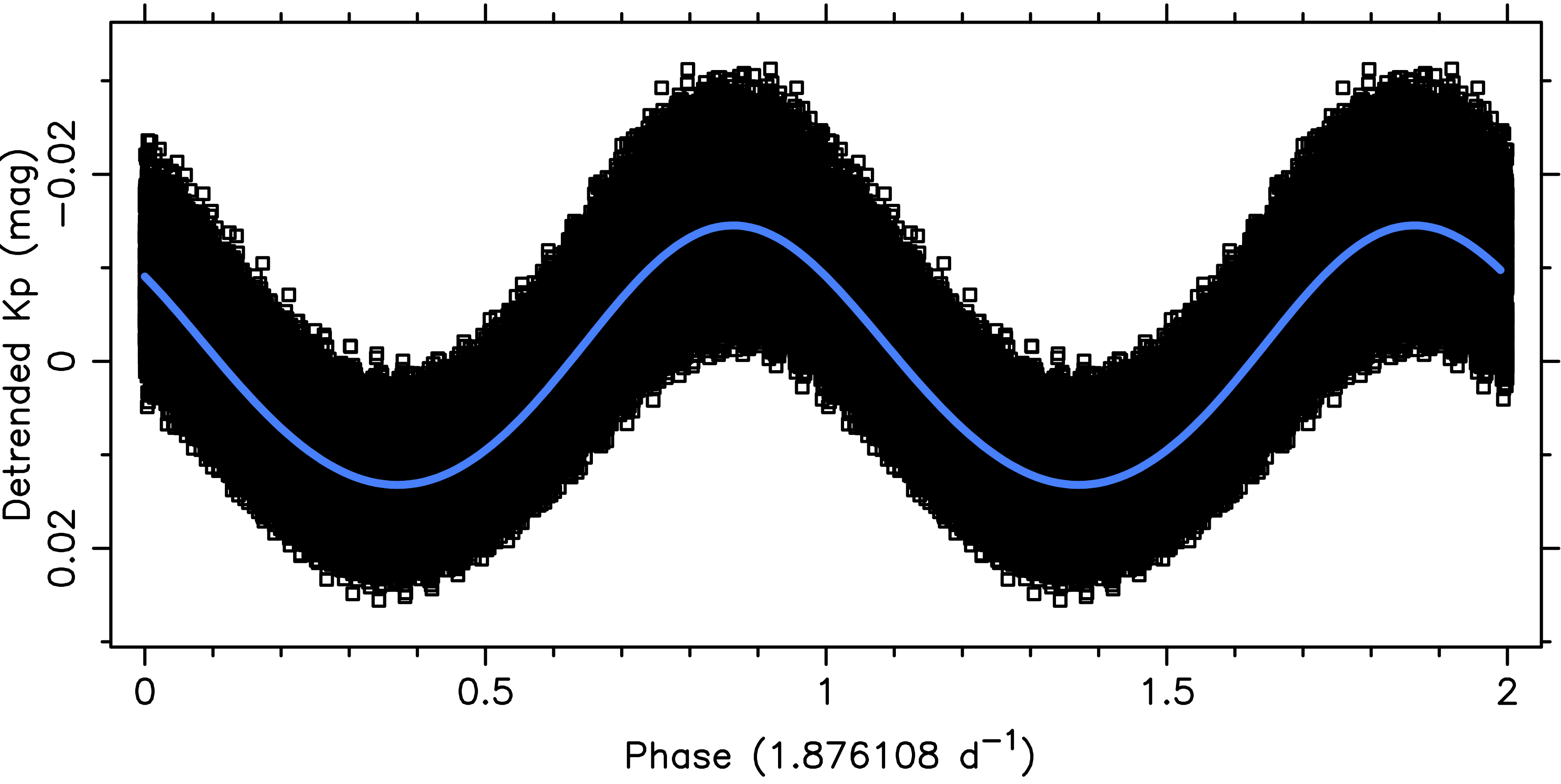} & \includegraphics[width=85mm]{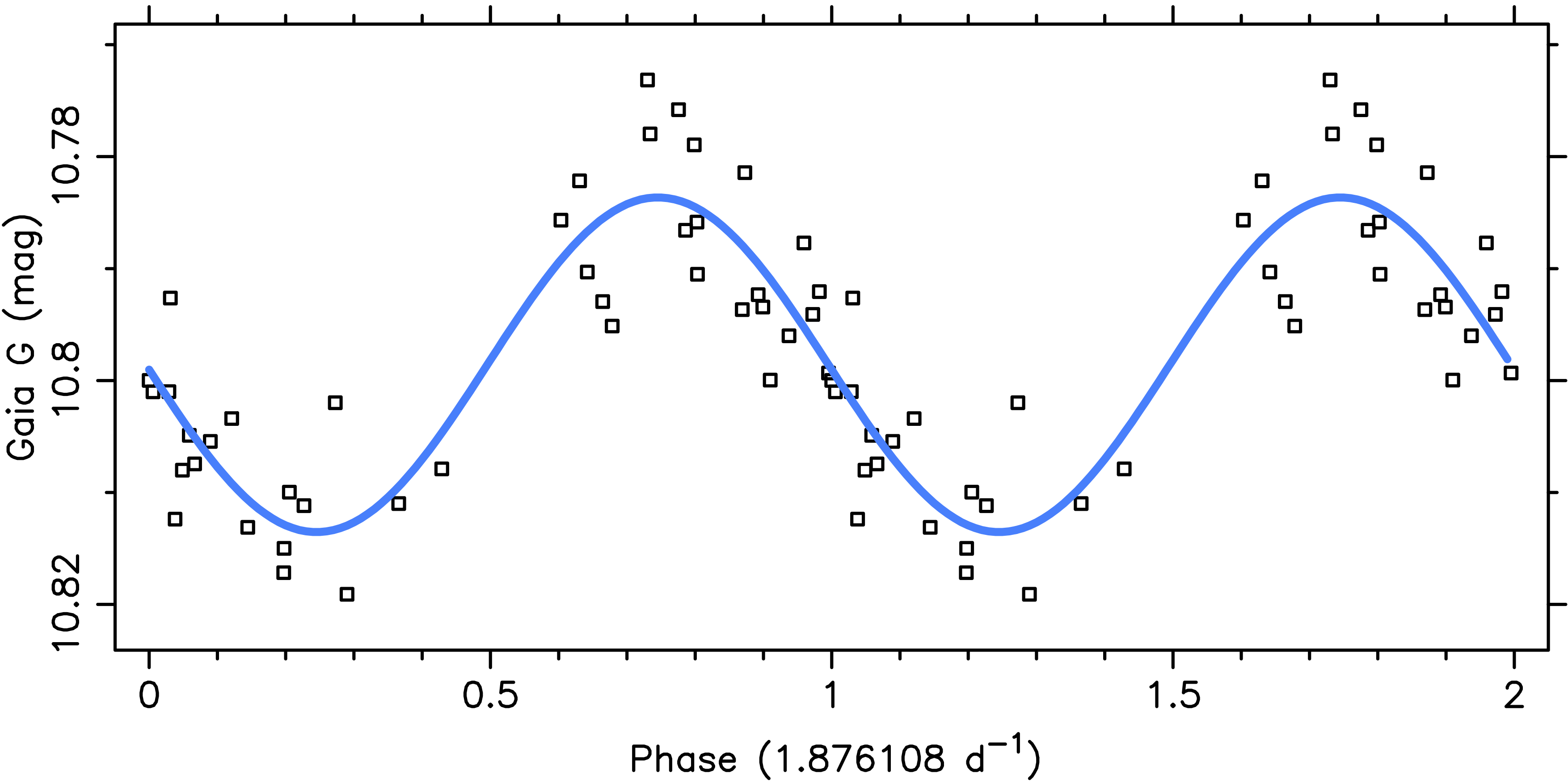} \\
   \includegraphics[width=85mm]{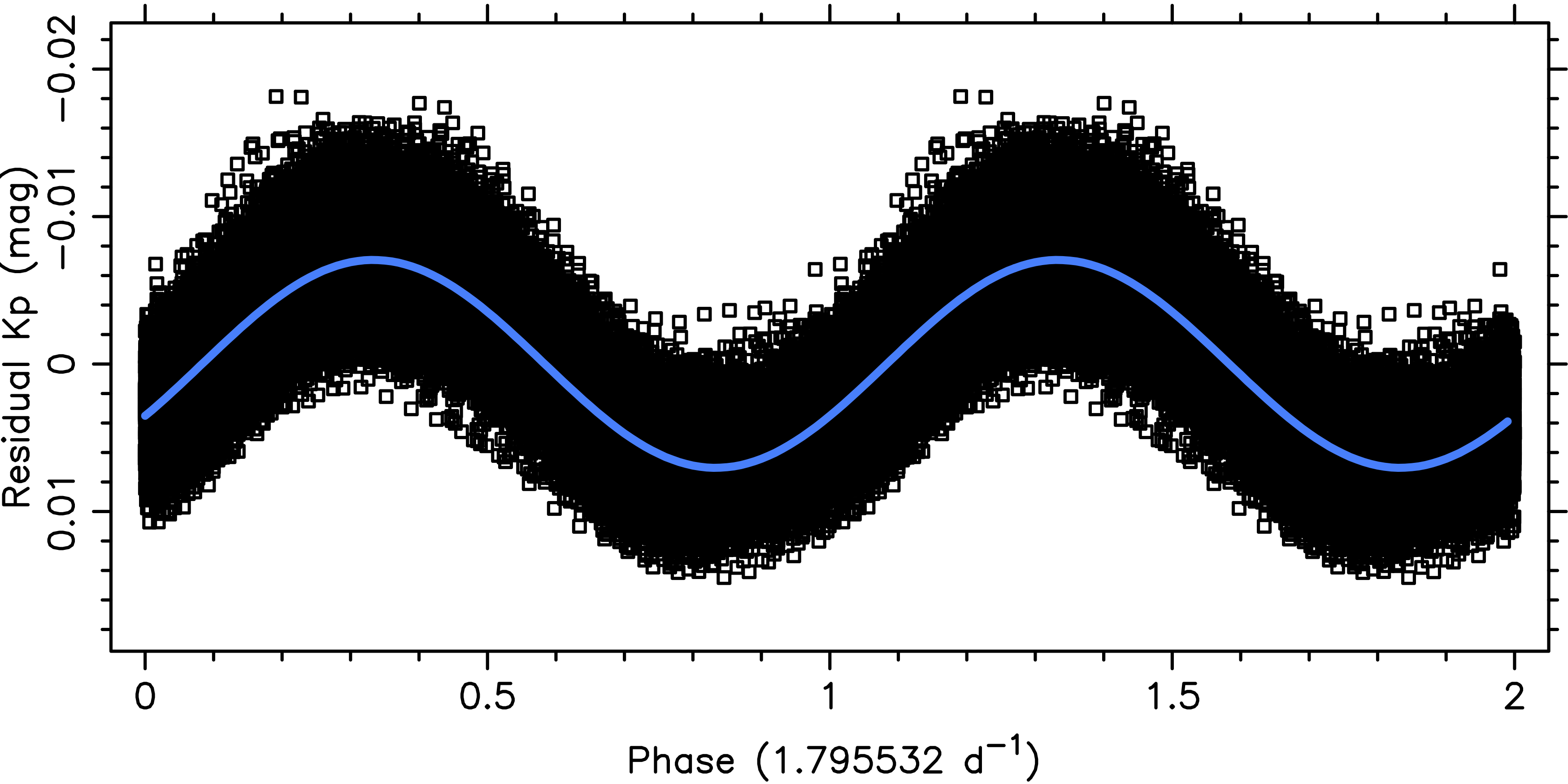}  & \includegraphics[width=85mm]{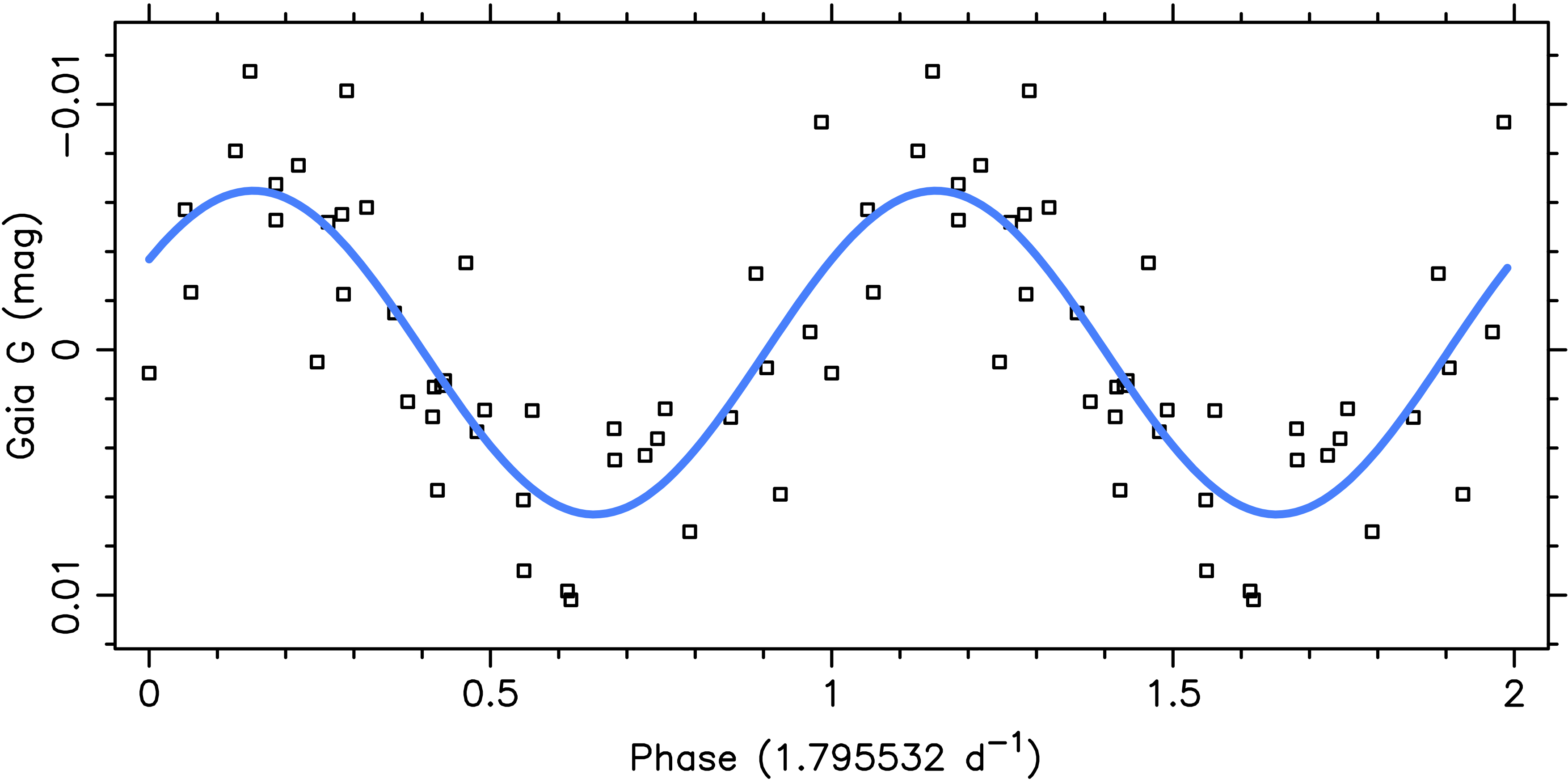}
  \end{tabular}
\end{center}
\caption{\label{kic7023122} Four phase diagrams for the $\gamma\,$Dor pulsator
  KIC\,7023122 whose two dominant g-mode frequencies (as listed in the legend of
  the $x-$axes) occur consistently in the periodograms of the Gaia DR3 G-band
  and {\it Kepler\/} photometry. The data are shown as black squares and the
  best harmonic fits for the fixed frequencies as blue lines.  For visibility
  purposes, the phases are shown for two cycles. }
\end{figure*}
\begin{figure*}
\begin{center} 
\begin{tabular}{cc}
  \includegraphics[width=85mm]{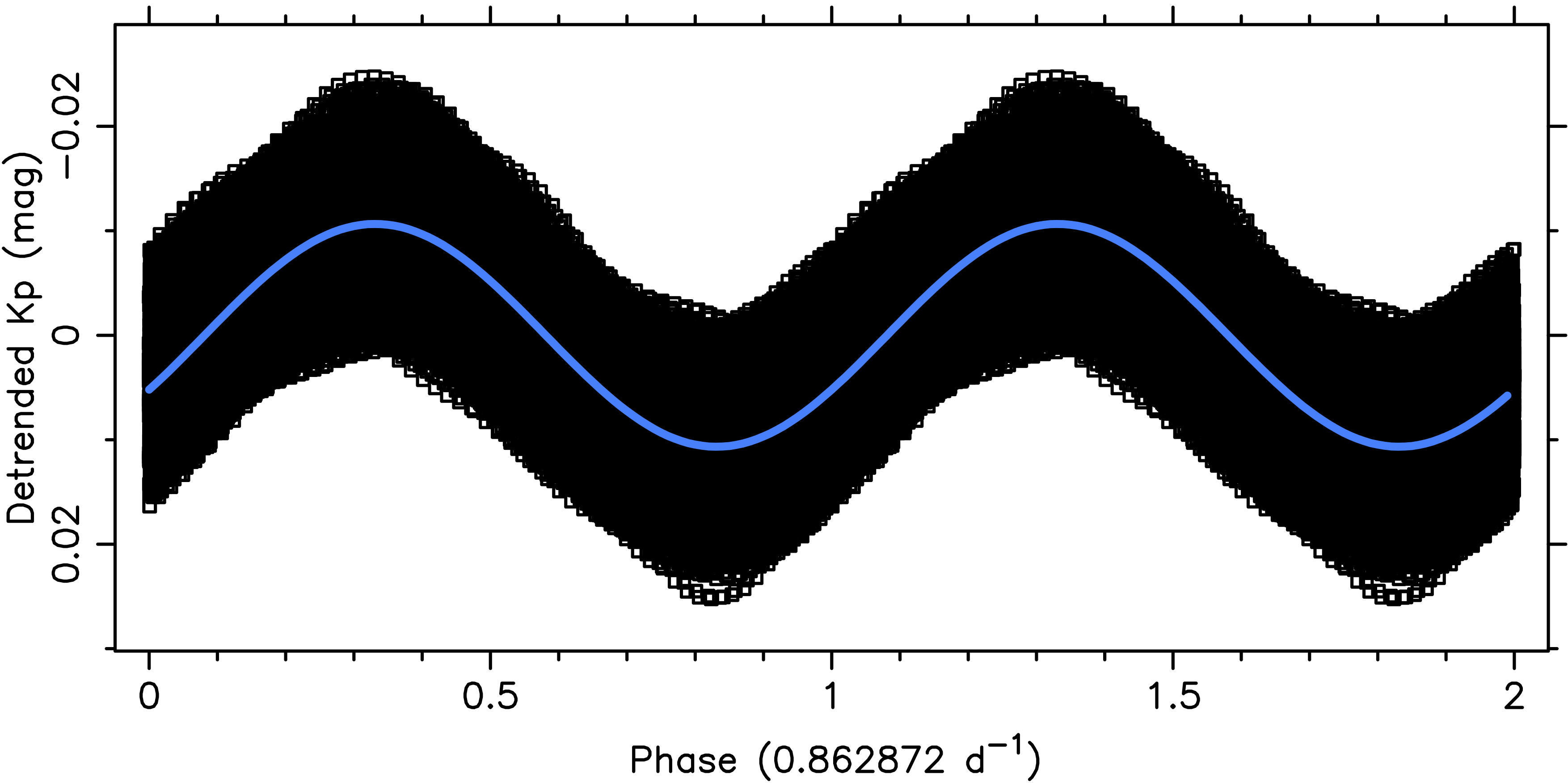} & \includegraphics[width=85mm]{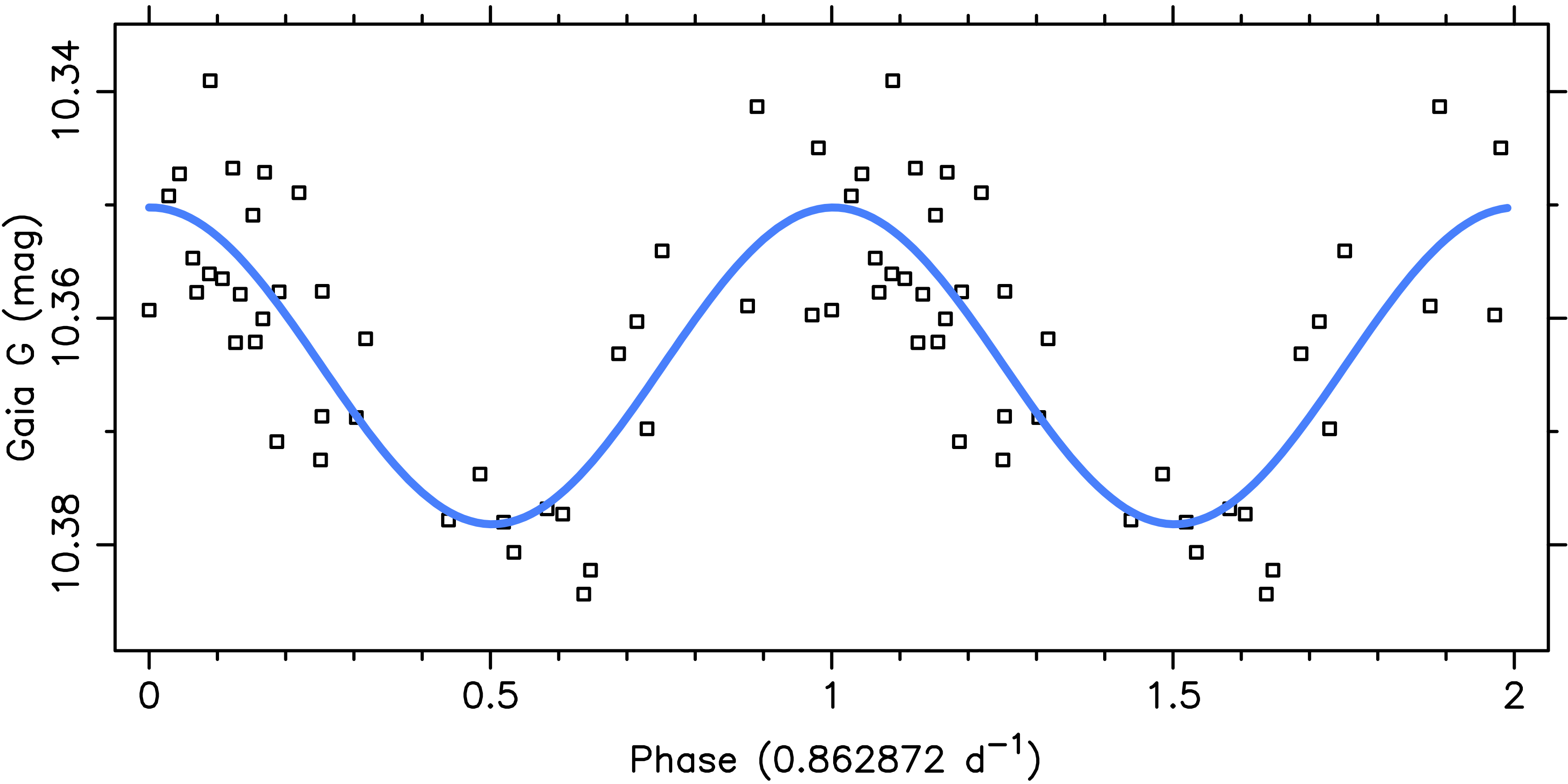} \\
  \includegraphics[width=85mm]{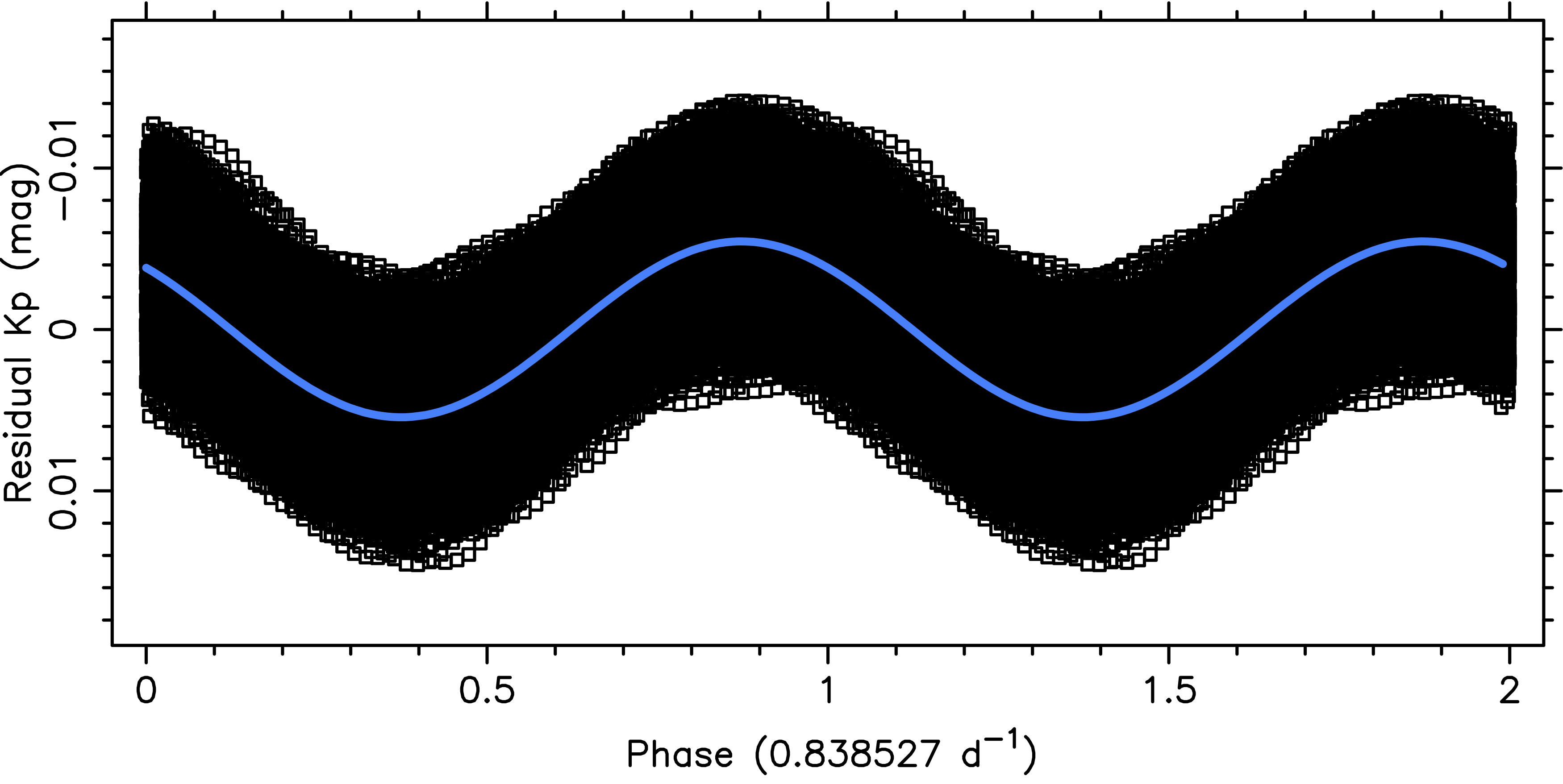} & \includegraphics[width=85mm]{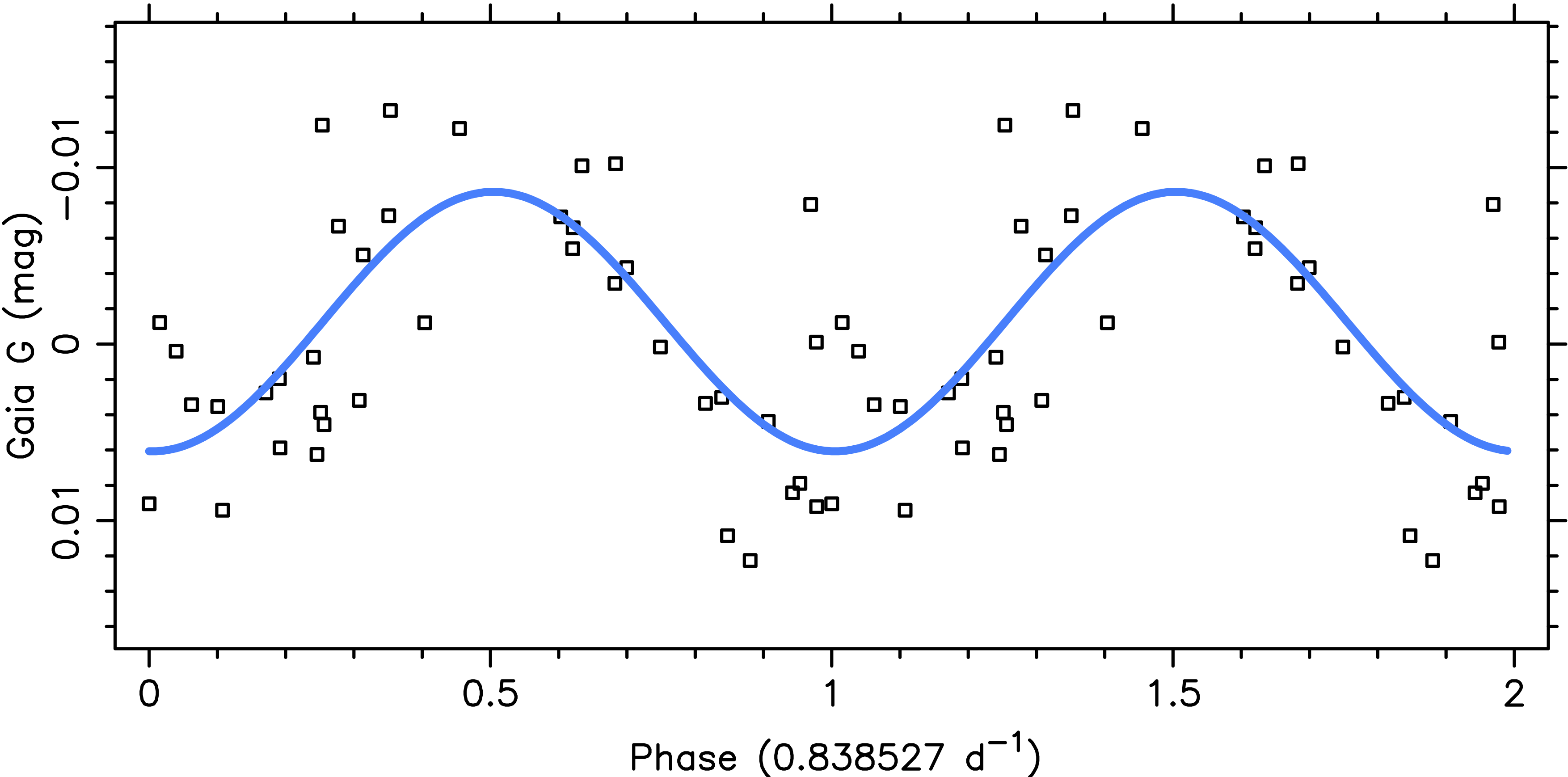} 
\end{tabular}
\end{center}
\caption{\label{kic7760680} Same as Fig.\,\ref{kic7023122} but for the SPB star
  KIC\,7760680.}
\end{figure*}

Given that the {\it Kepler\/} data are almost free of any aliasing, while
instrumental effects occur at the level of only a few $\mu$mag, we {know\/}
the two dominant frequencies of these 63 g-mode pulsating dwarfs up to high
precision. This allows us to assess the occurrence of the
instrumental effects present in the Gaia DR3 G-band epoch photometry, which
occur at mmag level and hence contaminate the pulsational g-mode signal quite
severely. Figure\,\ref{histogram} shows a histogram with the 63 pulsators
according to their dominant g-mode frequency deduced from the {\it Kepler\/}
data (in blue). The dominant frequency in the Gaia data is indicated by the
black hatched histogram and reveals that most of the detected dominant
frequencies occur above 6\,d$^{-1}$ indicating that they are either aliased 
frequencies or frequencies of instrumental origin.

The passband of the {\it Kepler\/} CCDs is somewhat bluer and narrower than the
Gaia \g-band. Moreover, the 63 g-mode pulsators each reveal tens of g modes with
accompanying multiperiodic beating in time \citep{VanReeth2015,Pedersen2021}.
The amplitude of the dominant signal in the {\it Kepler\/} and Gaia \g-band
are therefore not expected to be equal.  Nevertheless, they are of the same order, and so it is
meaningful to consider their distributions. This is revealed in
Fig.\,\ref{amplitudes}, which shows the dominant frequencies detected in the 126
light curves, in grey for the Gaia DR3 G-band data and in colour for the {\it
  Kepler\/} data (pink for the 37 $\gamma\,$Dor stars and blue for the 26 SPB
stars). It can be seen that the majority of the mode frequencies occurs in the
range $[0.2,3.1]$\,d$^{-1}$, as is well known for gravito-inertial modes in rotating
stars \citep{Aerts2021-IGW}. The dominant modes of these 63 well-known g-mode
pulsators have amplitudes covering the range $[0.5,33]$\,mmag, with the majority well below 10\,mmag. While several of
the dominant frequencies detected in the Gaia G-band occur in the appropriate
frequency range, Figure\,\ref{amplitudes} clearly shows the presence of peaks
at high frequencies, which are either aliased frequencies or periodic instrumental
artefacts.

Close inspection of Fig.\,\ref{amplitudes} reveals that 6 of the 37 $\gamma\,$Dor
stars (16\%) and 4 of the 26 SPB stars (15\%) share the same dominant intrinsic g-mode
frequency in their {\it Kepler\/} and Gaia G-band data. This is illustrated for
the $\gamma\,$Dor star KIC\,11080103 in Fig.\,\ref{kic11080103} and for the SPB
KIC\,4936089 in Fig.\,\ref{kic4936089}. KIC\,11080103 has a dominant prograde dipole
mode with frequency 1.241393\,d$^{-1}$ and amplitude of 12.13\,mmag in the {\it
Kepler\/} band \citep{VanReeth2015}. The SPB star KIC\,4936089, whose periodograms 
are shown in
Fig.\,\ref{kic4936089}, has a dominant g-mode frequency of 0.866263\,d$^{-1}$
with an amplitude of 6.26\,mmag. Its {\it Kepler\/} light curve revealed a
period spacing pattern consisting of 13 zonal dipole modes of consecutive radial
order.  This SPB star ranks number 8 of 26 in terms of dominant mode
amplitude, yet the 52 Gaia \g data points do allow to pick up the dominant mode,
thanks to relatively modest instrumental contamination for this star
(Fig.\,\ref{kic4936089}). These two examples show that the amplitude of the
dominant frequency peak in the periodogram is not a good criterion to select
g-mode pulsators (compare the upper panels of Figs\,\ref{kic11080103} and
\ref{kic4936089}). 

KIC\,11080103 and KIC\,4936089 are representative for six of the other cases
where {\it Kepler\/} and Gaia G-band data lead to the same dominant frequency,
with similar morphologies in the periodograms as those shown in
Figs\,\ref{kic11080103} and \ref{kic4936089}.  It concerns the four
$\gamma\,$Dor stars KIC\,3448365, KIC\,7365537, KIC\,7434470, KIC\,9480469, with
dominant g-mode frequencies of 1.500150\,d$^{-1}$, 2.925633\,d$^{-1}$,
1.698729\,d$^{-1}$, 1.994846\,d$^{-1}$, respectively, and the two SPB stars
KIC\,5941844 and KIC\,9020774 with dominant frequencies of 1.309558\,d$^{-1}$
and 1.900723\,d$^{-1}$.

For one single $\gamma\,$Dor star, KIC\,7023122, and one single SPB star, 
KIC\,7760680, also the second strongest frequencies coincide in the {\it Kepler\/} 
and Gaia G data.
The four phase diagrams of these two `best cases' among the g-mode pulsators are shown in
Figs\,\ref{kic7023122} and \ref{kic7760680}. These two examples show that,
despite the limited number of measurements in the DR3 time series, Gaia's G-band
data already allow us to detect multiperiodic non-radial oscillations at mmag
level, albeit it for a very small fraction (3\%) of pulsating dwarfs pulsators.

\begin{figure}
\begin{center} 
\resizebox{\hsize}{!}{\includegraphics{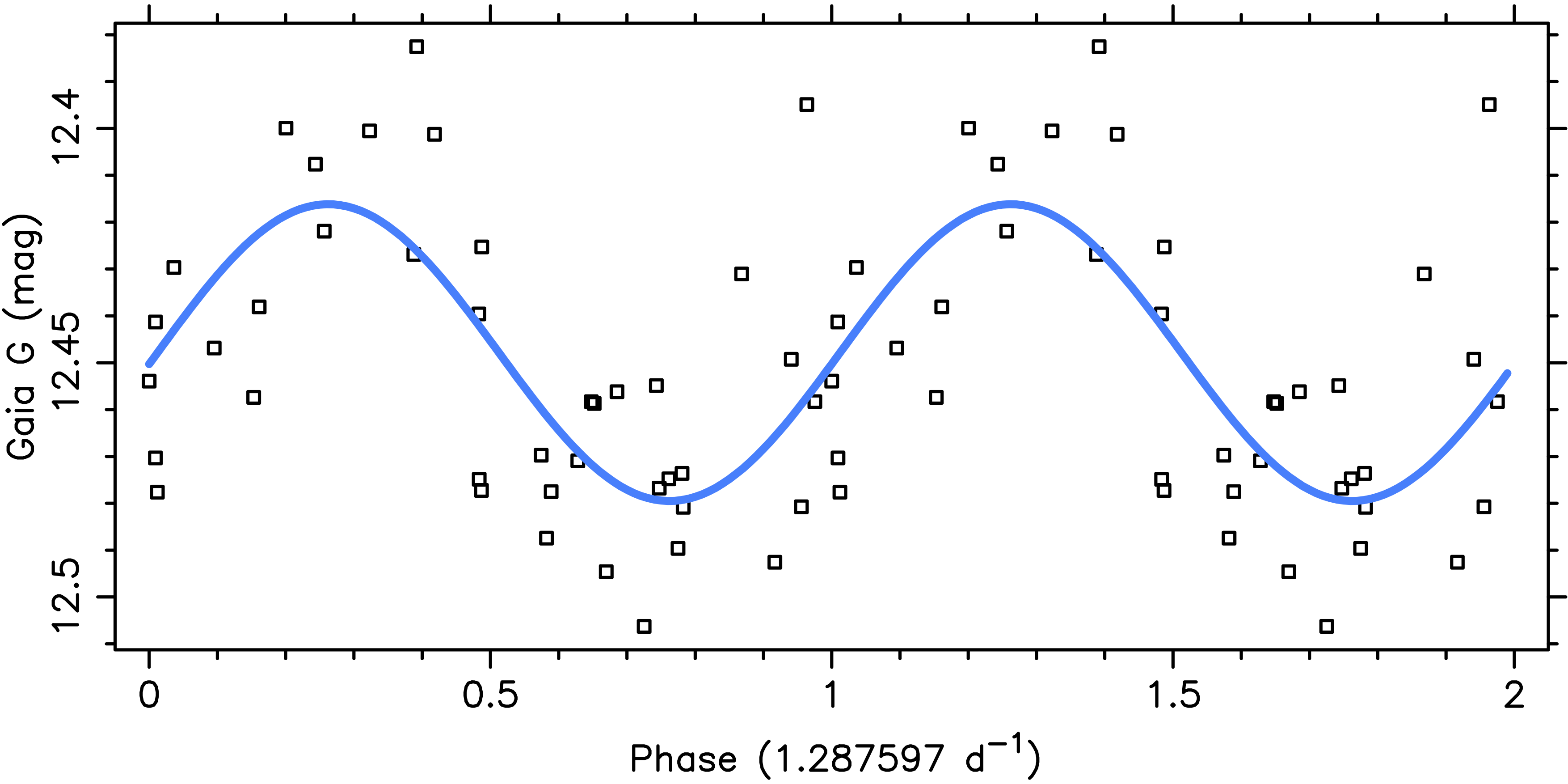}}
\resizebox{\hsize}{!}{\includegraphics{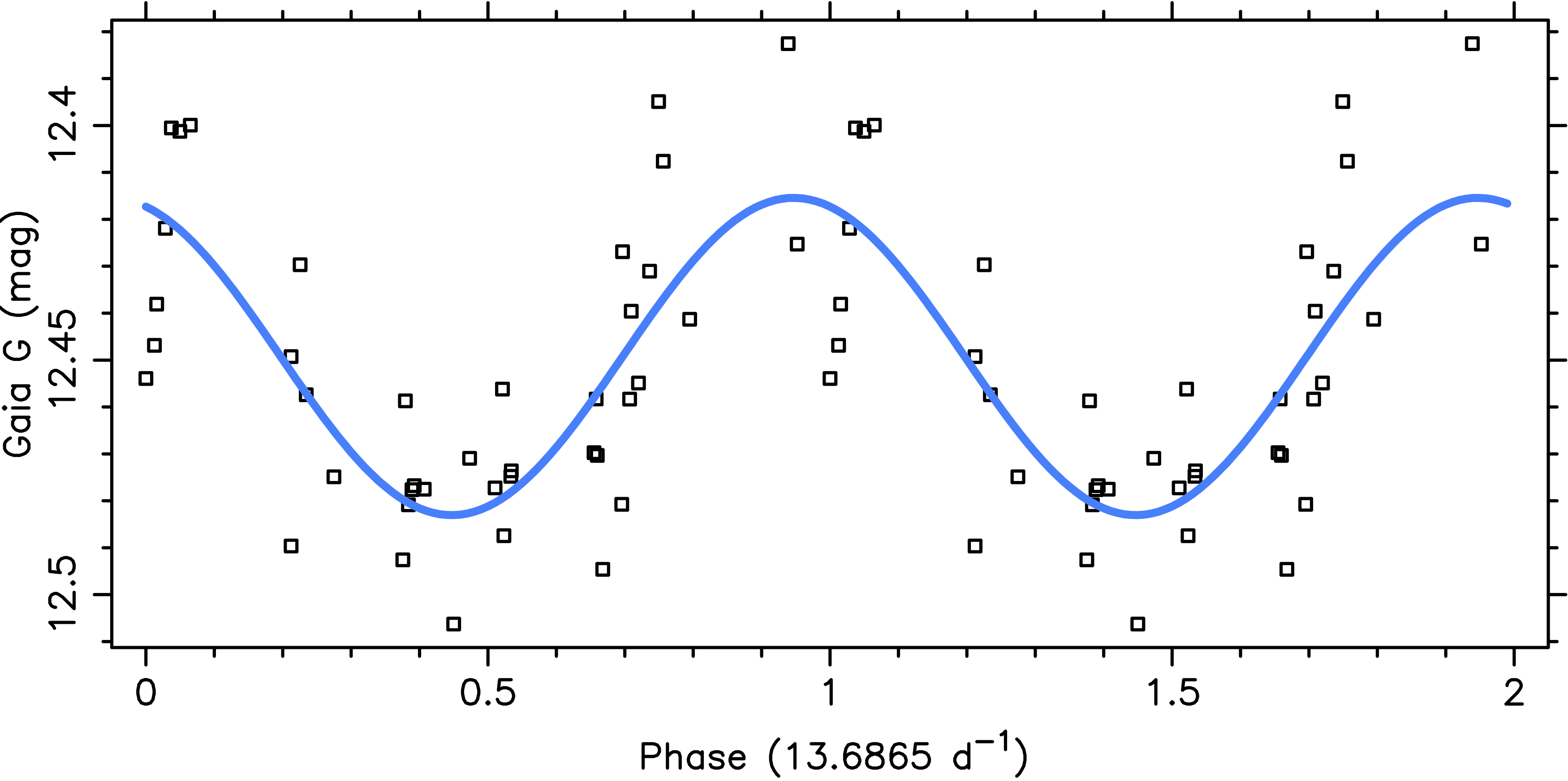}}
\end{center}
\caption{\label{kic6953103} Two phase diagrams for the largest-amplitude g-mode
  pulsator in the $\gamma\,$Dor sample. The upper graph is phase-folded with the
  true dominant g-mode frequency derived from {\it Kepler\/} data; its fit
  leads to a variance reduction in the Gaia DR3 G-band data of 47\%. The lower
  graph is phase-folded with the dominant frequency in the data, which is of
  instrumental origin and reduces the variance with 58\%.}
\end{figure}

We also checked for all 63 stars if it is meaningful to use the fraction of
the variance in the data explained by a harmonic fit with the dominant frequency
as a selection criterion to distinguish intrinsic g-mode frequencies from
instrumental frequencies. We found this not to be feasible, as this
merit function attains similar values for instrumental and intrinsic
frequencies. The corresponding phase diagrams also do not 
allow to distinguish between intrinsic and aliased/instrumental frequencies in a
meaningful way. This is illustrated  in Fig.\,\ref{kic6953103}
for the largest-amplitude $\gamma\,$Dor
pulsator of the sample. Figure\,\ref{kic6953103} shows two
phase diagrams based on the 40 Gaia DR3 G-band data points: one based on the
`true' intrinsic dominant g-mode frequency reducing the variance by 47\% and
another one using the dominant frequency in the data themselves, which is either 
aliased or of
instrumental origin. The latter frequency reduces the variance by 58\%. Hence
the variance reduction cannot be used as a criterion to distinguish between
instrumental and true frequencies \OK{in \gaia's DR3 time series}.

\end{document}

%% file: authors.tex
\author{
{\it Gaia} Collaboration
\and         J.~                     De Ridder\orcit{0000-0001-6726-2863}\inst{\ref{inst:0001}}
\and         V.~                        Ripepi\orcit{0000-0003-1801-426X}\inst{\ref{inst:0002}}
\and         C.~                         Aerts\orcit{0000-0003-1822-7126}\inst{\ref{inst:0001},\ref{inst:0004},\ref{inst:0005}}
\and         L.~                     Palaversa\orcit{0000-0003-3710-0331}\inst{\ref{inst:0006},\ref{inst:0007}}
\and         L.~                          Eyer\orcit{0000-0002-0182-8040}\inst{\ref{inst:0008}}
\and         B.~                          Holl\orcit{0000-0001-6220-3266}\inst{\ref{inst:0008},\ref{inst:0010}}
\and         M.~                        Audard\orcit{0000-0003-4721-034X}\inst{\ref{inst:0008},\ref{inst:0010}}
\and         L.~                     Rimoldini\orcit{0000-0002-0306-585X}\inst{\ref{inst:0010}}
\and     A.G.A.~                         Brown\orcit{0000-0002-7419-9679}\inst{\ref{inst:0014}}
\and         A.~                     Vallenari\orcit{0000-0003-0014-519X}\inst{\ref{inst:0015}}
\and         T.~                        Prusti\orcit{0000-0003-3120-7867}\inst{\ref{inst:0016}}
\and     J.H.J.~                    de Bruijne\orcit{0000-0001-6459-8599}\inst{\ref{inst:0016}}
\and         F.~                        Arenou\orcit{0000-0003-2837-3899}\inst{\ref{inst:0018}}
\and         C.~                     Babusiaux\orcit{0000-0002-7631-348X}\inst{\ref{inst:0019},\ref{inst:0018}}
\and         M.~                      Biermann\inst{\ref{inst:0021}}
\and       O.L.~                       Creevey\orcit{0000-0003-1853-6631}\inst{\ref{inst:0022}}
\and         C.~                     Ducourant\orcit{0000-0003-4843-8979}\inst{\ref{inst:0023}}
\and       D.W.~                         Evans\orcit{0000-0002-6685-5998}\inst{\ref{inst:0007}}
\and         R.~                        Guerra\orcit{0000-0002-9850-8982}\inst{\ref{inst:0025}}
\and         A.~                        Hutton\inst{\ref{inst:0026}}
\and         C.~                         Jordi\orcit{0000-0001-5495-9602}\inst{\ref{inst:0027}}
\and       S.A.~                       Klioner\orcit{0000-0003-4682-7831}\inst{\ref{inst:0028}}
\and       U.L.~                       Lammers\orcit{0000-0001-8309-3801}\inst{\ref{inst:0025}}
\and         L.~                     Lindegren\orcit{0000-0002-5443-3026}\inst{\ref{inst:0030}}
\and         X.~                          Luri\orcit{0000-0001-5428-9397}\inst{\ref{inst:0027}}
\and         F.~                       Mignard\inst{\ref{inst:0022}}
\and         C.~                         Panem\inst{\ref{inst:0033}}
\and         D.~            Pourbaix$^\dagger$\orcit{0000-0002-3020-1837}\inst{\ref{inst:0034},\ref{inst:0035}}
\and         S.~                       Randich\orcit{0000-0003-2438-0899}\inst{\ref{inst:0036}}
\and         P.~                    Sartoretti\inst{\ref{inst:0018}}
\and         C.~                      Soubiran\orcit{0000-0003-3304-8134}\inst{\ref{inst:0023}}
\and         P.~                         Tanga\orcit{0000-0002-2718-997X}\inst{\ref{inst:0022}}
\and       N.A.~                        Walton\orcit{0000-0003-3983-8778}\inst{\ref{inst:0007}}
\and     C.A.L.~                  Bailer-Jones\inst{\ref{inst:0005}}
\and         U.~                       Bastian\orcit{0000-0002-8667-1715}\inst{\ref{inst:0021}}
\and         R.~                       Drimmel\orcit{0000-0002-1777-5502}\inst{\ref{inst:0043}}
\and         F.~                        Jansen\inst{\ref{inst:0044}}
\and         D.~                          Katz\orcit{0000-0001-7986-3164}\inst{\ref{inst:0018}}
\and       M.G.~                      Lattanzi\orcit{0000-0003-0429-7748}\inst{\ref{inst:0043},\ref{inst:0047}}
\and         F.~                   van Leeuwen\inst{\ref{inst:0007}}
\and         J.~                        Bakker\inst{\ref{inst:0025}}
\and         C.~                      Cacciari\orcit{0000-0001-5174-3179}\inst{\ref{inst:0050}}
\and         J.~                 Casta\~{n}eda\orcit{0000-0001-7820-946X}\inst{\ref{inst:0051}}
\and         F.~                     De Angeli\orcit{0000-0003-1879-0488}\inst{\ref{inst:0007}}
\and         C.~                     Fabricius\orcit{0000-0003-2639-1372}\inst{\ref{inst:0027}}
\and         M.~                     Fouesneau\orcit{0000-0001-9256-5516}\inst{\ref{inst:0005}}
\and         Y.~                    Fr\'{e}mat\orcit{0000-0002-4645-6017}\inst{\ref{inst:0055}}
\and         L.~                     Galluccio\orcit{0000-0002-8541-0476}\inst{\ref{inst:0022}}
\and         A.~                      Guerrier\inst{\ref{inst:0033}}
\and         U.~                        Heiter\orcit{0000-0001-6825-1066}\inst{\ref{inst:0058}}
\and         E.~                        Masana\orcit{0000-0002-4819-329X}\inst{\ref{inst:0027}}
\and         R.~                      Messineo\inst{\ref{inst:0060}}
\and         N.~                       Mowlavi\orcit{0000-0003-1578-6993}\inst{\ref{inst:0008}}
\and         C.~                       Nicolas\inst{\ref{inst:0033}}
\and         K.~                  Nienartowicz\orcit{0000-0001-5415-0547}\inst{\ref{inst:0063},\ref{inst:0010}}
\and         F.~                       Pailler\orcit{0000-0002-4834-481X}\inst{\ref{inst:0033}}
\and         P.~                       Panuzzo\orcit{0000-0002-0016-8271}\inst{\ref{inst:0018}}
\and         F.~                        Riclet\inst{\ref{inst:0033}}
\and         W.~                          Roux\orcit{0000-0002-7816-1950}\inst{\ref{inst:0033}}
\and       G.M.~                      Seabroke\orcit{0000-0003-4072-9536}\inst{\ref{inst:0069}}
\and         R.~                         Sordo\orcit{0000-0003-4979-0659}\inst{\ref{inst:0015}}
\and         F.~                  Th\'{e}venin\inst{\ref{inst:0022}}
\and         G.~                  Gracia-Abril\inst{\ref{inst:0072},\ref{inst:0021}}
\and         J.~                       Portell\orcit{0000-0002-8886-8925}\inst{\ref{inst:0027}}
\and         D.~                      Teyssier\orcit{0000-0002-6261-5292}\inst{\ref{inst:0075}}
\and         M.~                       Altmann\orcit{0000-0002-0530-0913}\inst{\ref{inst:0021},\ref{inst:0077}}
\and         R.~                        Andrae\orcit{0000-0001-8006-6365}\inst{\ref{inst:0005}}
\and         I.~                Bellas-Velidis\inst{\ref{inst:0079}}
\and         K.~                        Benson\inst{\ref{inst:0069}}
\and         J.~                      Berthier\orcit{0000-0003-1846-6485}\inst{\ref{inst:0081}}
\and         R.~                        Blomme\orcit{0000-0002-2526-346X}\inst{\ref{inst:0055}}
\and       P.W.~                       Burgess\inst{\ref{inst:0007}}
\and         D.~                      Busonero\orcit{0000-0002-3903-7076}\inst{\ref{inst:0043}}
\and         G.~                         Busso\orcit{0000-0003-0937-9849}\inst{\ref{inst:0007}}
\and         H.~                   C\'{a}novas\orcit{0000-0001-7668-8022}\inst{\ref{inst:0075}}
\and         B.~                         Carry\orcit{0000-0001-5242-3089}\inst{\ref{inst:0022}}
\and         A.~                       Cellino\orcit{0000-0002-6645-334X}\inst{\ref{inst:0043}}
\and         N.~                         Cheek\inst{\ref{inst:0089}}
\and         G.~                    Clementini\orcit{0000-0001-9206-9723}\inst{\ref{inst:0050}}
\and         Y.~                      Damerdji\orcit{0000-0002-3107-4024}\inst{\ref{inst:0091},\ref{inst:0092}}
\and         M.~                      Davidson\inst{\ref{inst:0093}}
\and         P.~                    de Teodoro\inst{\ref{inst:0025}}
\and         M.~              Nu\~{n}ez Campos\inst{\ref{inst:0026}}
\and         L.~                    Delchambre\orcit{0000-0003-2559-408X}\inst{\ref{inst:0091}}
\and         A.~                      Dell'Oro\orcit{0000-0003-1561-9685}\inst{\ref{inst:0036}}
\and         P.~                        Esquej\orcit{0000-0001-8195-628X}\inst{\ref{inst:0098}}
\and         J.~   Fern\'{a}ndez-Hern\'{a}ndez\inst{\ref{inst:0099}}
\and         E.~                        Fraile\inst{\ref{inst:0098}}
\and         D.~                      Garabato\orcit{0000-0002-7133-6623}\inst{\ref{inst:0101}}
\and         P.~              Garc\'{i}a-Lario\orcit{0000-0003-4039-8212}\inst{\ref{inst:0025}}
\and         E.~                        Gosset\inst{\ref{inst:0091},\ref{inst:0035}}
\and         R.~                       Haigron\inst{\ref{inst:0018}}
\and      J.-L.~                     Halbwachs\orcit{0000-0003-2968-6395}\inst{\ref{inst:0106}}
\and       N.C.~                        Hambly\orcit{0000-0002-9901-9064}\inst{\ref{inst:0093}}
\and       D.L.~                      Harrison\orcit{0000-0001-8687-6588}\inst{\ref{inst:0007},\ref{inst:0109}}
\and         J.~                 Hern\'{a}ndez\orcit{0000-0002-0361-4994}\inst{\ref{inst:0025}}
\and         D.~                    Hestroffer\orcit{0000-0003-0472-9459}\inst{\ref{inst:0081}}
\and         T.~                        Hilger\orcit{0000-0003-1646-0063}\inst{\ref{inst:0028}} 
\and       S.T.~                       Hodgkin\orcit{0000-0002-5470-3962}\inst{\ref{inst:0007}}
\and         K.~                    Jan{\ss}en\orcit{0000-0002-8163-2493}\inst{\ref{inst:0113}}
\and         G.~          Jevardat de Fombelle\inst{\ref{inst:0008}}
\and         S.~                        Jordan\orcit{0000-0001-6316-6831}\inst{\ref{inst:0021}}
\and         A.~                 Krone-Martins\orcit{0000-0002-2308-6623}\inst{\ref{inst:0116},\ref{inst:0117}}
\and       A.C.~                     Lanzafame\orcit{0000-0002-2697-3607}\inst{\ref{inst:0118},\ref{inst:0119}}
\and         W.~                  L\"{ o}ffler\inst{\ref{inst:0021}}
\and         O.~                       Marchal\orcit{ 0000-0001-7461-892}\inst{\ref{inst:0106}}
\and       P.M.~                       Marrese\orcit{0000-0002-8162-3810}\inst{\ref{inst:0122},\ref{inst:0123}}
\and         A.~                      Moitinho\orcit{0000-0003-0822-5995}\inst{\ref{inst:0116}}
\and         K.~                      Muinonen\orcit{0000-0001-8058-2642}\inst{\ref{inst:0125},\ref{inst:0126}}
\and         P.~                       Osborne\inst{\ref{inst:0007}}
\and         E.~                       Pancino\orcit{0000-0003-0788-5879}\inst{\ref{inst:0036},\ref{inst:0123}}
\and         T.~                       Pauwels\inst{\ref{inst:0055}}
\and         A.~                  Recio-Blanco\orcit{0000-0002-6550-7377}\inst{\ref{inst:0022}}
\and         C.~                     Reyl\'{e}\orcit{0000-0003-2258-2403}\inst{\ref{inst:0132}}
\and         M.~                        Riello\orcit{0000-0002-3134-0935}\inst{\ref{inst:0007}}
\and         T.~                      Roegiers\orcit{0000-0002-1231-4440}\inst{\ref{inst:0134}}
\and         J.~                       Rybizki\orcit{0000-0002-0993-6089}\inst{\ref{inst:0005}}
\and       L.M.~                         Sarro\orcit{0000-0002-5622-5191}\inst{\ref{inst:0136}}
\and         C.~                        Siopis\orcit{0000-0002-6267-2924}\inst{\ref{inst:0034}}
\and         M.~                         Smith\inst{\ref{inst:0069}}
\and         A.~                      Sozzetti\orcit{0000-0002-7504-365X}\inst{\ref{inst:0043}}
\and         E.~                       Utrilla\inst{\ref{inst:0026}}
\and         M.~                   van Leeuwen\orcit{0000-0001-9698-2392}\inst{\ref{inst:0007}}
\and         U.~                         Abbas\orcit{0000-0002-5076-766X}\inst{\ref{inst:0043}}
\and         P.~               \'{A}brah\'{a}m\orcit{0000-0001-6015-646X}\inst{\ref{inst:0143},\ref{inst:0144}}
\and         A.~                Abreu Aramburu\inst{\ref{inst:0099}}
\and       J.J.~                        Aguado\inst{\ref{inst:0136}}
\and         M.~                          Ajaj\inst{\ref{inst:0018}}
\and         F.~                 Aldea-Montero\inst{\ref{inst:0025}}
\and         G.~                     Altavilla\orcit{0000-0002-9934-1352}\inst{\ref{inst:0122},\ref{inst:0123}}
\and       M.A.~                   \'{A}lvarez\orcit{0000-0002-6786-2620}\inst{\ref{inst:0101}}
\and         J.~                         Alves\orcit{0000-0002-4355-0921}\inst{\ref{inst:0152}}
\and         F.~                        Anders\inst{\ref{inst:0027}}
\and       R.I.~                      Anderson\orcit{0000-0001-8089-4419}\inst{\ref{inst:0154}}
\and         E.~                Anglada Varela\orcit{0000-0001-7563-0689}\inst{\ref{inst:0099}}
\and         T.~                        Antoja\orcit{0000-0003-2595-5148}\inst{\ref{inst:0027}}
\and         D.~                        Baines\orcit{0000-0002-6923-3756}\inst{\ref{inst:0075}}
\and       S.G.~                         Baker\orcit{0000-0002-6436-1257}\inst{\ref{inst:0069}}
\and         L.~        Balaguer-N\'{u}\~{n}ez\orcit{0000-0001-9789-7069}\inst{\ref{inst:0027}}
\and         E.~                      Balbinot\orcit{0000-0002-1322-3153}\inst{\ref{inst:0160}}
\and         Z.~                         Balog\orcit{0000-0003-1748-2926}\inst{\ref{inst:0021},\ref{inst:0005}}
\and         C.~                       Barache\inst{\ref{inst:0077}}
\and         D.~                       Barbato\inst{\ref{inst:0008},\ref{inst:0043}}
\and         M.~                        Barros\orcit{0000-0002-9728-9618}\inst{\ref{inst:0116}}
\and       M.A.~                       Barstow\orcit{0000-0002-7116-3259}\inst{\ref{inst:0167}}
\and         S.~                 Bartolom\'{e}\orcit{0000-0002-6290-6030}\inst{\ref{inst:0027}}
\and      J.-L.~                     Bassilana\inst{\ref{inst:0169}}
\and         N.~                       Bauchet\inst{\ref{inst:0018}}
\and         U.~                      Becciani\orcit{0000-0002-4389-8688}\inst{\ref{inst:0118}}
\and         M.~                    Bellazzini\orcit{0000-0001-8200-810X}\inst{\ref{inst:0050}}
\and         A.~                     Berihuete\orcit{0000-0002-8589-4423}\inst{\ref{inst:0173}}
\and         M.~                        Bernet\orcit{0000-0001-7503-1010}\inst{\ref{inst:0027}}
\and         S.~                       Bertone\orcit{0000-0001-9885-8440}\inst{\ref{inst:0175},\ref{inst:0176},\ref{inst:0043}}
\and         L.~                       Bianchi\orcit{0000-0002-7999-4372}\inst{\ref{inst:0178}}
\and         A.~                    Binnenfeld\orcit{0000-0002-9319-3838}\inst{\ref{inst:0179}}
\and         S.~               Blanco-Cuaresma\orcit{0000-0002-1584-0171}\inst{\ref{inst:0180}}
\and         T.~                          Boch\orcit{0000-0001-5818-2781}\inst{\ref{inst:0106}}
\and         A.~                       Bombrun\inst{\ref{inst:0182}}
\and         D.~                       Bossini\orcit{0000-0002-9480-8400}\inst{\ref{inst:0183}}
\and         S.~                    Bouquillon\inst{\ref{inst:0077},\ref{inst:0185}}
\and         A.~                     Bragaglia\orcit{0000-0002-0338-7883}\inst{\ref{inst:0050}}
\and         L.~                      Bramante\inst{\ref{inst:0060}}
\and         E.~                        Breedt\orcit{0000-0001-6180-3438}\inst{\ref{inst:0007}}
\and         A.~                       Bressan\orcit{0000-0002-7922-8440}\inst{\ref{inst:0189}}
\and         N.~                     Brouillet\orcit{0000-0002-3274-7024}\inst{\ref{inst:0023}}
\and         E.~                    Brugaletta\orcit{0000-0003-2598-6737}\inst{\ref{inst:0118}}
\and         B.~                   Bucciarelli\orcit{0000-0002-5303-0268}\inst{\ref{inst:0043},\ref{inst:0047}}
\and         A.~                       Burlacu\inst{\ref{inst:0194}}
\and       A.G.~                     Butkevich\orcit{0000-0002-4098-3588}\inst{\ref{inst:0043}}
\and         R.~                         Buzzi\orcit{0000-0001-9389-5701}\inst{\ref{inst:0043}}
\and         E.~                        Caffau\orcit{0000-0001-6011-6134}\inst{\ref{inst:0018}}
\and         R.~                   Cancelliere\orcit{0000-0002-9120-3799}\inst{\ref{inst:0198}}
\and         T.~                 Cantat-Gaudin\orcit{0000-0001-8726-2588}\inst{\ref{inst:0027},\ref{inst:0005}}
\and         R.~                      Carballo\orcit{0000-0001-7412-2498}\inst{\ref{inst:0201}}
\and         T.~                      Carlucci\inst{\ref{inst:0077}}
\and       M.I.~                     Carnerero\orcit{0000-0001-5843-5515}\inst{\ref{inst:0043}}
\and       J.M.~                      Carrasco\orcit{0000-0002-3029-5853}\inst{\ref{inst:0027}}
\and         L.~                   Casamiquela\orcit{0000-0001-5238-8674}\inst{\ref{inst:0023},\ref{inst:0018}}
\and         M.~                    Castellani\orcit{0000-0002-7650-7428}\inst{\ref{inst:0122}}
\and         A.~                 Castro-Ginard\orcit{0000-0002-9419-3725}\inst{\ref{inst:0014}}
\and         L.~                        Chaoul\inst{\ref{inst:0033}}
\and         P.~                       Charlot\orcit{0000-0002-9142-716X}\inst{\ref{inst:0023}}
\and         L.~                        Chemin\orcit{0000-0002-3834-7937}\inst{\ref{inst:0211}}
\and         V.~                    Chiaramida\inst{\ref{inst:0060}}
\and         A.~                     Chiavassa\orcit{0000-0003-3891-7554}\inst{\ref{inst:0022}}
\and         N.~                       Chornay\orcit{0000-0002-8767-3907}\inst{\ref{inst:0007}}
\and         G.~                     Comoretto\inst{\ref{inst:0075},\ref{inst:0216}}
\and         G.~                      Contursi\orcit{0000-0001-5370-1511}\inst{\ref{inst:0022}}
\and       W.J.~                        Cooper\orcit{0000-0003-3501-8967}\inst{\ref{inst:0218},\ref{inst:0043}}
\and         T.~                        Cornez\inst{\ref{inst:0169}}
\and         S.~                        Cowell\inst{\ref{inst:0007}}
\and         F.~                         Crifo\inst{\ref{inst:0018}}
\and         M.~                       Cropper\orcit{0000-0003-4571-9468}\inst{\ref{inst:0069}}
\and         M.~                        Crosta\orcit{0000-0003-4369-3786}\inst{\ref{inst:0043},\ref{inst:0225}}
\and         C.~                       Crowley\inst{\ref{inst:0182}}
\and         C.~                       Dafonte\orcit{0000-0003-4693-7555}\inst{\ref{inst:0101}}
\and         A.~                    Dapergolas\inst{\ref{inst:0079}}
\and         P.~                         David\inst{\ref{inst:0081}}
\and         P.~                    de Laverny\orcit{0000-0002-2817-4104}\inst{\ref{inst:0022}}
\and         F.~                      De Luise\orcit{0000-0002-6570-8208}\inst{\ref{inst:0231}}
\and         R.~                      De March\orcit{0000-0003-0567-842X}\inst{\ref{inst:0060}}
\and         R.~                      de Souza\inst{\ref{inst:0233}}
\and         A.~                     de Torres\inst{\ref{inst:0182}}
\and       E.F.~                    del Peloso\inst{\ref{inst:0021}}
\and         E.~                      del Pozo\inst{\ref{inst:0026}}
\and         M.~                         Delbo\orcit{0000-0002-8963-2404}\inst{\ref{inst:0022}}
\and         A.~                       Delgado\inst{\ref{inst:0098}}
\and      J.-B.~                       Delisle\orcit{0000-0001-5844-9888}\inst{\ref{inst:0008}}
\and         C.~                      Demouchy\inst{\ref{inst:0240}}
\and       T.E.~                 Dharmawardena\orcit{0000-0002-9583-5216}\inst{\ref{inst:0005}}
\and         S.~                       Diakite\inst{\ref{inst:0242}}
\and         C.~                        Diener\inst{\ref{inst:0007}}
\and         E.~                     Distefano\orcit{0000-0002-2448-2513}\inst{\ref{inst:0118}}
\and         C.~                       Dolding\inst{\ref{inst:0069}}
\and         H.~                          Enke\orcit{0000-0002-2366-8316}\inst{\ref{inst:0113}}
\and         C.~                         Fabre\inst{\ref{inst:0247}}
\and         M.~                      Fabrizio\orcit{0000-0001-5829-111X}\inst{\ref{inst:0122},\ref{inst:0123}}
\and         S.~                       Faigler\orcit{0000-0002-8368-5724}\inst{\ref{inst:0250}}
\and         G.~                      Fedorets\orcit{0000-0002-8418-4809}\inst{\ref{inst:0125},\ref{inst:0252}}
\and         P.~                      Fernique\orcit{0000-0002-3304-2923}\inst{\ref{inst:0106},\ref{inst:0254}}
\and         F.~                      Figueras\orcit{0000-0002-3393-0007}\inst{\ref{inst:0027}}
\and         Y.~                      Fournier\orcit{0000-0002-6633-9088}\inst{\ref{inst:0113}}
\and         C.~                        Fouron\inst{\ref{inst:0194}}
\and         F.~                     Fragkoudi\orcit{0000-0002-0897-3013}\inst{\ref{inst:0258},\ref{inst:0259},\ref{inst:0260}}
\and         M.~                           Gai\orcit{0000-0001-9008-134X}\inst{\ref{inst:0043}}
\and         A.~              Garcia-Gutierrez\inst{\ref{inst:0027}}
\and         M.~              Garcia-Reinaldos\inst{\ref{inst:0025}}
\and         M.~             Garc\'{i}a-Torres\orcit{0000-0002-6867-7080}\inst{\ref{inst:0264}}
\and         A.~                      Garofalo\orcit{0000-0002-5907-0375}\inst{\ref{inst:0050}}
\and         A.~                         Gavel\orcit{0000-0002-2963-722X}\inst{\ref{inst:0058}}
\and         P.~                        Gavras\orcit{0000-0002-4383-4836}\inst{\ref{inst:0098}}
\and         E.~                       Gerlach\orcit{0000-0002-9533-2168}\inst{\ref{inst:0028}}
\and         R.~                         Geyer\orcit{0000-0001-6967-8707}\inst{\ref{inst:0028}}
\and         P.~                      Giacobbe\orcit{0000-0001-7034-7024}\inst{\ref{inst:0043}}
\and         G.~                       Gilmore\orcit{0000-0003-4632-0213}\inst{\ref{inst:0007}}
\and         S.~                        Girona\orcit{0000-0002-1975-1918}\inst{\ref{inst:0272}}
\and         G.~                     Giuffrida\inst{\ref{inst:0122}}
\and         R.~                         Gomel\inst{\ref{inst:0250}}
\and         A.~                         Gomez\orcit{0000-0002-3796-3690}\inst{\ref{inst:0101}}
\and         J.~    Gonz\'{a}lez-N\'{u}\~{n}ez\orcit{0000-0001-5311-5555}\inst{\ref{inst:0089},\ref{inst:0277}}
\and         I.~   Gonz\'{a}lez-Santamar\'{i}a\orcit{0000-0002-8537-9384}\inst{\ref{inst:0101}}
\and       J.J.~            Gonz\'{a}lez-Vidal\inst{\ref{inst:0027}}
\and         M.~                       Granvik\orcit{0000-0002-5624-1888}\inst{\ref{inst:0125},\ref{inst:0281}}
\and         P.~                      Guillout\inst{\ref{inst:0106}}
\and         J.~                       Guiraud\inst{\ref{inst:0033}}
\and         R.~     Guti\'{e}rrez-S\'{a}nchez\inst{\ref{inst:0075}}
\and       L.P.~                           Guy\orcit{0000-0003-0800-8755}\inst{\ref{inst:0010},\ref{inst:0286}}
\and         D.~                Hatzidimitriou\orcit{0000-0002-5415-0464}\inst{\ref{inst:0287},\ref{inst:0079}}
\and         M.~                        Hauser\inst{\ref{inst:0005},\ref{inst:0290}}
\and         M.~                       Haywood\orcit{0000-0003-0434-0400}\inst{\ref{inst:0018}}
\and         A.~                        Helmer\inst{\ref{inst:0169}}
\and         A.~                         Helmi\orcit{0000-0003-3937-7641}\inst{\ref{inst:0160}}
\and       M.H.~                     Sarmiento\orcit{0000-0003-4252-5115}\inst{\ref{inst:0026}}
\and       S.L.~                       Hidalgo\orcit{0000-0002-0002-9298}\inst{\ref{inst:0295},\ref{inst:0296}}
\and         N.~                   H\l{}adczuk\orcit{0000-0001-9163-4209}\inst{\ref{inst:0025},\ref{inst:0298}}
\and         D.~                         Hobbs\orcit{0000-0002-2696-1366}\inst{\ref{inst:0030}}
\and         G.~                       Holland\inst{\ref{inst:0007}}
\and       H.E.~                        Huckle\inst{\ref{inst:0069}}
\and         K.~                       Jardine\inst{\ref{inst:0302}}
\and         G.~                    Jasniewicz\inst{\ref{inst:0303}}
\and         A.~          Jean-Antoine Piccolo\orcit{0000-0001-8622-212X}\inst{\ref{inst:0033}}
\and     \'{O}.~            Jim\'{e}nez-Arranz\orcit{0000-0001-7434-5165}\inst{\ref{inst:0027}}
\and         J.~             Juaristi Campillo\inst{\ref{inst:0021}}
\and         F.~                         Julbe\inst{\ref{inst:0027}}
\and         L.~                     Karbevska\inst{\ref{inst:0010},\ref{inst:0309}}
\and         P.~                      Kervella\orcit{0000-0003-0626-1749}\inst{\ref{inst:0310}}
\and         S.~                        Khanna\orcit{0000-0002-2604-4277}\inst{\ref{inst:0160},\ref{inst:0043}}
\and         G.~                    Kordopatis\orcit{0000-0002-9035-3920}\inst{\ref{inst:0022}}
\and       A.J.~                          Korn\orcit{0000-0002-3881-6756}\inst{\ref{inst:0058}}
\and      \'{A}~                K\'{o}sp\'{a}l\orcit{\'{u}t 15-17, 1121 }\inst{\ref{inst:0143},\ref{inst:0005},\ref{inst:0144}}
\and         Z.~           Kostrzewa-Rutkowska\inst{\ref{inst:0014},\ref{inst:0319}}
\and         K.~                Kruszy\'{n}ska\orcit{0000-0002-2729-5369}\inst{\ref{inst:0320}}
\and         M.~                           Kun\orcit{0000-0002-7538-5166}\inst{\ref{inst:0143}}
\and         P.~                       Laizeau\inst{\ref{inst:0322}}
\and         S.~                       Lambert\orcit{0000-0001-6759-5502}\inst{\ref{inst:0077}}
\and       A.F.~                         Lanza\orcit{0000-0001-5928-7251}\inst{\ref{inst:0118}}
\and         Y.~                         Lasne\inst{\ref{inst:0169}}
\and      J.-F.~                    Le Campion\inst{\ref{inst:0023}}
\and         Y.~                      Lebreton\orcit{0000-0002-4834-2144}\inst{\ref{inst:0310},\ref{inst:0328}}
\and         T.~                     Lebzelter\orcit{0000-0002-0702-7551}\inst{\ref{inst:0152}}
\and         S.~                        Leccia\orcit{0000-0001-5685-6930}\inst{\ref{inst:0002}}
\and         N.~                       Leclerc\inst{\ref{inst:0018}}
\and         I.~                 Lecoeur-Taibi\orcit{0000-0003-0029-8575}\inst{\ref{inst:0010}}
\and         S.~                          Liao\orcit{0000-0002-9346-0211}\inst{\ref{inst:0333},\ref{inst:0043},\ref{inst:0335}}
\and       E.L.~                        Licata\orcit{0000-0002-5203-0135}\inst{\ref{inst:0043}}
\and     H.E.P.~                  Lindstr{\o}m\inst{\ref{inst:0043},\ref{inst:0338},\ref{inst:0339}}
\and       T.A.~                        Lister\orcit{0000-0002-3818-7769}\inst{\ref{inst:0340}}
\and         E.~                       Livanou\orcit{0000-0003-0628-2347}\inst{\ref{inst:0287}}
\and         A.~                         Lobel\orcit{0000-0001-5030-019X}\inst{\ref{inst:0055}}
\and         A.~                         Lorca\inst{\ref{inst:0026}}
\and         C.~                          Loup\inst{\ref{inst:0106}}
\and         P.~                 Madrero Pardo\inst{\ref{inst:0027}}
\and         A.~               Magdaleno Romeo\inst{\ref{inst:0194}}
\and         S.~                       Managau\inst{\ref{inst:0169}}
\and       R.G.~                          Mann\orcit{0000-0002-0194-325X}\inst{\ref{inst:0093}}
\and         M.~                      Manteiga\orcit{0000-0002-7711-5581}\inst{\ref{inst:0349}}
\and       J.M.~                      Marchant\orcit{0000-0002-3678-3145}\inst{\ref{inst:0350}}
\and         M.~                       Marconi\orcit{0000-0002-1330-2927}\inst{\ref{inst:0002}}
\and         J.~                        Marcos\inst{\ref{inst:0075}}
\and     M.M.S.~                 Marcos Santos\inst{\ref{inst:0089}}
\and         D.~                Mar\'{i}n Pina\orcit{0000-0001-6482-1842}\inst{\ref{inst:0027}}
\and         S.~                      Marinoni\orcit{0000-0001-7990-6849}\inst{\ref{inst:0122},\ref{inst:0123}}
\and         F.~                       Marocco\orcit{0000-0001-7519-1700}\inst{\ref{inst:0357}}
\and       D.J.~                      Marshall\orcit{0000-0003-3956-3524}\inst{\ref{inst:0358}}
\and         L.~                   Martin Polo\inst{\ref{inst:0089}}
\and       J.M.~            Mart\'{i}n-Fleitas\orcit{0000-0002-8594-569X}\inst{\ref{inst:0026}}
\and         G.~                        Marton\orcit{0000-0002-1326-1686}\inst{\ref{inst:0143}}
\and         N.~                          Mary\inst{\ref{inst:0169}}
\and         A.~                         Masip\orcit{0000-0003-1419-0020}\inst{\ref{inst:0027}}
\and         D.~                       Massari\orcit{0000-0001-8892-4301}\inst{\ref{inst:0050}}
\and         A.~          Mastrobuono-Battisti\orcit{0000-0002-2386-9142}\inst{\ref{inst:0018}}
\and         T.~                         Mazeh\orcit{0000-0002-3569-3391}\inst{\ref{inst:0250}}
\and       P.J.~                      McMillan\orcit{0000-0002-8861-2620}\inst{\ref{inst:0030}}
\and         S.~                       Messina\orcit{0000-0002-2851-2468}\inst{\ref{inst:0118}}
\and         D.~                      Michalik\orcit{0000-0002-7618-6556}\inst{\ref{inst:0016}}
\and       N.R.~                        Millar\inst{\ref{inst:0007}}
\and         A.~                         Mints\orcit{0000-0002-8440-1455}\inst{\ref{inst:0113}}
\and         D.~                        Molina\orcit{0000-0003-4814-0275}\inst{\ref{inst:0027}}
\and         R.~                      Molinaro\orcit{0000-0003-3055-6002}\inst{\ref{inst:0002}}
\and         L.~                    Moln\'{a}r\orcit{0000-0002-8159-1599}\inst{\ref{inst:0143},\ref{inst:0375},\ref{inst:0144}}
\and         G.~                        Monari\orcit{0000-0002-6863-0661}\inst{\ref{inst:0106}}
\and         M.~                   Mongui\'{o}\orcit{0000-0002-4519-6700}\inst{\ref{inst:0027}}
\and         P.~                   Montegriffo\orcit{0000-0001-5013-5948}\inst{\ref{inst:0050}}
\and         A.~                       Montero\inst{\ref{inst:0026}}
\and         R.~                           Mor\orcit{0000-0002-8179-6527}\inst{\ref{inst:0027}}
\and         A.~                          Mora\inst{\ref{inst:0026}}
\and         R.~                    Morbidelli\orcit{0000-0001-7627-4946}\inst{\ref{inst:0043}}
\and         T.~                         Morel\orcit{0000-0002-8176-4816}\inst{\ref{inst:0091}}
\and         D.~                        Morris\inst{\ref{inst:0093}}
\and         T.~                      Muraveva\orcit{0000-0002-0969-1915}\inst{\ref{inst:0050}}
\and       C.P.~                        Murphy\inst{\ref{inst:0025}}
\and         I.~                       Musella\orcit{0000-0001-5909-6615}\inst{\ref{inst:0002}}
\and         Z.~                          Nagy\orcit{0000-0002-3632-1194}\inst{\ref{inst:0143}}
\and         L.~                         Noval\inst{\ref{inst:0169}}
\and         F.~                     Oca\~{n}a\inst{\ref{inst:0075},\ref{inst:0391}}
\and         A.~                         Ogden\inst{\ref{inst:0007}}
\and         C.~                     Ordenovic\inst{\ref{inst:0022}}
\and       J.O.~                        Osinde\inst{\ref{inst:0098}}
\and         C.~                        Pagani\orcit{0000-0001-5477-4720}\inst{\ref{inst:0167}}
\and         I.~                        Pagano\orcit{0000-0001-9573-4928}\inst{\ref{inst:0118}}
\and       P.A.~                       Palicio\orcit{0000-0002-7432-8709}\inst{\ref{inst:0022}}
\and         L.~               Pallas-Quintela\orcit{0000-0001-9296-3100}\inst{\ref{inst:0101}}
\and         A.~                        Panahi\orcit{0000-0001-5850-4373}\inst{\ref{inst:0250}}
\and         S.~               Payne-Wardenaar\inst{\ref{inst:0021}}
\and         X.~         Pe\~{n}alosa Esteller\inst{\ref{inst:0027}}
\and         A.~                 Penttil\"{ a}\orcit{0000-0001-7403-1721}\inst{\ref{inst:0125}}
\and         B.~                        Pichon\orcit{0000 0000 0062 1449}\inst{\ref{inst:0022}}
\and       A.M.~                    Piersimoni\orcit{0000-0002-8019-3708}\inst{\ref{inst:0231}}
\and      F.-X.~                        Pineau\orcit{0000-0002-2335-4499}\inst{\ref{inst:0106}}
\and         E.~                        Plachy\orcit{0000-0002-5481-3352}\inst{\ref{inst:0143},\ref{inst:0375},\ref{inst:0144}}
\and         G.~                          Plum\inst{\ref{inst:0018}}
\and         E.~                        Poggio\orcit{0000-0003-3793-8505}\inst{\ref{inst:0022},\ref{inst:0043}}
\and         A.~                      Pr\v{s}a\orcit{0000-0002-1913-0281}\inst{\ref{inst:0411}}
\and         L.~                        Pulone\orcit{0000-0002-5285-998X}\inst{\ref{inst:0122}}
\and         E.~                        Racero\orcit{0000-0002-6101-9050}\inst{\ref{inst:0089},\ref{inst:0391}}
\and         S.~                       Ragaini\inst{\ref{inst:0050}}
\and         M.~                        Rainer\orcit{0000-0002-8786-2572}\inst{\ref{inst:0036},\ref{inst:0417}}
\and       C.M.~                       Raiteri\orcit{0000-0003-1784-2784}\inst{\ref{inst:0043}}
\and         P.~                         Ramos\orcit{0000-0002-5080-7027}\inst{\ref{inst:0027},\ref{inst:0106}}
\and         M.~                  Ramos-Lerate\inst{\ref{inst:0075}}
\and         P.~                  Re Fiorentin\orcit{0000-0002-4995-0475}\inst{\ref{inst:0043}}
\and         S.~                        Regibo\inst{\ref{inst:0001}}
\and       P.J.~                      Richards\inst{\ref{inst:0424}}
\and         C.~                     Rios Diaz\inst{\ref{inst:0098}}
\and         A.~                          Riva\orcit{0000-0002-6928-8589}\inst{\ref{inst:0043}}
\and      H.-W.~                           Rix\orcit{0000-0003-4996-9069}\inst{\ref{inst:0005}}
\and         G.~                         Rixon\orcit{0000-0003-4399-6568}\inst{\ref{inst:0007}}
\and         N.~                      Robichon\orcit{0000-0003-4545-7517}\inst{\ref{inst:0018}}
\and       A.C.~                         Robin\orcit{0000-0001-8654-9499}\inst{\ref{inst:0132}}
\and         C.~                         Robin\inst{\ref{inst:0169}}
\and         M.~                       Roelens\orcit{0000-0003-0876-4673}\inst{\ref{inst:0008}}
\and     H.R.O.~                        Rogues\inst{\ref{inst:0240}}
\and         L.~                    Rohrbasser\inst{\ref{inst:0010}}
\and         M.~              Romero-G\'{o}mez\orcit{0000-0003-3936-1025}\inst{\ref{inst:0027}}
\and         N.~                        Rowell\orcit{0000-0003-3809-1895}\inst{\ref{inst:0093}}
\and         F.~                         Royer\orcit{0000-0002-9374-8645}\inst{\ref{inst:0018}}
\and         D.~                    Ruz Mieres\orcit{0000-0002-9455-157X}\inst{\ref{inst:0007}}
\and       K.A.~                       Rybicki\orcit{0000-0002-9326-9329}\inst{\ref{inst:0320}}
\and         G.~                      Sadowski\orcit{0000-0002-3411-1003}\inst{\ref{inst:0034}}
\and         A.~        S\'{a}ez N\'{u}\~{n}ez\inst{\ref{inst:0027}}
\and         A.~       Sagrist\`{a} Sell\'{e}s\orcit{0000-0001-6191-2028}\inst{\ref{inst:0021}}
\and         J.~                      Sahlmann\orcit{0000-0001-9525-3673}\inst{\ref{inst:0098}}
\and         E.~                      Salguero\inst{\ref{inst:0099}}
\and         N.~                       Samaras\orcit{0000-0001-8375-6652}\inst{\ref{inst:0055},\ref{inst:0446}}
\and         V.~               Sanchez Gimenez\orcit{0000-0003-1797-3557}\inst{\ref{inst:0027}}
\and         N.~                         Sanna\orcit{0000-0001-9275-9492}\inst{\ref{inst:0036}}
\and         R.~                 Santove\~{n}a\orcit{0000-0002-9257-2131}\inst{\ref{inst:0101}}
\and         M.~                       Sarasso\orcit{0000-0001-5121-0727}\inst{\ref{inst:0043}}
\and         M.~                    Schultheis\orcit{0000-0002-6590-1657}\inst{\ref{inst:0022}}
\and         E.~                       Sciacca\orcit{0000-0002-5574-2787}\inst{\ref{inst:0118}}
\and         M.~                         Segol\inst{\ref{inst:0240}}
\and       J.C.~                       Segovia\inst{\ref{inst:0089}}
\and         D.~                 S\'{e}gransan\orcit{0000-0003-2355-8034}\inst{\ref{inst:0008}}
\and         D.~                        Semeux\inst{\ref{inst:0247}}
\and         S.~                        Shahaf\orcit{0000-0001-9298-8068}\inst{\ref{inst:0457}}
\and       H.I.~                      Siddiqui\orcit{0000-0003-1853-6033}\inst{\ref{inst:0458}}
\and         A.~                       Siebert\orcit{0000-0001-8059-2840}\inst{\ref{inst:0106},\ref{inst:0254}}
\and         L.~                       Siltala\orcit{0000-0002-6938-794X}\inst{\ref{inst:0125}}
\and         A.~                       Silvelo\orcit{0000-0002-5126-6365}\inst{\ref{inst:0101}}
\and         E.~                        Slezak\inst{\ref{inst:0022}}
\and         I.~                        Slezak\inst{\ref{inst:0022}}
\and       R.L.~                         Smart\orcit{0000-0002-4424-4766}\inst{\ref{inst:0043}}
\and       O.N.~                        Snaith\inst{\ref{inst:0018}}
\and         E.~                        Solano\inst{\ref{inst:0467}}
\and         F.~                       Solitro\inst{\ref{inst:0060}}
\and         D.~                        Souami\orcit{0000-0003-4058-0815}\inst{\ref{inst:0310},\ref{inst:0470}}
\and         J.~                       Souchay\inst{\ref{inst:0077}}
\and         A.~                        Spagna\orcit{0000-0003-1732-2412}\inst{\ref{inst:0043}}
\and         L.~                         Spina\orcit{0000-0002-9760-6249}\inst{\ref{inst:0015}}
\and         F.~                         Spoto\orcit{0000-0001-7319-5847}\inst{\ref{inst:0180}}
\and       I.A.~                        Steele\orcit{0000-0001-8397-5759}\inst{\ref{inst:0350}}
\and         H.~            Steidelm\"{ u}ller\inst{\ref{inst:0028}}
\and       C.A.~                    Stephenson\inst{\ref{inst:0075},\ref{inst:0478}}
\and         M.~                  S\"{ u}veges\orcit{0000-0003-3017-5322}\inst{\ref{inst:0479}}
\and         J.~                        Surdej\orcit{0000-0002-7005-1976}\inst{\ref{inst:0091},\ref{inst:0481}}
\and         L.~                      Szabados\orcit{0000-0002-2046-4131}\inst{\ref{inst:0143}}
\and         E.~                  Szegedi-Elek\orcit{0000-0001-7807-6644}\inst{\ref{inst:0143}}
\and         F.~                         Taris\inst{\ref{inst:0077}}
\and       M.B.~                        Taylor\orcit{0000-0002-4209-1479}\inst{\ref{inst:0485}}
\and         R.~                      Teixeira\orcit{0000-0002-6806-6626}\inst{\ref{inst:0233}}
\and         L.~                       Tolomei\orcit{0000-0002-3541-3230}\inst{\ref{inst:0060}}
\and         N.~                       Tonello\orcit{0000-0003-0550-1667}\inst{\ref{inst:0272}}
\and         F.~                         Torra\orcit{0000-0002-8429-299X}\inst{\ref{inst:0051}}
\and         J.~               Torra$^\dagger$\inst{\ref{inst:0027}}
\and         G.~                Torralba Elipe\orcit{0000-0001-8738-194X}\inst{\ref{inst:0101}}
\and         M.~                     Trabucchi\orcit{0000-0002-1429-2388}\inst{\ref{inst:0492},\ref{inst:0008}}
\and       A.T.~                       Tsounis\inst{\ref{inst:0494}}
\and         C.~                         Turon\orcit{0000-0003-1236-5157}\inst{\ref{inst:0018}}
\and         A.~                          Ulla\orcit{0000-0001-6424-5005}\inst{\ref{inst:0496}}
\and         N.~                         Unger\orcit{0000-0003-3993-7127}\inst{\ref{inst:0008}}
\and       M.V.~                      Vaillant\inst{\ref{inst:0169}}
\and         E.~                    van Dillen\inst{\ref{inst:0240}}
\and         W.~                    van Reeven\inst{\ref{inst:0500}}
\and         O.~                         Vanel\orcit{0000-0002-7898-0454}\inst{\ref{inst:0018}}
\and         A.~                     Vecchiato\orcit{0000-0003-1399-5556}\inst{\ref{inst:0043}}
\and         Y.~                         Viala\inst{\ref{inst:0018}}
\and         D.~                       Vicente\orcit{0000-0002-1584-1182}\inst{\ref{inst:0272}}
\and         S.~                     Voutsinas\inst{\ref{inst:0093}}
\and         M.~                        Weiler\inst{\ref{inst:0027}}
\and         T.~                        Wevers\orcit{0000-0002-4043-9400}\inst{\ref{inst:0007},\ref{inst:0508}}
\and      \L{}.~                   Wyrzykowski\orcit{0000-0002-9658-6151}\inst{\ref{inst:0320}}
\and         A.~                        Yoldas\inst{\ref{inst:0007}}
\and         P.~                         Yvard\inst{\ref{inst:0240}}
\and         H.~                          Zhao\orcit{0000-0003-2645-6869}\inst{\ref{inst:0022}}
\and         J.~                         Zorec\inst{\ref{inst:0513}}
\and         S.~                        Zucker\orcit{0000-0003-3173-3138}\inst{\ref{inst:0179}}
\and         T.~                       Zwitter\orcit{0000-0002-2325-8763}\inst{\ref{inst:0515}}
}
\institute{
     Instituut voor Sterrenkunde, KU Leuven, Celestijnenlaan 200D, 3001 Leuven, Belgium\relax                                                                                                                                                                                                                                                                      \label{inst:0001}
\and INAF - Osservatorio Astronomico di Capodimonte, Via Moiariello 16, 80131, Napoli, Italy\relax                                                                                                                                                                                                                                                                 \label{inst:0002}\vfill
\and Department of Astrophysics/IMAPP, Radboud University, P.O.Box 9010, 6500 GL Nijmegen, The Netherlands\relax                                                                                                                                                                                                                                                   \label{inst:0004}\vfill
\and Max Planck Institute for Astronomy, K\"{ o}nigstuhl 17, 69117 Heidelberg, Germany\relax                                                                                                                                                                                                                                                                       \label{inst:0005}\vfill
\and Ru{\dj}er Bo\v{s}kovi\'{c} Institute, Bijeni\v{c}ka cesta 54, 10000 Zagreb, Croatia\relax                                                                                                                                                                                                                                                                     \label{inst:0006}\vfill
\and Institute of Astronomy, University of Cambridge, Madingley Road, Cambridge CB3 0HA, United Kingdom\relax                                                                                                                                                                                                                                                      \label{inst:0007}\vfill
\and Department of Astronomy, University of Geneva, Chemin Pegasi 51, 1290 Versoix, Switzerland\relax                                                                                                                                                                                                                                                              \label{inst:0008}\vfill
\and Department of Astronomy, University of Geneva, Chemin d'Ecogia 16, 1290 Versoix, Switzerland\relax                                                                                                                                                                                                                                                            \label{inst:0010}\vfill
\and Leiden Observatory, Leiden University, Niels Bohrweg 2, 2333 CA Leiden, The Netherlands\relax                                                                                                                                                                                                                                                                 \label{inst:0014}\vfill
\and INAF - Osservatorio astronomico di Padova, Vicolo Osservatorio 5, 35122 Padova, Italy\relax                                                                                                                                                                                                                                                                   \label{inst:0015}\vfill
\and European Space Agency (ESA), European Space Research and Technology Centre (ESTEC), Keplerlaan 1, 2201AZ, Noordwijk, The Netherlands\relax                                                                                                                                                                                                                    \label{inst:0016}\vfill
\and GEPI, Observatoire de Paris, Universit\'{e} PSL, CNRS, 5 Place Jules Janssen, 92190 Meudon, France\relax                                                                                                                                                                                                                                                      \label{inst:0018}\vfill
\and Univ. Grenoble Alpes, CNRS, IPAG, 38000 Grenoble, France\relax                                                                                                                                                                                                                                                                                                \label{inst:0019}\vfill
\and Astronomisches Rechen-Institut, Zentrum f\"{ u}r Astronomie der Universit\"{ a}t Heidelberg, M\"{ o}nchhofstr. 12-14, 69120 Heidelberg, Germany\relax                                                                                                                                                                                                         \label{inst:0021}\vfill
\and Universit\'{e} C\^{o}te d'Azur, Observatoire de la C\^{o}te d'Azur, CNRS, Laboratoire Lagrange, Bd de l'Observatoire, CS 34229, 06304 Nice Cedex 4, France\relax                                                                                                                                                                                              \label{inst:0022}\vfill
\and Laboratoire d'astrophysique de Bordeaux, Univ. Bordeaux, CNRS, B18N, all{\'e}e Geoffroy Saint-Hilaire, 33615 Pessac, France\relax                                                                                                                                                                                                                             \label{inst:0023}\vfill
\and European Space Agency (ESA), European Space Astronomy Centre (ESAC), Camino bajo del Castillo, s/n, Urbanizacion Villafranca del Castillo, Villanueva de la Ca\~{n}ada, 28692 Madrid, Spain\relax                                                                                                                                                             \label{inst:0025}\vfill
\and Aurora Technology for European Space Agency (ESA), Camino bajo del Castillo, s/n, Urbanizacion Villafranca del Castillo, Villanueva de la Ca\~{n}ada, 28692 Madrid, Spain\relax                                                                                                                                                                               \label{inst:0026}\vfill
\and Institut de Ci\`{e}ncies del Cosmos (ICCUB), Universitat  de  Barcelona  (IEEC-UB), Mart\'{i} i  Franqu\`{e}s  1, 08028 Barcelona, Spain\relax                                                                                                                                                                                                                \label{inst:0027}\vfill
\and Lohrmann Observatory, Technische Universit\"{ a}t Dresden, Mommsenstra{\ss}e 13, 01062 Dresden, Germany\relax                                                                                                                                                                                                                                                 \label{inst:0028}\vfill
\and Lund Observatory, Department of Astronomy and Theoretical Physics, Lund University, Box 43, 22100 Lund, Sweden\relax                                                                                                                                                                                                                                          \label{inst:0030}\vfill
\and CNES Centre Spatial de Toulouse, 18 avenue Edouard Belin, 31401 Toulouse Cedex 9, France\relax                                                                                                                                                                                                                                                                \label{inst:0033}\vfill
\and Institut d'Astronomie et d'Astrophysique, Universit\'{e} Libre de Bruxelles CP 226, Boulevard du Triomphe, 1050 Brussels, Belgium\relax                                                                                                                                                                                                                       \label{inst:0034}\vfill
\and F.R.S.-FNRS, Rue d'Egmont 5, 1000 Brussels, Belgium\relax                                                                                                                                                                                                                                                                                                     \label{inst:0035}\vfill
\and INAF - Osservatorio Astrofisico di Arcetri, Largo Enrico Fermi 5, 50125 Firenze, Italy\relax                                                                                                                                                                                                                                                                  \label{inst:0036}\vfill
\and INAF - Osservatorio Astrofisico di Torino, via Osservatorio 20, 10025 Pino Torinese (TO), Italy\relax                                                                                                                                                                                                                                                         \label{inst:0043}\vfill
\and European Space Agency (ESA, retired)\relax                                                                                                                                                                                                                                                                                                                    \label{inst:0044}\vfill
\and University of Turin, Department of Physics, Via Pietro Giuria 1, 10125 Torino, Italy\relax                                                                                                                                                                                                                                                                    \label{inst:0047}\vfill
\and INAF - Osservatorio di Astrofisica e Scienza dello Spazio di Bologna, via Piero Gobetti 93/3, 40129 Bologna, Italy\relax                                                                                                                                                                                                                                      \label{inst:0050}\vfill
\and DAPCOM for Institut de Ci\`{e}ncies del Cosmos (ICCUB), Universitat  de  Barcelona  (IEEC-UB), Mart\'{i} i  Franqu\`{e}s  1, 08028 Barcelona, Spain\relax                                                                                                                                                                                                     \label{inst:0051}\vfill
\and Royal Observatory of Belgium, Ringlaan 3, 1180 Brussels, Belgium\relax                                                                                                                                                                                                                                                                                        \label{inst:0055}\vfill
\and Observational Astrophysics, Division of Astronomy and Space Physics, Department of Physics and Astronomy, Uppsala University, Box 516, 751 20 Uppsala, Sweden\relax                                                                                                                                                                                           \label{inst:0058}\vfill
\and ALTEC S.p.a, Corso Marche, 79,10146 Torino, Italy\relax                                                                                                                                                                                                                                                                                                       \label{inst:0060}\vfill
\and S\`{a}rl, Geneva, Switzerland\relax                                                                                                                                                                                                                                                                                                                           \label{inst:0063}\vfill
\and Mullard Space Science Laboratory, University College London, Holmbury St Mary, Dorking, Surrey RH5 6NT, United Kingdom\relax                                                                                                                                                                                                                                  \label{inst:0069}\vfill
\and Gaia DPAC Project Office, ESAC, Camino bajo del Castillo, s/n, Urbanizacion Villafranca del Castillo, Villanueva de la Ca\~{n}ada, 28692 Madrid, Spain\relax                                                                                                                                                                                                  \label{inst:0072}\vfill
\and Telespazio UK S.L. for European Space Agency (ESA), Camino bajo del Castillo, s/n, Urbanizacion Villafranca del Castillo, Villanueva de la Ca\~{n}ada, 28692 Madrid, Spain\relax                                                                                                                                                                              \label{inst:0075}\vfill
\and SYRTE, Observatoire de Paris, Universit\'{e} PSL, CNRS,  Sorbonne Universit\'{e}, LNE, 61 avenue de l'Observatoire 75014 Paris, France\relax                                                                                                                                                                                                                  \label{inst:0077}\vfill
\and National Observatory of Athens, I. Metaxa and Vas. Pavlou, Palaia Penteli, 15236 Athens, Greece\relax                                                                                                                                                                                                                                                         \label{inst:0079}\vfill
\and IMCCE, Observatoire de Paris, Universit\'{e} PSL, CNRS, Sorbonne Universit{\'e}, Univ. Lille, 77 av. Denfert-Rochereau, 75014 Paris, France\relax                                                                                                                                                                                                             \label{inst:0081}\vfill
\and Serco Gesti\'{o}n de Negocios for European Space Agency (ESA), Camino bajo del Castillo, s/n, Urbanizacion Villafranca del Castillo, Villanueva de la Ca\~{n}ada, 28692 Madrid, Spain\relax                                                                                                                                                                   \label{inst:0089}\vfill
\and Institut d'Astrophysique et de G\'{e}ophysique, Universit\'{e} de Li\`{e}ge, 19c, All\'{e}e du 6 Ao\^{u}t, B-4000 Li\`{e}ge, Belgium\relax                                                                                                                                                                                                                    \label{inst:0091}\vfill
\and CRAAG - Centre de Recherche en Astronomie, Astrophysique et G\'{e}ophysique, Route de l'Observatoire Bp 63 Bouzareah 16340 Algiers, Algeria\relax                                                                                                                                                                                                             \label{inst:0092}\vfill
\and Institute for Astronomy, University of Edinburgh, Royal Observatory, Blackford Hill, Edinburgh EH9 3HJ, United Kingdom\relax                                                                                                                                                                                                                                  \label{inst:0093}\vfill
\and RHEA for European Space Agency (ESA), Camino bajo del Castillo, s/n, Urbanizacion Villafranca del Castillo, Villanueva de la Ca\~{n}ada, 28692 Madrid, Spain\relax                                                                                                                                                                                            \label{inst:0098}\vfill
\and ATG Europe for European Space Agency (ESA), Camino bajo del Castillo, s/n, Urbanizacion Villafranca del Castillo, Villanueva de la Ca\~{n}ada, 28692 Madrid, Spain\relax                                                                                                                                                                                      \label{inst:0099}\vfill
\and CIGUS CITIC - Department of Computer Science and Information Technologies, University of A Coru\~{n}a, Campus de Elvi\~{n}a s/n, A Coru\~{n}a, 15071, Spain\relax                                                                                                                                                                                             \label{inst:0101}\vfill
\and Universit\'{e} de Strasbourg, CNRS, Observatoire astronomique de Strasbourg, UMR 7550,  11 rue de l'Universit\'{e}, 67000 Strasbourg, France\relax                                                                                                                                                                                                            \label{inst:0106}\vfill
\and Kavli Institute for Cosmology Cambridge, Institute of Astronomy, Madingley Road, Cambridge, CB3 0HA\relax                                                                                                                                                                                                                                                     \label{inst:0109}\vfill
\and Leibniz Institute for Astrophysics Potsdam (AIP), An der Sternwarte 16, 14482 Potsdam, Germany\relax                                                                                                                                                                                                                                                          \label{inst:0113}\vfill
\and CENTRA, Faculdade de Ci\^{e}ncias, Universidade de Lisboa, Edif. C8, Campo Grande, 1749-016 Lisboa, Portugal\relax                                                                                                                                                                                                                                            \label{inst:0116}\vfill
\and Department of Informatics, Donald Bren School of Information and Computer Sciences, University of California, Irvine, 5226 Donald Bren Hall, 92697-3440 CA Irvine, United States\relax                                                                                                                                                                        \label{inst:0117}\vfill
\and INAF - Osservatorio Astrofisico di Catania, via S. Sofia 78, 95123 Catania, Italy\relax                                                                                                                                                                                                                                                                       \label{inst:0118}\vfill
\and Dipartimento di Fisica e Astronomia ""Ettore Majorana"", Universit\`{a} di Catania, Via S. Sofia 64, 95123 Catania, Italy\relax                                                                                                                                                                                                                               \label{inst:0119}\vfill
\and INAF - Osservatorio Astronomico di Roma, Via Frascati 33, 00078 Monte Porzio Catone (Roma), Italy\relax                                                                                                                                                                                                                                                       \label{inst:0122}\vfill
\and Space Science Data Center - ASI, Via del Politecnico SNC, 00133 Roma, Italy\relax                                                                                                                                                                                                                                                                             \label{inst:0123}\vfill
\and Department of Physics, University of Helsinki, P.O. Box 64, 00014 Helsinki, Finland\relax                                                                                                                                                                                                                                                                     \label{inst:0125}\vfill
\and Finnish Geospatial Research Institute FGI, Geodeetinrinne 2, 02430 Masala, Finland\relax                                                                                                                                                                                                                                                                      \label{inst:0126}\vfill
\and Institut UTINAM CNRS UMR6213, Universit\'{e} Bourgogne Franche-Comt\'{e}, OSU THETA Franche-Comt\'{e} Bourgogne, Observatoire de Besan\c{c}on, BP1615, 25010 Besan\c{c}on Cedex, France\relax                                                                                                                                                                 \label{inst:0132}\vfill
\and HE Space Operations BV for European Space Agency (ESA), Keplerlaan 1, 2201AZ, Noordwijk, The Netherlands\relax                                                                                                                                                                                                                                                \label{inst:0134}\vfill
\and Dpto. de Inteligencia Artificial, UNED, c/ Juan del Rosal 16, 28040 Madrid, Spain\relax                                                                                                                                                                                                                                                                       \label{inst:0136}\vfill
\and Konkoly Observatory, Research Centre for Astronomy and Earth Sciences, E\"{ o}tv\"{ o}s Lor{\'a}nd Research Network (ELKH), MTA Centre of Excellence, Konkoly Thege Mikl\'{o}s \'{u}t 15-17, 1121 Budapest, Hungary\relax                                                                                                                                     \label{inst:0143}\vfill
\and ELTE E\"{ o}tv\"{ o}s Lor\'{a}nd University, Institute of Physics, 1117, P\'{a}zm\'{a}ny P\'{e}ter s\'{e}t\'{a}ny 1A, Budapest, Hungary\relax                                                                                                                                                                                                                 \label{inst:0144}\vfill
\and University of Vienna, Department of Astrophysics, T\"{ u}rkenschanzstra{\ss}e 17, A1180 Vienna, Austria\relax                                                                                                                                                                                                                                                 \label{inst:0152}\vfill
\and Institute of Physics, Laboratory of Astrophysics, Ecole Polytechnique F\'ed\'erale de Lausanne (EPFL), Observatoire de Sauverny, 1290 Versoix, Switzerland\relax                                                                                                                                                                                              \label{inst:0154}\vfill
\and Kapteyn Astronomical Institute, University of Groningen, Landleven 12, 9747 AD Groningen, The Netherlands\relax                                                                                                                                                                                                                                               \label{inst:0160}\vfill
\and School of Physics and Astronomy / Space Park Leicester, University of Leicester, University Road, Leicester LE1 7RH, United Kingdom\relax                                                                                                                                                                                                                     \label{inst:0167}\vfill
\and Thales Services for CNES Centre Spatial de Toulouse, 18 avenue Edouard Belin, 31401 Toulouse Cedex 9, France\relax                                                                                                                                                                                                                                            \label{inst:0169}\vfill
\and Depto. Estad\'istica e Investigaci\'on Operativa. Universidad de C\'adiz, Avda. Rep\'ublica Saharaui s/n, 11510 Puerto Real, C\'adiz, Spain\relax                                                                                                                                                                                                             \label{inst:0173}\vfill
\and Center for Research and Exploration in Space Science and Technology, University of Maryland Baltimore County, 1000 Hilltop Circle, Baltimore MD, USA\relax                                                                                                                                                                                                    \label{inst:0175}\vfill
\and GSFC - Goddard Space Flight Center, Code 698, 8800 Greenbelt Rd, 20771 MD Greenbelt, United States\relax                                                                                                                                                                                                                                                      \label{inst:0176}\vfill
\and EURIX S.r.l., Corso Vittorio Emanuele II 61, 10128, Torino, Italy\relax                                                                                                                                                                                                                                                                                       \label{inst:0178}\vfill
\and Porter School of the Environment and Earth Sciences, Tel Aviv University, Tel Aviv 6997801, Israel\relax                                                                                                                                                                                                                                                      \label{inst:0179}\vfill
\and Harvard-Smithsonian Center for Astrophysics, 60 Garden St., MS 15, Cambridge, MA 02138, USA\relax                                                                                                                                                                                                                                                             \label{inst:0180}\vfill
\and HE Space Operations BV for European Space Agency (ESA), Camino bajo del Castillo, s/n, Urbanizacion Villafranca del Castillo, Villanueva de la Ca\~{n}ada, 28692 Madrid, Spain\relax                                                                                                                                                                          \label{inst:0182}\vfill
\and Instituto de Astrof\'{i}sica e Ci\^{e}ncias do Espa\c{c}o, Universidade do Porto, CAUP, Rua das Estrelas, PT4150-762 Porto, Portugal\relax                                                                                                                                                                                                                    \label{inst:0183}\vfill
\and LFCA/DAS,Universidad de Chile,CNRS,Casilla 36-D, Santiago, Chile\relax                                                                                                                                                                                                                                                                                        \label{inst:0185}\vfill
\and SISSA - Scuola Internazionale Superiore di Studi Avanzati, via Bonomea 265, 34136 Trieste, Italy\relax                                                                                                                                                                                                                                                        \label{inst:0189}\vfill
\and Telespazio for CNES Centre Spatial de Toulouse, 18 avenue Edouard Belin, 31401 Toulouse Cedex 9, France\relax                                                                                                                                                                                                                                                 \label{inst:0194}\vfill
\and University of Turin, Department of Computer Sciences, Corso Svizzera 185, 10149 Torino, Italy\relax                                                                                                                                                                                                                                                           \label{inst:0198}\vfill
\and Dpto. de Matem\'{a}tica Aplicada y Ciencias de la Computaci\'{o}n, Univ. de Cantabria, ETS Ingenieros de Caminos, Canales y Puertos, Avda. de los Castros s/n, 39005 Santander, Spain\relax                                                                                                                                                                   \label{inst:0201}\vfill
\and Centro de Astronom\'{i}a - CITEVA, Universidad de Antofagasta, Avenida Angamos 601, Antofagasta 1270300, Chile\relax                                                                                                                                                                                                                                          \label{inst:0211}\vfill
\and DLR Gesellschaft f\"{ u}r Raumfahrtanwendungen (GfR) mbH M\"{ u}nchener Stra{\ss}e 20 , 82234 We{\ss}ling\relax                                                                                                                                                                                                                                               \label{inst:0216}\vfill
\and Centre for Astrophysics Research, University of Hertfordshire, College Lane, AL10 9AB, Hatfield, United Kingdom\relax                                                                                                                                                                                                                                         \label{inst:0218}\vfill
\and University of Turin, Mathematical Department ""G.Peano"", Via Carlo Alberto 10, 10123 Torino, Italy\relax                                                                                                                                                                                                                                                     \label{inst:0225}\vfill
\and INAF - Osservatorio Astronomico d'Abruzzo, Via Mentore Maggini, 64100 Teramo, Italy\relax                                                                                                                                                                                                                                                                     \label{inst:0231}\vfill
\and Instituto de Astronomia, Geof\`{i}sica e Ci\^{e}ncias Atmosf\'{e}ricas, Universidade de S\~{a}o Paulo, Rua do Mat\~{a}o, 1226, Cidade Universitaria, 05508-900 S\~{a}o Paulo, SP, Brazil\relax                                                                                                                                                                \label{inst:0233}\vfill
\and APAVE SUDEUROPE SAS for CNES Centre Spatial de Toulouse, 18 avenue Edouard Belin, 31401 Toulouse Cedex 9, France\relax                                                                                                                                                                                                                                        \label{inst:0240}\vfill
\and M\'{e}socentre de calcul de Franche-Comt\'{e}, Universit\'{e} de Franche-Comt\'{e}, 16 route de Gray, 25030 Besan\c{c}on Cedex, France\relax                                                                                                                                                                                                                  \label{inst:0242}\vfill
\and ATOS for CNES Centre Spatial de Toulouse, 18 avenue Edouard Belin, 31401 Toulouse Cedex 9, France\relax                                                                                                                                                                                                                                                       \label{inst:0247}\vfill
\and School of Physics and Astronomy, Tel Aviv University, Tel Aviv 6997801, Israel\relax                                                                                                                                                                                                                                                                          \label{inst:0250}\vfill
\and Astrophysics Research Centre, School of Mathematics and Physics, Queen's University Belfast, Belfast BT7 1NN, UK\relax                                                                                                                                                                                                                                        \label{inst:0252}\vfill
\and Centre de Donn\'{e}es Astronomique de Strasbourg, Strasbourg, France\relax                                                                                                                                                                                                                                                                                    \label{inst:0254}\vfill
\and Institute for Computational Cosmology, Department of Physics, Durham University, Durham DH1 3LE, UK\relax                                                                                                                                                                                                                                                     \label{inst:0258}\vfill
\and European Southern Observatory, Karl-Schwarzschild-Str. 2, 85748 Garching, Germany\relax                                                                                                                                                                                                                                                                       \label{inst:0259}\vfill
\and Max-Planck-Institut f\"{ u}r Astrophysik, Karl-Schwarzschild-Stra{\ss}e 1, 85748 Garching, Germany\relax                                                                                                                                                                                                                                                      \label{inst:0260}\vfill
\and Data Science and Big Data Lab, Pablo de Olavide University, 41013, Seville, Spain\relax                                                                                                                                                                                                                                                                       \label{inst:0264}\vfill
\and Barcelona Supercomputing Center (BSC), Pla\c{c}a Eusebi G\"{ u}ell 1-3, 08034-Barcelona, Spain\relax                                                                                                                                                                                                                                                          \label{inst:0272}\vfill
\and ETSE Telecomunicaci\'{o}n, Universidade de Vigo, Campus Lagoas-Marcosende, 36310 Vigo, Galicia, Spain\relax                                                                                                                                                                                                                                                   \label{inst:0277}\vfill
\and Asteroid Engineering Laboratory, Space Systems, Lule\aa{} University of Technology, Box 848, S-981 28 Kiruna, Sweden\relax                                                                                                                                                                                                                                    \label{inst:0281}\vfill
\and Vera C Rubin Observatory,  950 N. Cherry Avenue, Tucson, AZ 85719, USA\relax                                                                                                                                                                                                                                                                                  \label{inst:0286}\vfill
\and Department of Astrophysics, Astronomy and Mechanics, National and Kapodistrian University of Athens, Panepistimiopolis, Zografos, 15783 Athens, Greece\relax                                                                                                                                                                                                  \label{inst:0287}\vfill
\and TRUMPF Photonic Components GmbH, Lise-Meitner-Stra{\ss}e 13,  89081 Ulm, Germany\relax                                                                                                                                                                                                                                                                        \label{inst:0290}\vfill
\and IAC - Instituto de Astrofisica de Canarias, Via L\'{a}ctea s/n, 38200 La Laguna S.C., Tenerife, Spain\relax                                                                                                                                                                                                                                                   \label{inst:0295}\vfill
\and Department of Astrophysics, University of La Laguna, Via L\'{a}ctea s/n, 38200 La Laguna S.C., Tenerife, Spain\relax                                                                                                                                                                                                                                          \label{inst:0296}\vfill
\and Faculty of Aerospace Engineering, Delft University of Technology, Kluyverweg 1, 2629 HS Delft, The Netherlands\relax                                                                                                                                                                                                                                          \label{inst:0298}\vfill
\and Radagast Solutions\relax                                                                                                                                                                                                                                                                                                                                      \label{inst:0302}\vfill
\and Laboratoire Univers et Particules de Montpellier, CNRS Universit\'{e} Montpellier, Place Eug\`{e}ne Bataillon, CC72, 34095 Montpellier Cedex 05, France\relax                                                                                                                                                                                                 \label{inst:0303}\vfill
\and Universit\'{e} de Caen Normandie, C\^{o}te de Nacre Boulevard Mar\'{e}chal Juin, 14032 Caen, France\relax                                                                                                                                                                                                                                                     \label{inst:0309}\vfill
\and LESIA, Observatoire de Paris, Universit\'{e} PSL, CNRS, Sorbonne Universit\'{e}, Universit\'{e} de Paris, 5 Place Jules Janssen, 92190 Meudon, France\relax                                                                                                                                                                                                   \label{inst:0310}\vfill
\and SRON Netherlands Institute for Space Research, Niels Bohrweg 4, 2333 CA Leiden, The Netherlands\relax                                                                                                                                                                                                                                                         \label{inst:0319}\vfill
\and Astronomical Observatory, University of Warsaw,  Al. Ujazdowskie 4, 00-478 Warszawa, Poland\relax                                                                                                                                                                                                                                                             \label{inst:0320}\vfill
\and Scalian for CNES Centre Spatial de Toulouse, 18 avenue Edouard Belin, 31401 Toulouse Cedex 9, France\relax                                                                                                                                                                                                                                                    \label{inst:0322}\vfill
\and Universit\'{e} Rennes, CNRS, IPR (Institut de Physique de Rennes) - UMR 6251, 35000 Rennes, France\relax                                                                                                                                                                                                                                                      \label{inst:0328}\vfill
\and Shanghai Astronomical Observatory, Chinese Academy of Sciences, 80 Nandan Road, Shanghai 200030, People's Republic of China\relax                                                                                                                                                                                                                             \label{inst:0333}\vfill
\and University of Chinese Academy of Sciences, No.19(A) Yuquan Road, Shijingshan District, Beijing 100049, People's Republic of China\relax                                                                                                                                                                                                                       \label{inst:0335}\vfill
\and Niels Bohr Institute, University of Copenhagen, Juliane Maries Vej 30, 2100 Copenhagen {\O}, Denmark\relax                                                                                                                                                                                                                                                    \label{inst:0338}\vfill
\and DXC Technology, Retortvej 8, 2500 Valby, Denmark\relax                                                                                                                                                                                                                                                                                                        \label{inst:0339}\vfill
\and Las Cumbres Observatory, 6740 Cortona Drive Suite 102, Goleta, CA 93117, USA\relax                                                                                                                                                                                                                                                                            \label{inst:0340}\vfill
\and CIGUS CITIC, Department of Nautical Sciences and Marine Engineering, University of A Coru\~{n}a, Paseo de Ronda 51, 15071, A Coru\~{n}a, Spain\relax                                                                                                                                                                                                          \label{inst:0349}\vfill
\and Astrophysics Research Institute, Liverpool John Moores University, 146 Brownlow Hill, Liverpool L3 5RF, United Kingdom\relax                                                                                                                                                                                                                                  \label{inst:0350}\vfill
\and IPAC, Mail Code 100-22, California Institute of Technology, 1200 E. California Blvd., Pasadena, CA 91125, USA\relax                                                                                                                                                                                                                                           \label{inst:0357}\vfill
\and IRAP, Universit\'{e} de Toulouse, CNRS, UPS, CNES, 9 Av. colonel Roche, BP 44346, 31028 Toulouse Cedex 4, France\relax                                                                                                                                                                                                                                        \label{inst:0358}\vfill
\and MTA CSFK Lend\"{ u}let Near-Field Cosmology Research Group, Konkoly Observatory, MTA Research Centre for Astronomy and Earth Sciences, Konkoly Thege Mikl\'{o}s \'{u}t 15-17, 1121 Budapest, Hungary\relax                                                                                                                                                    \label{inst:0375}\vfill
\and Departmento de F\'{i}sica de la Tierra y Astrof\'{i}sica, Universidad Complutense de Madrid, 28040 Madrid, Spain\relax                                                                                                                                                                                                                                        \label{inst:0391}\vfill
\and Villanova University, Department of Astrophysics and Planetary Science, 800 E Lancaster Avenue, Villanova PA 19085, USA\relax                                                                                                                                                                                                                                 \label{inst:0411}\vfill
\and INAF - Osservatorio Astronomico di Brera, via E. Bianchi, 46, 23807 Merate (LC), Italy\relax                                                                                                                                                                                                                                                                  \label{inst:0417}\vfill
\and STFC, Rutherford Appleton Laboratory, Harwell, Didcot, OX11 0QX, United Kingdom\relax                                                                                                                                                                                                                                                                         \label{inst:0424}\vfill
\and Charles University, Faculty of Mathematics and Physics, Astronomical Institute of Charles University, V Holesovickach 2, 18000 Prague, Czech Republic\relax                                                                                                                                                                                                   \label{inst:0446}\vfill
\and Department of Particle Physics and Astrophysics, Weizmann Institute of Science, Rehovot 7610001, Israel\relax                                                                                                                                                                                                                                                 \label{inst:0457}\vfill
\and Department of Astrophysical Sciences, 4 Ivy Lane, Princeton University, Princeton NJ 08544, USA\relax                                                                                                                                                                                                                                                         \label{inst:0458}\vfill
\and Departamento de Astrof\'{i}sica, Centro de Astrobiolog\'{i}a (CSIC-INTA), ESA-ESAC. Camino Bajo del Castillo s/n. 28692 Villanueva de la Ca\~{n}ada, Madrid, Spain\relax                                                                                                                                                                                      \label{inst:0467}\vfill
\and naXys, University of Namur, Rempart de la Vierge, 5000 Namur, Belgium\relax                                                                                                                                                                                                                                                                                   \label{inst:0470}\vfill
\and CGI Deutschland B.V. \& Co. KG, Mornewegstr. 30, 64293 Darmstadt, Germany\relax                                                                                                                                                                                                                                                                               \label{inst:0478}\vfill
\and Institute of Global Health, University of Geneva\relax                                                                                                                                                                                                                                                                                                        \label{inst:0479}\vfill
\and Astronomical Observatory Institute, Faculty of Physics, Adam Mickiewicz University, Pozna\'{n}, Poland\relax                                                                                                                                                                                                                                                  \label{inst:0481}\vfill
\and H H Wills Physics Laboratory, University of Bristol, Tyndall Avenue, Bristol BS8 1TL, United Kingdom\relax                                                                                                                                                                                                                                                    \label{inst:0485}\vfill
\and Department of Physics and Astronomy G. Galilei, University of Padova, Vicolo dell'Osservatorio 3, 35122, Padova, Italy\relax                                                                                                                                                                                                                                  \label{inst:0492}\vfill
\and CERN, Geneva, Switzerland\relax                                                                                                                                                                                                                                                                                                                               \label{inst:0494}\vfill
\and Applied Physics Department, Universidade de Vigo, 36310 Vigo, Spain\relax                                                                                                                                                                                                                                                                                     \label{inst:0496}\vfill
\and Association of Universities for Research in Astronomy, 1331 Pennsylvania Ave. NW, Washington, DC 20004, USA\relax                                                                                                                                                                                                                                             \label{inst:0500}\vfill
\and European Southern Observatory, Alonso de C\'ordova 3107, Casilla 19, Santiago, Chile\relax                                                                                                                                                                                                                                                                    \label{inst:0508}\vfill
\and Sorbonne Universit\'{e}, CNRS, UMR7095, Institut d'Astrophysique de Paris, 98bis bd. Arago, 75014 Paris, France\relax                                                                                                                                                                                                                                         \label{inst:0513}\vfill
\and Faculty of Mathematics and Physics, University of Ljubljana, Jadranska ulica 19, 1000 Ljubljana, Slovenia\relax                                                                                                                                                                                                                                               \label{inst:0515}\vfill
}

%% file: article.bbl
\begin{thebibliography}{145}
\expandafter\ifx\csname natexlab\endcsname\relax\def\natexlab#1{#1}\fi

\bibitem[{{Aerts}(2021)}]{Aerts2021}
{Aerts}, C. 2021, Reviews of Modern Physics, 93, 015001

\bibitem[{{Aerts} {et~al.}(2021){Aerts}, {Augustson}, {Mathis}, {Pedersen},
  {Mombarg}, {Vanlaer}, {Van Beeck}, \& {Van Reeth}}]{Aerts2021-IGW}
{Aerts}, C., {Augustson}, K., {Mathis}, S., {et~al.} 2021, \aap, 656, A121

\bibitem[{{Aerts} \& {Kolenberg}(2005)}]{AertsKolenberg2005}
{Aerts}, C. \& {Kolenberg}, K. 2005, \aap, 431, 615

\bibitem[{{Aerts} {et~al.}(2019){Aerts}, {Mathis}, \& {Rogers}}]{Aerts2019}
{Aerts}, C., {Mathis}, S., \& {Rogers}, T.~M. 2019, \araa, 57, 35

\bibitem[{{Aerts} {et~al.}(2014){Aerts}, {Molenberghs}, {Kenward}, \&
  {Neiner}}]{Aerts2014}
{Aerts}, C., {Molenberghs}, G., {Kenward}, M.~G., \& {Neiner}, C. 2014, \apj,
  781, 88

\bibitem[{{Aerts} {et~al.}(2018){Aerts}, {Molenberghs}, {Michielsen},
  {Pedersen}, {Bj{\"o}rklund}, {Johnston}, {Mombarg}, {Bowman}, {Buysschaert},
  {P{\'a}pics}, {Sekaran}, {Sundqvist}, {Tkachenko}, {Truyaert}, {Van Reeth},
  \& {Vermeyen}}]{Aerts2018}
{Aerts}, C., {Molenberghs}, G., {Michielsen}, M., {et~al.} 2018, \apjs, 237, 15

\bibitem[{{Andrae}(2022)}]{Andrae2022}
{Andrae}, R. e.~a. 2022, \aap

\bibitem[{{Antoci} {et~al.}(2019){Antoci}, {Cunha}, {Bowman}, {Murphy},
  {Kurtz}, {Bedding}, {Borre}, {Christophe}, {Daszy{\'n}ska-Daszkiewicz},
  {Fox-Machado}, {Garc{\'\i}a Hern{\'a}ndez}, {Ghasemi}, {Handberg}, {Hansen},
  {Hasanzadeh}, {Houdek}, {Johnston}, {Justesen}, {Kahraman Alicavus},
  {Kotysz}, {Latham}, {Matthews}, {M{\o}nster}, {Niemczura}, {Paunzen},
  {S{\'a}nchez Arias}, {Pigulski}, {Pepper}, {Richey-Yowell}, {Safari},
  {Seager}, {Smalley}, {Shutt}, {S{\'o}dor}, {Su{\'a}rez}, {Tkachenko}, {Wu},
  {Zwintz}, {Barcel{\'o} Forteza}, {Brunsden}, {Bogn{\'a}r}, {Buzasi},
  {Chowdhury}, {De Cat}, {Evans}, {Guo}, {Guzik}, {Jevtic}, {Lampens}, {Lares
  Martiz}, {Lovekin}, {Li}, {Mirouh}, {Mkrtichian}, {Monteiro}, {Nemec},
  {Ouazzani}, {Pascual-Granado}, {Reese}, {Rieutord}, {Rodon}, {Skarka},
  {Sowicka}, {Stateva}, {Szab{\'o}}, \& {Weiss}}]{Antoci2019}
{Antoci}, V., {Cunha}, M.~S., {Bowman}, D.~M., {et~al.} 2019, \mnras, 490, 4040

\bibitem[{{Astropy Collaboration} {et~al.}(2018){Astropy Collaboration},
  {Price-Whelan}, {Sip{\H{o}}cz}, {G{\"u}nther}, {Lim}, {Crawford}, {Conseil},
  {Shupe}, {Craig}, {Dencheva}, {Ginsburg}, {Vand erPlas}, {Bradley},
  {P{\'e}rez-Su{\'a}rez}, {de Val-Borro}, {Aldcroft}, {Cruz}, {Robitaille},
  {Tollerud}, {Ardelean}, {Babej}, {Bach}, {Bachetti}, {Bakanov}, {Bamford},
  {Barentsen}, {Barmby}, {Baumbach}, {Berry}, {Biscani}, {Boquien}, {Bostroem},
  {Bouma}, {Brammer}, {Bray}, {Breytenbach}, {Buddelmeijer}, {Burke},
  {Calderone}, {Cano Rodr{\'\i}guez}, {Cara}, {Cardoso}, {Cheedella}, {Copin},
  {Corrales}, {Crichton}, {D'Avella}, {Deil}, {Depagne}, {Dietrich}, {Donath},
  {Droettboom}, {Earl}, {Erben}, {Fabbro}, {Ferreira}, {Finethy}, {Fox},
  {Garrison}, {Gibbons}, {Goldstein}, {Gommers}, {Greco}, {Greenfield},
  {Groener}, {Grollier}, {Hagen}, {Hirst}, {Homeier}, {Horton}, {Hosseinzadeh},
  {Hu}, {Hunkeler}, {Ivezi{\'c}}, {Jain}, {Jenness}, {Kanarek}, {Kendrew},
  {Kern}, {Kerzendorf}, {Khvalko}, {King}, {Kirkby}, {Kulkarni}, {Kumar},
  {Lee}, {Lenz}, {Littlefair}, {Ma}, {Macleod}, {Mastropietro}, {McCully},
  {Montagnac}, {Morris}, {Mueller}, {Mumford}, {Muna}, {Murphy}, {Nelson},
  {Nguyen}, {Ninan}, {N{\"o}the}, {Ogaz}, {Oh}, {Parejko}, {Parley}, {Pascual},
  {Patil}, {Patil}, {Plunkett}, {Prochaska}, {Rastogi}, {Reddy Janga},
  {Sabater}, {Sakurikar}, {Seifert}, {Sherbert}, {Sherwood-Taylor}, {Shih},
  {Sick}, {Silbiger}, {Singanamalla}, {Singer}, {Sladen}, {Sooley},
  {Sornarajah}, {Streicher}, {Teuben}, {Thomas}, {Tremblay}, {Turner},
  {Terr{\'o}n}, {van Kerkwijk}, {de la Vega}, {Watkins}, {Weaver}, {Whitmore},
  {Woillez}, {Zabalza}, \& {Astropy Contributors}}]{Astropy2018}
{Astropy Collaboration}, {Price-Whelan}, A.~M., {Sip{\H{o}}cz}, B.~M., {et~al.}
  2018, \aj, 156, 123

\bibitem[{{Astropy Collaboration} {et~al.}(2013){Astropy Collaboration},
  {Robitaille}, {Tollerud}, {Greenfield}, {Droettboom}, {Bray}, {Aldcroft},
  {Davis}, {Ginsburg}, {Price-Whelan}, {Kerzendorf}, {Conley}, {Crighton},
  {Barbary}, {Muna}, {Ferguson}, {Grollier}, {Parikh}, {Nair}, {Unther},
  {Deil}, {Woillez}, {Conseil}, {Kramer}, {Turner}, {Singer}, {Fox}, {Weaver},
  {Zabalza}, {Edwards}, {Azalee Bostroem}, {Burke}, {Casey}, {Crawford},
  {Dencheva}, {Ely}, {Jenness}, {Labrie}, {Lim}, {Pierfederici}, {Pontzen},
  {Ptak}, {Refsdal}, {Servillat}, \& {Streicher}}]{Astropy2013}
{Astropy Collaboration}, {Robitaille}, T.~P., {Tollerud}, E.~J., {et~al.} 2013,
  \aap, 558, A33

\bibitem[{{Auvergne} {et~al.}(2009){Auvergne}, {Bodin}, {Boisnard}, {Buey},
  {Chaintreuil}, {Epstein}, {Jouret}, {Lam-Trong}, {Levacher}, {Magnan},
  {Perez}, {Plasson}, {Plesseria}, {Peter}, {Steller}, {Tiph{\`e}ne}, {Baglin},
  {Agogu{\'e}}, {Appourchaux}, {Barbet}, {Beaufort}, {Bellenger}, {Berlin},
  {Bernardi}, {Blouin}, {Boumier}, {Bonneau}, {Briet}, {Butler}, {Cautain},
  {Chiavassa}, {Costes}, {Cuvilho}, {Cunha-Parro}, {de Oliveira Fialho},
  {Decaudin}, {Defise}, {Djalal}, {Docclo}, {Drummond}, {Dupuis}, {Exil},
  {Faur{\'e}}, {Gaboriaud}, {Gamet}, {Gavalda}, {Grolleau}, {Gueguen},
  {Guivarc'h}, {Guterman}, {Hasiba}, {Huntzinger}, {Hustaix}, {Imbert},
  {Jeanville}, {Johlander}, {Jorda}, {Journoud}, {Karioty}, {Kerjean},
  {Lafond}, {Lapeyrere}, {Landiech}, {Larqu{\'e}}, {Laudet}, {Le Merrer},
  {Leporati}, {Leruyet}, {Levieuge}, {Llebaria}, {Martin}, {Mazy}, {Mesnager},
  {Michel}, {Moalic}, {Monjoin}, {Naudet}, {Neukirchner}, {Nguyen-Kim},
  {Ollivier}, {Orcesi}, {Ottacher}, {Oulali}, {Parisot}, {Perruchot},
  {Piacentino}, {Pinheiro da Silva}, {Platzer}, {Pontet}, {Pradines},
  {Quentin}, {Rohbeck}, {Rolland}, {Rollenhagen}, {Romagnan}, {Russ}, {Samadi},
  {Schmidt}, {Schwartz}, {Sebbag}, {Smit}, {Sunter}, {Tello}, {Toulouse},
  {Ulmer}, {Vandermarcq}, {Vergnault}, {Wallner}, {Waultier}, \&
  {Zanatta}}]{Auvergne2009}
{Auvergne}, M., {Bodin}, P., {Boisnard}, L., {et~al.} 2009, \aap, 506, 411

\bibitem[{{Baade} {et~al.}(2018{\natexlab{a}}){Baade}, {Pigulski}, {Rivinius},
  {Carciofi}, {Panoglou}, {Ghoreyshi}, {Handler}, {Kuschnig}, {Moffat},
  {Pablo}, {Popowicz}, {Wade}, {Weiss}, \& {Zwintz}}]{Baade2018a}
{Baade}, D., {Pigulski}, A., {Rivinius}, T., {et~al.} 2018{\natexlab{a}}, \aap,
  610, A70

\bibitem[{{Baade} {et~al.}(2018{\natexlab{b}}){Baade}, {Pigulski}, {Rivinius},
  {Wang}, {Martayan}, {Handler}, {Panoglou}, {Carciofi}, {Kuschnig}, {Mehner},
  {Moffat}, {Pablo}, {Rucinski}, {Wade}, {Weiss}, \& {Zwintz}}]{Baade2018b}
{Baade}, D., {Pigulski}, A., {Rivinius}, T., {et~al.} 2018{\natexlab{b}}, \aap,
  620, A145

\bibitem[{{Baade} {et~al.}(2016){Baade}, {Rivinius}, {Pigulski}, {Carciofi},
  {Martayan}, {Moffat}, {Wade}, {Weiss}, {Grunhut}, {Handler}, {Kuschnig},
  {Mehner}, {Pablo}, {Popowicz}, {Rucinski}, \& {Whittaker}}]{Baade2016}
{Baade}, D., {Rivinius}, T., {Pigulski}, A., {et~al.} 2016, \aap, 588, A56

\bibitem[{{Bailer-Jones} {et~al.}(2021){Bailer-Jones}, {Rybizki}, {Fouesneau},
  {Demleitner}, \& {Andrae}}]{BailerJones2021}
{Bailer-Jones}, C.~A.~L., {Rybizki}, J., {Fouesneau}, M., {Demleitner}, M., \&
  {Andrae}, R. 2021, \aj, 161, 147

\bibitem[{{Balona}(2016)}]{Balona2016}
{Balona}, L.~A. 2016, \mnras, 457, 3724

\bibitem[{{Balona} {et~al.}(2019){Balona}, {Handler}, {Chowdhury}, {Ozuyar},
  {Engelbrecht}, {Mirouh}, {Wade}, {David-Uraz}, \& {Cantiello}}]{Balona2019}
{Balona}, L.~A., {Handler}, G., {Chowdhury}, S., {et~al.} 2019, \mnras, 485,
  3457

\bibitem[{{Balona} \& {Ozuyar}(2020)}]{Balona2020}
{Balona}, L.~A. \& {Ozuyar}, D. 2020, \mnras, 493, 5871

\bibitem[{{Balona} {et~al.}(2011{\natexlab{a}}){Balona}, {Pigulski}, {De Cat},
  {Handler}, {Guti{\'e}rrez-Soto}, {Engelbrecht}, {Frescura}, {Briquet},
  {Cuypers}, {Daszy{\'n}ska-Daszkiewicz}, {Degroote}, {Dukes}, {Garcia},
  {Green}, {Heber}, {Kawaler}, {Lehmann}, {Leroy}, {Molenda-{\.Z}aaowicz},
  {Neiner}, {Noels}, {Nuspl}, {{\O}stensen}, {Pricopi}, {Roxburgh}, {Salmon},
  {Smith}, {Su{\'a}rez}, {Suran}, {Szab{\'o}}, {Uytterhoeven},
  {Christensen-Dalsgaard}, {Kjeldsen}, {Caldwell}, {Girouard}, \&
  {Sanderfer}}]{Balona2011a}
{Balona}, L.~A., {Pigulski}, A., {De Cat}, P., {et~al.} 2011{\natexlab{a}},
  \mnras, 413, 2403

\bibitem[{{Balona} {et~al.}(2011{\natexlab{b}}){Balona}, {Ripepi}, {Catanzaro},
  {Kurtz}, {Smalley}, {De Cat}, {Eyer}, {Grigahc{\`e}ne}, {Leccia},
  {Southworth}, {Uytterhoeven}, {van Winckel}, {Christensen-Dalsgaard},
  {Kjeldsen}, {Caldwell}, {van Cleve}, \& {Girouard}}]{Balona2011b}
{Balona}, L.~A., {Ripepi}, V., {Catanzaro}, G., {et~al.} 2011{\natexlab{b}},
  \mnras, 414, 792

\bibitem[{{Baluev}(2008)}]{Baluev2008}
{Baluev}, R.~V. 2008, \mnras, 385, 1279

\bibitem[{{Bouchaud} {et~al.}(2020){Bouchaud}, {Domiciano de Souza},
  {Rieutord}, {Reese}, \& {Kervella}}]{Bouchaud2020}
{Bouchaud}, K., {Domiciano de Souza}, A., {Rieutord}, M., {Reese}, D.~R., \&
  {Kervella}, P. 2020, \aap, 633, A78

\bibitem[{{Bowman} {et~al.}(2018){Bowman}, {Buysschaert}, {Neiner},
  {P{\'a}pics}, {Oksala}, \& {Aerts}}]{Bowman2018a}
{Bowman}, D.~M., {Buysschaert}, B., {Neiner}, C., {et~al.} 2018, \aap, 616, A77

\bibitem[{{Bowman} {et~al.}(2016){Bowman}, {Kurtz}, {Breger}, {Murphy}, \&
  {Holdsworth}}]{Bowman2016}
{Bowman}, D.~M., {Kurtz}, D.~W., {Breger}, M., {Murphy}, S.~J., \&
  {Holdsworth}, D.~L. 2016, \mnras, 460, 1970

\bibitem[{{Breger}(1979)}]{Breger1979}
{Breger}, M. 1979, \pasp, 91, 5

\bibitem[{{Breger}(1990)}]{Breger1990}
{Breger}, M. 1990, Delta Scuti Star Newsletter, 2, 13

\bibitem[{{Breger}(2000)}]{Breger2000}
{Breger}, M. 2000, in Astronomical Society of the Pacific Conference Series,
  Vol. 210, Delta Scuti and Related Stars, ed. M.~{Breger} \& M.~{Montgomery},
  3

\bibitem[{{Briquet} {et~al.}(2007){Briquet}, {Hubrig}, {De Cat}, {Aerts},
  {North}, \& {Sch{\"o}ller}}]{Briquet2007}
{Briquet}, M., {Hubrig}, S., {De Cat}, P., {et~al.} 2007, \aap, 466, 269

\bibitem[{{Burssens} {et~al.}(2020){Burssens}, {Sim{\'o}n-D{\'\i}az}, {Bowman},
  {Holgado}, {Michielsen}, {de Burgos}, {Castro}, {Barb{\'a}}, \&
  {Aerts}}]{Burssens2020}
{Burssens}, S., {Sim{\'o}n-D{\'\i}az}, S., {Bowman}, D.~M., {et~al.} 2020,
  \aap, 639, A81

\bibitem[{{Buysschaert} {et~al.}(2018){Buysschaert}, {Aerts}, {Bowman},
  {Johnston}, {Van Reeth}, {Pedersen}, {Mathis}, \& {Neiner}}]{Buysschaert2018}
{Buysschaert}, B., {Aerts}, C., {Bowman}, D.~M., {et~al.} 2018, \aap, 616, A148

\bibitem[{{Chen} {et~al.}(2020){Chen}, {Wang}, {Deng}, {de Grijs}, {Yang}, \&
  {Tian}}]{Chen2020}
{Chen}, X., {Wang}, S., {Deng}, L., {et~al.} 2020, \apjs, 249, 18

\bibitem[{{Clementini} {et~al.}(2019){Clementini}, {Ripepi}, {Molinaro},
  {Garofalo}, {Muraveva}, {Rimoldini}, {Guy}, {Jevardat de Fombelle},
  {Nienartowicz}, {Marchal}, {Audard}, {Holl}, {Leccia}, {Marconi}, {Musella},
  {Mowlavi}, {Lecoeur-Taibi}, {Eyer}, {De Ridder}, {Regibo}, {Sarro},
  {Szabados}, {Evans}, \& {Riello}}]{Clementini2019}
{Clementini}, G., {Ripepi}, V., {Molinaro}, R., {et~al.} 2019, \aap, 622, A60

\bibitem[{{Daszy\'{n}ska-Daszkiewicz}
  {et~al.}(2007){Daszy\'{n}ska-Daszkiewicz}, {Dziembowski}, \&
  {Pamyatnykh}}]{Daszynska-Daszkiewicz2007}
{Daszy\'{n}ska-Daszkiewicz}, J., {Dziembowski}, W.~A., \& {Pamyatnykh}, A.~A.
  2007, \actaa, 57, 11

\bibitem[{{Daszy{\'n}ska-Daszkiewicz}
  {et~al.}(2002){Daszy{\'n}ska-Daszkiewicz}, {Dziembowski}, {Pamyatnykh}, \&
  {Goupil}}]{Daszynska2002}
{Daszy{\'n}ska-Daszkiewicz}, J., {Dziembowski}, W.~A., {Pamyatnykh}, A.~A., \&
  {Goupil}, M.~J. 2002, \aap, 392, 151

\bibitem[{{Daszy{\'n}ska-Daszkiewicz}
  {et~al.}(2013){Daszy{\'n}ska-Daszkiewicz}, {Ostrowski}, \&
  {Pamyatnykh}}]{Daszynska-Daszkiewicz2013}
{Daszy{\'n}ska-Daszkiewicz}, J., {Ostrowski}, J., \& {Pamyatnykh}, A.~A. 2013,
  \mnras, 432, 3153

\bibitem[{{Daszy{\'n}ska-Daszkiewicz}
  {et~al.}(2017){Daszy{\'n}ska-Daszkiewicz}, {Pamyatnykh}, {Walczak}, {Colgan},
  {Fontes}, \& {Kilcrease}}]{Daszynska2017}
{Daszy{\'n}ska-Daszkiewicz}, J., {Pamyatnykh}, A.~A., {Walczak}, P., {et~al.}
  2017, \mnras, 466, 2284

\bibitem[{{Degroote} {et~al.}(2011){Degroote}, {Acke}, {Samadi}, {Aerts},
  {Kurtz}, {Noels}, {Miglio}, {Montalb{\'a}n}, {Bloemen}, {Baglin}, {Baudin},
  {Catala}, {Michel}, \& {Auvergne}}]{Degroote2011}
{Degroote}, P., {Acke}, B., {Samadi}, R., {et~al.} 2011, \aap, 536, A82

\bibitem[{{Degroote} {et~al.}(2010){Degroote}, {Aerts}, {Baglin}, {Miglio},
  {Briquet}, {Noels}, {Niemczura}, {Montalban}, {Bloemen}, {Oreiro},
  {Vu{\v{c}}kovi{\'c}}, {Smolders}, {Auvergne}, {Baudin}, {Catala}, \&
  {Michel}}]{Degroote2010}
{Degroote}, P., {Aerts}, C., {Baglin}, A., {et~al.} 2010, \nat, 464, 259

\bibitem[{{Degroote} {et~al.}(2009){Degroote}, {Aerts}, {Ollivier}, {Miglio},
  {Debosscher}, {Cuypers}, {Briquet}, {Montalb{\'a}n}, {Thoul}, {Noels}, {De
  Cat}, {Balaguer-N{\'u}{\~n}ez}, {Maceroni}, {Ribas}, {Auvergne}, {Baglin},
  {Deleuil}, {Weiss}, {Jorda}, {Baudin}, \& {Samadi}}]{Degroote2009}
{Degroote}, P., {Aerts}, C., {Ollivier}, M., {et~al.} 2009, \aap, 506, 471

\bibitem[{{Deng} {et~al.}(2018){Deng}, {Li}, {Wu}, \& {Chen}}]{Deng2018}
{Deng}, Z.~M., {Li}, Y., {Wu}, T., \& {Chen}, X.~H. 2018, Acta Astronomica
  Sinica, 59, 49

\bibitem[{{Dhouib} {et~al.}(2021{\natexlab{a}}){Dhouib}, {Prat}, {Van Reeth},
  \& {Mathis}}]{Dhouib2021a}
{Dhouib}, H., {Prat}, V., {Van Reeth}, T., \& {Mathis}, S. 2021{\natexlab{a}},
  \aap, 652, A154

\bibitem[{{Dhouib} {et~al.}(2021{\natexlab{b}}){Dhouib}, {Prat}, {Van Reeth},
  \& {Mathis}}]{Dhouib2021b}
{Dhouib}, H., {Prat}, V., {Van Reeth}, T., \& {Mathis}, S. 2021{\natexlab{b}},
  \aap, 656, A122

\bibitem[{{Diago} {et~al.}(2009){Diago}, {Guti{\'e}rrez-Soto}, {Auvergne},
  {Fabregat}, {Hubert}, {Floquet}, {Fr{\'e}mat}, {Garrido}, {Andrade}, {de
  Batz}, {Emilio}, {Espinosa Lara}, {Huat}, {Janot-Pacheco}, {Leroy},
  {Martayan}, {Neiner}, {Semaan}, {Suso}, {Catala}, {Poretti}, {Rainer},
  {Uytterhoeven}, {Michel}, \& {Samadi}}]{Diago2009}
{Diago}, P.~D., {Guti{\'e}rrez-Soto}, J., {Auvergne}, M., {et~al.} 2009, \aap,
  506, 125

\bibitem[{{Distefano et al.}(2022)}]{Distefano2022}
{Distefano et al.} 2022, \aap\ in prep.

\bibitem[{{Drake} {et~al.}(2017){Drake}, {Djorgovski}, {Catelan}, {Graham},
  {Mahabal}, {Larson}, {Christensen}, {Torrealba}, {Beshore}, {McNaught},
  {Garradd}, {Belokurov}, \& {Koposov}}]{Drake2017}
{Drake}, A.~J., {Djorgovski}, S.~G., {Catelan}, M., {et~al.} 2017, \mnras, 469,
  3688

\bibitem[{{Dupret} {et~al.}(2005){Dupret}, {Grigahc{\`e}ne}, {Garrido},
  {Gabriel}, \& {Scuflaire}}]{Dupret2005}
{Dupret}, M.~A., {Grigahc{\`e}ne}, A., {Garrido}, R., {Gabriel}, M., \&
  {Scuflaire}, R. 2005, \aap, 435, 927

\bibitem[{{Eker} {et~al.}(2020){Eker}, {Soydugan}, {Bilir}, {Bak{\i}{\c{s}}},
  {Ali{\c{c}}avu{\c{s}}}, {{\"O}zer}, {Aslan}, {Alpsoy}, \&
  {K{\"o}se}}]{Eker2020}
{Eker}, Z., {Soydugan}, F., {Bilir}, S., {et~al.} 2020, \mnras, 496, 3887

\bibitem[{{Eyer et al.}(2022)}]{Eyer2022}
{Eyer et al.} 2022, \aap\ in prep.

\bibitem[{{Gaia Collaboration} {et~al.}(2019){Gaia Collaboration}, {Eyer},
  {Rimoldini}, {Audard}, {Anderson}, {Nienartowicz}, {Glass}, {Marchal},
  {Grenon}, {Mowlavi}, {Holl}, {Clementini}, {Aerts}, {Mazeh}, {Evans},
  {Szabados}, {Brown}, {Vallenari}, {Prusti}, {de Bruijne}, {Babusiaux},
  {Bailer-Jones}, {Biermann}, {Jansen}, {Jordi}, {Klioner}, {Lammers},
  {Lindegren}, {Luri}, {Mignard}, {Panem}, {Pourbaix}, {Randich}, {Sartoretti},
  {Siddiqui}, {Soubiran}, {van Leeuwen}, {Walton}, {Arenou}, {Bastian},
  {Cropper}, {Drimmel}, {Katz}, {Lattanzi}, {Bakker}, {Cacciari},
  {Casta{\~n}eda}, {Chaoul}, {Cheek}, {De Angeli}, {Fabricius}, {Guerra},
  {Masana}, {Messineo}, {Panuzzo}, {Portell}, {Riello}, {Seabroke}, {Tanga},
  {Th{\'e}venin}, {Gracia-Abril}, {Comoretto}, {Garcia-Reinaldos}, {Teyssier},
  {Altmann}, {Andrae}, {Bellas-Velidis}, {Benson}, {Berthier}, {Blomme},
  {Burgess}, {Busso}, {Carry}, {Cellino}, {Clotet}, {Creevey}, {Davidson}, {De
  Ridder}, {Delchambre}, {Dell'Oro}, {Ducourant},
  {Fern{\'a}ndez-Hern{\'a}ndez}, {Fouesneau}, {Fr{\'e}mat}, {Galluccio},
  {Garc{\'\i}a-Torres}, {Gonz{\'a}lez-N{\'u}{\~n}ez}, {Gonz{\'a}lez-Vidal},
  {Gosset}, {Guy}, {Halbwachs}, {Hambly}, {Harrison}, {Hern{\'a}ndez},
  {Hestroffer}, {Hodgkin}, {Hutton}, {Jasniewicz}, {Jean-Antoine-Piccolo},
  {Jordan}, {Korn}, {Krone-Martins}, {Lanzafame}, {Lebzelter}, {L{\"o}ffler},
  {Manteiga}, {Marrese}, {Mart{\'\i}n-Fleitas}, {Moitinho}, {Mora}, {Muinonen},
  {Osinde}, {Pancino}, {Pauwels}, {Petit}, {Recio-Blanco}, {Richards}, {Robin},
  {Sarro}, {Siopis}, {Smith}, {Sozzetti}, {S{\"u}veges}, {Torra}, {van Reeven},
  {Abbas}, {Abreu Aramburu}, {Accart}, {Altavilla}, {{\'A}lvarez}, {Alvarez},
  {Alves}, {Andrei}, {Anglada Varela}, {Antiche}, {Antoja}, {Arcay},
  {Astraatmadja}, {Bach}, {Baker}, {Balaguer-N{\'u}{\~n}ez}, {Balm}, {Barache},
  {Barata}, {Barbato}, {Barblan}, {Barklem}, {Barrado}, {Barros}, {Barstow},
  {Bartholom{\'e} Mu{\~n}oz}, {Bassilana}, {Becciani}, {Bellazzini},
  {Berihuete}, {Bertone}, {Bianchi}, {Bienaym{\'e}}, {Blanco-Cuaresma}, {Boch},
  {Boeche}, {Bombrun}, {Borrachero}, {Bossini}, {Bouquillon}, {Bourda},
  {Bragaglia}, {Bramante}, {Breddels}, {Bressan}, {Brouillet},
  {Br{\"u}semeister}, {Brugaletta}, {Bucciarelli}, {Burlacu}, {Busonero},
  {Butkevich}, {Buzzi}, {Caffau}, {Cancelliere}, {Cannizzaro}, {Cantat-Gaudin},
  {Carballo}, {Carlucci}, {Carrasco}, {Casamiquela}, {Castellani},
  {Castro-Ginard}, {Charlot}, {Chemin}, {Chiavassa}, {Cocozza}, {Costigan},
  {Cowell}, {Crifo}, {Crosta}, {Crowley}, {Cuypers}, {Dafonte}, {Damerdji},
  {Dapergolas}, {David}, {David}, {de Laverny}, {De Luise}, {De March}, {de
  Martino}, {de Souza}, {de Torres}, {Debosscher}, {del Pozo}, {Delbo},
  {Delgado}, {Delgado}, {Diakite}, {Diener}, {Distefano}, {Dolding},
  {Drazinos}, {Dur{\'a}n}, {Edvardsson}, {Enke}, {Eriksson}, {Esquej}, {Eynard
  Bontemps}, {Fabre}, {Fabrizio}, {Faigler}, {Falc{\~a}o}, {Farr{\`a}s Casas},
  {Federici}, {Fedorets}, {Fernique}, {Figueras}, {Filippi}, {Findeisen},
  {Fonti}, {Fraile}, {Fraser}, {Fr{\'e}zouls}, {Gai}, {Galleti}, {Garabato},
  {Garc{\'\i}a-Sedano}, {Garofalo}, {Garralda}, {Gavel}, {Gavras}, {Gerssen},
  {Geyer}, {Giacobbe}, {Gilmore}, {Girona}, {Giuffrida}, {Gomes}, {Granvik},
  {Gueguen}, {Guerrier}, {Guiraud}, {Guti{\'e}rrez-S{\'a}nchez}, {Haigron},
  {Hatzidimitriou}, {Hauser}, {Haywood}, {Heiter}, {Helmi}, {Heu}, {Hilger},
  {Hobbs}, {Hofmann}, {Holland}, {Huckle}, {Hypki}, {Icardi}, {Jan{\ss}en},
  {Jevardat de Fombelle}, {Jonker}, {Juh{\'a}sz}, {Julbe}, {Karampelas},
  {Kewley}, {Klar}, {Kochoska}, {Kohley}, {Kolenberg}, {Kontizas}, {Kontizas},
  {Koposov}, {Kordopatis}, {Kostrzewa-Rutkowska}, {Koubsky}, {Lambert},
  {Lanza}, {Lasne}, {Lavigne}, {Le Fustec}, {Le Poncin-Lafitte}, {Lebreton},
  {Leccia}, {Leclerc}, {Lecoeur-Taibi}, {Lenhardt}, {Leroux}, {Liao}, {Licata},
  {Lindstr{\o}m}, {Lister}, {Livanou}, {Lobel}, {L{\'o}pez}, {Lorenz},
  {Managau}, {Mann}, {Mantelet}, {Marchant}, {Marconi}, {Marinoni},
  {Marschalk{\'o}}, {Marshall}, {Martino}, {Marton}, {Mary}, {Massari},
  {Matijevi{\v{c}}}, {McMillan}, {Messina}, {Michalik}, {Millar}, {Molina},
  {Molinaro}, {Moln{\'a}r}, {Montegriffo}, {Mor}, {Morbidelli}, {Morel},
  {Morgenthaler}, {Morris}, {Mulone}, {Muraveva}, {Musella}, {Nelemans},
  {Nicastro}, {Noval}, {O'Mullane}, {Ord{\'e}novic}, {Ord{\'o}{\~n}ez-Blanco},
  {Osborne}, {Pagani}, {Pagano}, {Pailler}, {Palacin}, {Palaversa}, {Panahi},
  {Pawlak}, {Piersimoni}, {Pineau}, {Plachy}, {Plum}, {Poggio}, {Poujoulet},
  {Pr{\v{s}}a}, {Pulone}, {Racero}, {Ragaini}, {Rambaux}, {Ramos-Lerate},
  {Regibo}, {Reyl{\'e}}, {Riclet}, {Ripepi}, {Riva}, {Rivard}, {Rixon},
  {Roegiers}, {Roelens}, {Romero-G{\'o}mez}, {Rowell}, {Royer}, {Ruiz-Dern},
  {Sadowski}, {Sagrist{\`a} Sell{\'e}s}, {Sahlmann}, {Salgado}, {Salguero},
  {Sanna}, {Santana-Ros}, {Sarasso}, {Savietto}, {Schultheis}, {Sciacca},
  {Segol}, {Segovia}, {S{\'e}gransan}, {Shih}, {Siltala}, {Silva}, {Smart},
  {Smith}, {Solano}, {Solitro}, {Sordo}, {Soria Nieto}, {Souchay}, {Spagna},
  {Spoto}, {Stampa}, {Steele}, {Steidelm{\"u}ller}, {Stephenson}, {Stoev},
  {Suess}, {Surdej}, {Szegedi-Elek}, {Tapiador}, {Taris}, {Tauran}, {Taylor},
  {Teixeira}, {Terrett}, {Teyssandier}, {Thuillot}, {Titarenko}, {Torra
  Clotet}, {Turon}, {Ulla}, {Utrilla}, {Uzzi}, {Vaillant}, {Valentini},
  {Valette}, {van Elteren}, {Van Hemelryck}, {van Leeuwen}, {Vaschetto},
  {Vecchiato}, {Veljanoski}, {Viala}, {Vicente}, {Vogt}, {von Essen}, {Voss},
  {Votruba}, {Voutsinas}, {Walmsley}, {Weiler}, {Wertz}, {Wevers},
  {Wyrzykowski}, {Yoldas}, {{\v{Z}}erjal}, {Ziaeepour}, {Zorec}, {Zschocke},
  {Zucker}, {Zurbach}, \& {Zwitter}}]{Eyer2019}
{Gaia Collaboration}, {Eyer}, L., {Rimoldini}, L., {et~al.} 2019, \aap, 623,
  A110

\bibitem[{{Gaia Collaboration} {et~al.}(2016){Gaia Collaboration}, {Prusti},
  {de Bruijne}, {Brown}, {Vallenari}, {Babusiaux}, {Bailer-Jones}, {Bastian},
  {Biermann}, {Evans}, {Eyer}, {Jansen}, {Jordi}, {Klioner}, {Lammers},
  {Lindegren}, {Luri}, {Mignard}, {Milligan}, {Panem}, {Poinsignon},
  {Pourbaix}, {Randich}, {Sarri}, {Sartoretti}, {Siddiqui}, {Soubiran},
  {Valette}, {van Leeuwen}, {Walton}, {Aerts}, {Arenou}, {Cropper}, {Drimmel},
  {H{\o}g}, {Katz}, {Lattanzi}, {O'Mullane}, {Grebel}, {Holland}, {Huc},
  {Passot}, {Bramante}, {Cacciari}, {Casta{\~n}eda}, {Chaoul}, {Cheek}, {De
  Angeli}, {Fabricius}, {Guerra}, {Hern{\'a}ndez}, {Jean-Antoine-Piccolo},
  {Masana}, {Messineo}, {Mowlavi}, {Nienartowicz}, {Ord{\'o}{\~n}ez-Blanco},
  {Panuzzo}, {Portell}, {Richards}, {Riello}, {Seabroke}, {Tanga},
  {Th{\'e}venin}, {Torra}, {Els}, {Gracia-Abril}, {Comoretto},
  {Garcia-Reinaldos}, {Lock}, {Mercier}, {Altmann}, {Andrae}, {Astraatmadja},
  {Bellas-Velidis}, {Benson}, {Berthier}, {Blomme}, {Busso}, {Carry},
  {Cellino}, {Clementini}, {Cowell}, {Creevey}, {Cuypers}, {Davidson}, {De
  Ridder}, {de Torres}, {Delchambre}, {Dell'Oro}, {Ducourant}, {Fr{\'e}mat},
  {Garc{\'\i}a-Torres}, {Gosset}, {Halbwachs}, {Hambly}, {Harrison}, {Hauser},
  {Hestroffer}, {Hodgkin}, {Huckle}, {Hutton}, {Jasniewicz}, {Jordan},
  {Kontizas}, {Korn}, {Lanzafame}, {Manteiga}, {Moitinho}, {Muinonen},
  {Osinde}, {Pancino}, {Pauwels}, {Petit}, {Recio-Blanco}, {Robin}, {Sarro},
  {Siopis}, {Smith}, {Smith}, {Sozzetti}, {Thuillot}, {van Reeven}, {Viala},
  {Abbas}, {Abreu Aramburu}, {Accart}, {Aguado}, {Allan}, {Allasia},
  {Altavilla}, {{\'A}lvarez}, {Alves}, {Anderson}, {Andrei}, {Anglada Varela},
  {Antiche}, {Antoja}, {Ant{\'o}n}, {Arcay}, {Atzei}, {Ayache}, {Bach},
  {Baker}, {Balaguer-N{\'u}{\~n}ez}, {Barache}, {Barata}, {Barbier}, {Barblan},
  {Baroni}, {Barrado y Navascu{\'e}s}, {Barros}, {Barstow}, {Becciani},
  {Bellazzini}, {Bellei}, {Bello Garc{\'\i}a}, {Belokurov}, {Bendjoya},
  {Berihuete}, {Bianchi}, {Bienaym{\'e}}, {Billebaud}, {Blagorodnova},
  {Blanco-Cuaresma}, {Boch}, {Bombrun}, {Borrachero}, {Bouquillon}, {Bourda},
  {Bouy}, {Bragaglia}, {Breddels}, {Brouillet}, {Br{\"u}semeister},
  {Bucciarelli}, {Budnik}, {Burgess}, {Burgon}, {Burlacu}, {Busonero}, {Buzzi},
  {Caffau}, {Cambras}, {Campbell}, {Cancelliere}, {Cantat-Gaudin}, {Carlucci},
  {Carrasco}, {Castellani}, {Charlot}, {Charnas}, {Charvet}, {Chassat},
  {Chiavassa}, {Clotet}, {Cocozza}, {Collins}, {Collins}, {Costigan}, {Crifo},
  {Cross}, {Crosta}, {Crowley}, {Dafonte}, {Damerdji}, {Dapergolas}, {David},
  {David}, {De Cat}, {de Felice}, {de Laverny}, {De Luise}, {De March}, {de
  Martino}, {de Souza}, {Debosscher}, {del Pozo}, {Delbo}, {Delgado},
  {Delgado}, {di Marco}, {Di Matteo}, {Diakite}, {Distefano}, {Dolding}, {Dos
  Anjos}, {Drazinos}, {Dur{\'a}n}, {Dzigan}, {Ecale}, {Edvardsson}, {Enke},
  {Erdmann}, {Escolar}, {Espina}, {Evans}, {Eynard Bontemps}, {Fabre},
  {Fabrizio}, {Faigler}, {Falc{\~a}o}, {Farr{\`a}s Casas}, {Faye}, {Federici},
  {Fedorets}, {Fern{\'a}ndez-Hern{\'a}ndez}, {Fernique}, {Fienga}, {Figueras},
  {Filippi}, {Findeisen}, {Fonti}, {Fouesneau}, {Fraile}, {Fraser}, {Fuchs},
  {Furnell}, {Gai}, {Galleti}, {Galluccio}, {Garabato}, {Garc{\'\i}a-Sedano},
  {Gar{\'e}}, {Garofalo}, {Garralda}, {Gavras}, {Gerssen}, {Geyer}, {Gilmore},
  {Girona}, {Giuffrida}, {Gomes}, {Gonz{\'a}lez-Marcos},
  {Gonz{\'a}lez-N{\'u}{\~n}ez}, {Gonz{\'a}lez-Vidal}, {Granvik}, {Guerrier},
  {Guillout}, {Guiraud}, {G{\'u}rpide}, {Guti{\'e}rrez-S{\'a}nchez}, {Guy},
  {Haigron}, {Hatzidimitriou}, {Haywood}, {Heiter}, {Helmi}, {Hobbs},
  {Hofmann}, {Holl}, {Holland}, {Hunt}, {Hypki}, {Icardi}, {Irwin}, {Jevardat
  de Fombelle}, {Jofr{\'e}}, {Jonker}, {Jorissen}, {Julbe}, {Karampelas},
  {Kochoska}, {Kohley}, {Kolenberg}, {Kontizas}, {Koposov}, {Kordopatis},
  {Koubsky}, {Kowalczyk}, {Krone-Martins}, {Kudryashova}, {Kull}, {Bachchan},
  {Lacoste-Seris}, {Lanza}, {Lavigne}, {Le Poncin-Lafitte}, {Lebreton},
  {Lebzelter}, {Leccia}, {Leclerc}, {Lecoeur-Taibi}, {Lemaitre}, {Lenhardt},
  {Leroux}, {Liao}, {Licata}, {Lindstr{\o}m}, {Lister}, {Livanou}, {Lobel},
  {L{\"o}ffler}, {L{\'o}pez}, {Lopez-Lozano}, {Lorenz}, {Loureiro},
  {MacDonald}, {Magalh{\~a}es Fernandes}, {Managau}, {Mann}, {Mantelet},
  {Marchal}, {Marchant}, {Marconi}, {Marie}, {Marinoni}, {Marrese},
  {Marschalk{\'o}}, {Marshall}, {Mart{\'\i}n-Fleitas}, {Martino}, {Mary},
  {Matijevi{\v{c}}}, {Mazeh}, {McMillan}, {Messina}, {Mestre}, {Michalik},
  {Millar}, {Miranda}, {Molina}, {Molinaro}, {Molinaro}, {Moln{\'a}r},
  {Moniez}, {Montegriffo}, {Monteiro}, {Mor}, {Mora}, {Morbidelli}, {Morel},
  {Morgenthaler}, {Morley}, {Morris}, {Mulone}, {Muraveva}, {Musella},
  {Narbonne}, {Nelemans}, {Nicastro}, {Noval}, {Ord{\'e}novic},
  {Ordieres-Mer{\'e}}, {Osborne}, {Pagani}, {Pagano}, {Pailler}, {Palacin},
  {Palaversa}, {Parsons}, {Paulsen}, {Pecoraro}, {Pedrosa}, {Pentik{\"a}inen},
  {Pereira}, {Pichon}, {Piersimoni}, {Pineau}, {Plachy}, {Plum}, {Poujoulet},
  {Pr{\v{s}}a}, {Pulone}, {Ragaini}, {Rago}, {Rambaux}, {Ramos-Lerate},
  {Ranalli}, {Rauw}, {Read}, {Regibo}, {Renk}, {Reyl{\'e}}, {Ribeiro},
  {Rimoldini}, {Ripepi}, {Riva}, {Rixon}, {Roelens}, {Romero-G{\'o}mez},
  {Rowell}, {Royer}, {Rudolph}, {Ruiz-Dern}, {Sadowski}, {Sagrist{\`a}
  Sell{\'e}s}, {Sahlmann}, {Salgado}, {Salguero}, {Sarasso}, {Savietto},
  {Schnorhk}, {Schultheis}, {Sciacca}, {Segol}, {Segovia}, {Segransan},
  {Serpell}, {Shih}, {Smareglia}, {Smart}, {Smith}, {Solano}, {Solitro},
  {Sordo}, {Soria Nieto}, {Souchay}, {Spagna}, {Spoto}, {Stampa}, {Steele},
  {Steidelm{\"u}ller}, {Stephenson}, {Stoev}, {Suess}, {S{\"u}veges}, {Surdej},
  {Szabados}, {Szegedi-Elek}, {Tapiador}, {Taris}, {Tauran}, {Taylor},
  {Teixeira}, {Terrett}, {Tingley}, {Trager}, {Turon}, {Ulla}, {Utrilla},
  {Valentini}, {van Elteren}, {Van Hemelryck}, {van Leeuwen}, {Varadi},
  {Vecchiato}, {Veljanoski}, {Via}, {Vicente}, {Vogt}, {Voss}, {Votruba},
  {Voutsinas}, {Walmsley}, {Weiler}, {Weingrill}, {Werner}, {Wevers},
  {Whitehead}, {Wyrzykowski}, {Yoldas}, {{\v{Z}}erjal}, {Zucker}, {Zurbach},
  {Zwitter}, {Alecu}, {Allen}, {Allende Prieto}, {Amorim},
  {Anglada-Escud{\'e}}, {Arsenijevic}, {Azaz}, {Balm}, {Beck}, {Bernstein},
  {Bigot}, {Bijaoui}, {Blasco}, {Bonfigli}, {Bono}, {Boudreault}, {Bressan},
  {Brown}, {Brunet}, {Bunclark}, {Buonanno}, {Butkevich}, {Carret}, {Carrion},
  {Chemin}, {Ch{\'e}reau}, {Corcione}, {Darmigny}, {de Boer}, {de Teodoro}, {de
  Zeeuw}, {Delle Luche}, {Domingues}, {Dubath}, {Fodor}, {Fr{\'e}zouls},
  {Fries}, {Fustes}, {Fyfe}, {Gallardo}, {Gallegos}, {Gardiol}, {Gebran},
  {Gomboc}, {G{\'o}mez}, {Grux}, {Gueguen}, {Heyrovsky}, {Hoar}, {Iannicola},
  {Isasi Parache}, {Janotto}, {Joliet}, {Jonckheere}, {Keil}, {Kim},
  {Klagyivik}, {Klar}, {Knude}, {Kochukhov}, {Kolka}, {Kos}, {Kutka}, {Lainey},
  {LeBouquin}, {Liu}, {Loreggia}, {Makarov}, {Marseille}, {Martayan},
  {Martinez-Rubi}, {Massart}, {Meynadier}, {Mignot}, {Munari}, {Nguyen},
  {Nordlander}, {Ocvirk}, {O'Flaherty}, {Olias Sanz}, {Ortiz}, {Osorio},
  {Oszkiewicz}, {Ouzounis}, {Palmer}, {Park}, {Pasquato}, {Peltzer}, {Peralta},
  {P{\'e}turaud}, {Pieniluoma}, {Pigozzi}, {Poels}, {Prat}, {Prod'homme},
  {Raison}, {Rebordao}, {Risquez}, {Rocca-Volmerange}, {Rosen}, {Ruiz-Fuertes},
  {Russo}, {Sembay}, {Serraller Vizcaino}, {Short}, {Siebert}, {Silva},
  {Sinachopoulos}, {Slezak}, {Soffel}, {Sosnowska}, {Strai{\v{z}}ys}, {ter
  Linden}, {Terrell}, {Theil}, {Tiede}, {Troisi}, {Tsalmantza}, {Tur},
  {Vaccari}, {Vachier}, {Valles}, {Van Hamme}, {Veltz}, {Virtanen}, {Wallut},
  {Wichmann}, {Wilkinson}, {Ziaeepour}, \& {Zschocke}}]{Prusti2016}
{Gaia Collaboration}, {Prusti}, T., {de Bruijne}, J.~H.~J., {et~al.} 2016,
  \aap, 595, A1

\bibitem[{{Garc{\'\i}a} \& {Ballot}(2019)}]{Garcia2019}
{Garc{\'\i}a}, R.~A. \& {Ballot}, J. 2019, Living Reviews in Solar Physics, 16,
  4

\bibitem[{{Gavras et al.}(2022)}]{Gavras2022}
{Gavras et al.} 2022, \aap\ in prep.

\bibitem[{{Gebruers} {et~al.}(2021){Gebruers}, {Straumit}, {Tkachenko},
  {Mombarg}, {Pedersen}, {Van Reeth}, {Li}, {Lampens}, {Escorza}, {Bowman}, {De
  Cat}, {Vermeylen}, {Bodensteiner}, {Rix}, \& {Aerts}}]{Gebruers2021}
{Gebruers}, S., {Straumit}, I., {Tkachenko}, A., {et~al.} 2021, \aap, 650, A151

\bibitem[{{Grigahc{\`e}ne} {et~al.}(2010){Grigahc{\`e}ne}, {Antoci}, {Balona},
  {Catanzaro}, {Daszy{\'n}ska-Daszkiewicz}, {Guzik}, {Handler}, {Houdek},
  {Kurtz}, {Marconi}, {Monteiro}, {Moya}, {Ripepi}, {Su{\'a}rez},
  {Uytterhoeven}, {Borucki}, {Brown}, {Christensen-Dalsgaard}, {Gilliland},
  {Jenkins}, {Kjeldsen}, {Koch}, {Bernabei}, {Bradley}, {Breger}, {Di
  Criscienzo}, {Dupret}, {Garc{\'\i}a}, {Garc{\'\i}a Hern{\'a}ndez},
  {Jackiewicz}, {Kaiser}, {Lehmann}, {Mart{\'\i}n-Ruiz}, {Mathias},
  {Molenda-{\.Z}akowicz}, {Nemec}, {Nuspl}, {Papar{\'o}}, {Roth}, {Szab{\'o}},
  {Suran}, \& {Ventura}}]{Grigahcene2010}
{Grigahc{\`e}ne}, A., {Antoci}, V., {Balona}, L., {et~al.} 2010, \apjl, 713,
  L192

\bibitem[{{Guzik} {et~al.}(2000){Guzik}, {Kaye}, {Bradley}, {Cox}, \&
  {Neuforge}}]{Guzik2000}
{Guzik}, J.~A., {Kaye}, A.~B., {Bradley}, P.~A., {Cox}, A.~N., \& {Neuforge},
  C. 2000, \apjl, 542, L57

\bibitem[{Harris {et~al.}(2020)Harris, Millman, van~der Walt, Gommers,
  Virtanen, Cournapeau, Wieser, Taylor, Berg, Smith, Kern, Picus, Hoyer, van
  Kerkwijk, Brett, Haldane, del R{\'{i}}o, Wiebe, Peterson,
  G{\'{e}}rard-Marchant, Sheppard, Reddy, Weckesser, Abbasi, Gohlke, \&
  Oliphant}]{harris2020}
Harris, C.~R., Millman, K.~J., van~der Walt, S.~J., {et~al.} 2020, Nature, 585,
  357

\bibitem[{{Hekker} \& {Christensen-Dalsgaard}(2017)}]{HekkerJCD2017}
{Hekker}, S. \& {Christensen-Dalsgaard}, J. 2017, \aapr, 25, 1

\bibitem[{{Henneco} {et~al.}(2021){Henneco}, {Van Reeth}, {Prat}, {Mathis},
  {Mombarg}, \& {Aerts}}]{Henneco2021}
{Henneco}, J., {Van Reeth}, T., {Prat}, V., {et~al.} 2021, \aap, 648, A97

\bibitem[{{Holl et al.}(2022)}]{Holl2022}
{Holl et al.} 2022, \aap\ in prep.

\bibitem[{{Hon} {et~al.}(2018){Hon}, {Stello}, \& {Yu}}]{Hon2018}
{Hon}, M., {Stello}, D., \& {Yu}, J. 2018, \mnras, 476, 3233

\bibitem[{{Huat} {et~al.}(2009){Huat}, {Hubert}, {Baudin}, {Floquet}, {Neiner},
  {Fr{\'e}mat}, {Guti{\'e}rrez-Soto}, {Andrade}, {de Batz}, {Diago}, {Emilio},
  {Espinosa Lara}, {Fabregat}, {Janot-Pacheco}, {Leroy}, {Martayan}, {Semaan},
  {Suso}, {Auvergne}, {Catala}, {Michel}, \& {Samadi}}]{Huat2009}
{Huat}, A.~L., {Hubert}, A.~M., {Baudin}, F., {et~al.} 2009, \aap, 506, 95

\bibitem[{Hunter(2007)}]{Hunter2007}
Hunter, J.~D. 2007, Computing in Science \& Engineering, 9, 90

\bibitem[{{Jayasinghe} {et~al.}(2020){Jayasinghe}, {Stanek}, {Kochanek},
  {Vallely}, {Shappee}, {Holoien}, {Thompson}, {Prieto}, {Pejcha}, {Fausnaugh},
  {Otero}, {Hurst}, \& {Will}}]{Jayasinghe2020}
{Jayasinghe}, T., {Stanek}, K.~Z., {Kochanek}, C.~S., {et~al.} 2020, \mnras,
  493, 4186

\bibitem[{{Koch} {et~al.}(2010){Koch}, {Borucki}, {Basri}, {Batalha}, {Brown},
  {Caldwell}, {Christensen-Dalsgaard}, {Cochran}, {DeVore}, {Dunham},
  {Gautier}, {Geary}, {Gilliland}, {Gould}, {Jenkins}, {Kondo}, {Latham},
  {Lissauer}, {Marcy}, {Monet}, {Sasselov}, {Boss}, {Brownlee}, {Caldwell},
  {Dupree}, {Howell}, {Kjeldsen}, {Meibom}, {Morrison}, {Owen}, {Reitsema},
  {Tarter}, {Bryson}, {Dotson}, {Gazis}, {Haas}, {Kolodziejczak}, {Rowe}, {Van
  Cleve}, {Allen}, {Chandrasekaran}, {Clarke}, {Li}, {Quintana}, {Tenenbaum},
  {Twicken}, \& {Wu}}]{Koch2010}
{Koch}, D.~G., {Borucki}, W.~J., {Basri}, G., {et~al.} 2010, \apjl, 713, L79

\bibitem[{{Kurtz}(2022)}]{Kurtz2022}
{Kurtz}, D. 2022, \araa, 60, in press, arXiv:2201.11629

\bibitem[{{Kurtz}(2000)}]{Kurtz2000}
{Kurtz}, D.~W. 2000, in Astronomical Society of the Pacific Conference Series,
  Vol. 210, Delta Scuti and Related Stars, ed. M.~{Breger} \& M.~{Montgomery},
  287

\bibitem[{{Kurtz} {et~al.}(2014){Kurtz}, {Saio}, {Takata}, {Shibahashi},
  {Murphy}, \& {Sekii}}]{Kurtz2014}
{Kurtz}, D.~W., {Saio}, H., {Takata}, M., {et~al.} 2014, \mnras, 444, 102

\bibitem[{{Labadie-Bartz} {et~al.}(2021){Labadie-Bartz}, {Baade}, {Carciofi},
  {Rubio}, {Rivinius}, {Borre}, {Martayan}, \& {Siverd}}]{Labadie2021}
{Labadie-Bartz}, J., {Baade}, D., {Carciofi}, A.~C., {et~al.} 2021, \mnras,
  502, 242

\bibitem[{{Le Diz{\`e}s} {et~al.}(2021){Le Diz{\`e}s}, {Rieutord}, \&
  {Charpinet}}]{Altair2021}
{Le Diz{\`e}s}, C., {Rieutord}, M., \& {Charpinet}, S. 2021, \aap, 653, A26

\bibitem[{{Lee}(2021)}]{Lee2021}
{Lee}, U. 2021, \mnras, 505, 1495

\bibitem[{{Lee} \& {Saio}(2020)}]{Lee2020}
{Lee}, U. \& {Saio}, H. 2020, \mnras, 497, 4117

\bibitem[{{Li} {et~al.}(2020){Li}, {Van Reeth}, {Bedding}, {Murphy}, {Antoci},
  {Ouazzani}, \& {Barbara}}]{GangLi2020}
{Li}, G., {Van Reeth}, T., {Bedding}, T.~R., {et~al.} 2020, \mnras, 491, 3586

\bibitem[{{Lindegren} {et~al.}(2021){Lindegren}, {Bastian}, {Biermann},
  {Bombrun}, {de Torres}, {Gerlach}, {Geyer}, {Hern{\'a}ndez}, {Hilger},
  {Hobbs}, {Klioner}, {Lammers}, {McMillan}, {Ramos-Lerate},
  {Steidelm{\"u}ller}, {Stephenson}, \& {van Leeuwen}}]{Lindegren2021}
{Lindegren}, L., {Bastian}, U., {Biermann}, M., {et~al.} 2021, \aap, 649, A4

\bibitem[{{McKinney}(2010)}]{Mckinney2010}
{McKinney}, W. 2010, in {P}roceedings of the 9th {P}ython in {S}cience
  {C}onference, ed. {S}t\'efan van~der {W}alt \& {J}arrod {M}illman, 56 -- 61

\bibitem[{{McNamara}(2011)}]{McNamara2011}
{McNamara}, D.~H. 2011, \aj, 142, 110

\bibitem[{{McNamara} {et~al.}(2007){McNamara}, {Clementini}, \&
  {Marconi}}]{McNamara2007}
{McNamara}, D.~H., {Clementini}, G., \& {Marconi}, M. 2007, \aj, 133, 2752

\bibitem[{{Miglio} {et~al.}(2007){Miglio}, {Montalb{\'a}n}, \&
  {Dupret}}]{Miglio2007}
{Miglio}, A., {Montalb{\'a}n}, J., \& {Dupret}, M.-A. 2007, \mnras, 375, L21

\bibitem[{{Mombarg} {et~al.}(2020){Mombarg}, {Dotter}, {Van Reeth},
  {Tkachenko}, {Gebruers}, \& {Aerts}}]{Mombarg2020}
{Mombarg}, J. S.~G., {Dotter}, A., {Van Reeth}, T., {et~al.} 2020, \apj, 895,
  51

\bibitem[{{Mombarg} {et~al.}(2021){Mombarg}, {Van Reeth}, \&
  {Aerts}}]{Mombarg2021}
{Mombarg}, J.~S.~G., {Van Reeth}, T., \& {Aerts}, C. 2021, \aap, 650, A58

\bibitem[{{Mombarg} {et~al.}(2019){Mombarg}, {Van Reeth}, {Pedersen},
  {Molenberghs}, {Bowman}, {Johnston}, {Tkachenko}, \& {Aerts}}]{Mombarg2019}
{Mombarg}, J.~S.~G., {Van Reeth}, T., {Pedersen}, M.~G., {et~al.} 2019, \mnras,
  485, 3248

\bibitem[{{Moravveji} {et~al.}(2015){Moravveji}, {Aerts}, {P{\'a}pics},
  {Triana}, \& {Vandoren}}]{Moravveji2015}
{Moravveji}, E., {Aerts}, C., {P{\'a}pics}, P.~I., {Triana}, S.~A., \&
  {Vandoren}, B. 2015, \aap, 580, A27

\bibitem[{{Mowlavi} {et~al.}(2013){Mowlavi}, {Barblan}, {Saesen}, \&
  {Eyer}}]{Mowlavi2013}
{Mowlavi}, N., {Barblan}, F., {Saesen}, S., \& {Eyer}, L. 2013, \aap, 554, A108

\bibitem[{{Mowlavi} {et~al.}(2018){Mowlavi}, {Lecoeur-Ta{\"\i}bi}, {Lebzelter},
  {Rimoldini}, {Lorenz}, {Audard}, {De Ridder}, {Eyer}, {Guy}, {Holl},
  {Jevardat de Fombelle}, {Marchal}, {Nienartowicz}, {Regibo}, {Roelens}, \&
  {Sarro}}]{Mowlavi2018}
{Mowlavi}, N., {Lecoeur-Ta{\"\i}bi}, I., {Lebzelter}, T., {et~al.} 2018, \aap,
  618, A58

\bibitem[{{Mowlavi} {et~al.}(2016){Mowlavi}, {Saesen}, {Semaan}, {Eggenberger},
  {Barblan}, {Eyer}, {Ekstr{\"o}m}, \& {Georgy}}]{Mowlavi2016}
{Mowlavi}, N., {Saesen}, S., {Semaan}, T., {et~al.} 2016, \aap, 595, L1

\bibitem[{{Mo{\'z}dzierski} {et~al.}(2019){Mo{\'z}dzierski}, {Pigulski},
  {Ko{\l}aczkowski}, {Michalska}, {Kopacki}, {Carrier}, {Walczak}, {Narwid},
  {St{\k{e}}{\'s}licki}, {Fu}, {Jiang}, {Zhang}, {Jackiewicz}, {Telting},
  {Morel}, {Saesen}, {Zahajkiewicz}, {Bru{\'s}}, {{\'S}r{\'o}dka},
  {Vu{\v{c}}kovi{\'c}}, {Verhoelst}, {Van Helshoecht}, {Lefever}, {Gielen},
  {Decin}, {Vanautgaerden}, \& {Aerts}}]{Mozdzierski2019}
{Mo{\'z}dzierski}, D., {Pigulski}, A., {Ko{\l}aczkowski}, Z., {et~al.} 2019,
  \aap, 632, A95

\bibitem[{{Mo{\'z}dzierski} {et~al.}(2014){Mo{\'z}dzierski}, {Pigulski},
  {Kopacki}, {Ko{\l}aczkowski}, \& {St{\k{e}}{\'s}licki}}]{Mozdzierski2014}
{Mo{\'z}dzierski}, D., {Pigulski}, A., {Kopacki}, G., {Ko{\l}aczkowski}, Z., \&
  {St{\k{e}}{\'s}licki}, M. 2014, \actaa, 64, 89

\bibitem[{{Murphy} {et~al.}(2016){Murphy}, {Fossati}, {Bedding}, {Saio},
  {Kurtz}, {Grassitelli}, \& {Wang}}]{Murphy2016}
{Murphy}, S.~J., {Fossati}, L., {Bedding}, T.~R., {et~al.} 2016, \mnras, 459,
  1201

\bibitem[{{Murphy} {et~al.}(2019){Murphy}, {Hey}, {Van Reeth}, \&
  {Bedding}}]{Murphy2019}
{Murphy}, S.~J., {Hey}, D., {Van Reeth}, T., \& {Bedding}, T.~R. 2019, \mnras,
  485, 2380

\bibitem[{{Neiner} {et~al.}(2012){Neiner}, {Floquet}, {Samadi}, {Espinosa
  Lara}, {Fr{\'e}mat}, {Mathis}, {Leroy}, {de Batz}, {Rainer}, {Poretti},
  {Mathias}, {Guarro Fl{\'o}}, {Buil}, {Ribeiro}, {Alecian}, {Andrade},
  {Briquet}, {Diago}, {Emilio}, {Fabregat}, {Guti{\'e}rrez-Soto}, {Hubert},
  {Janot-Pacheco}, {Martayan}, {Semaan}, {Suso}, \& {Zorec}}]{Neiner2012}
{Neiner}, C., {Floquet}, M., {Samadi}, R., {et~al.} 2012, \aap, 546, A47

\bibitem[{{Neiner} {et~al.}(2009){Neiner}, {Guti{\'e}rrez-Soto}, {Baudin}, {de
  Batz}, {Fr{\'e}mat}, {Huat}, {Floquet}, {Hubert}, {Leroy}, {Diago},
  {Poretti}, {Carrier}, {Rainer}, {Catala}, {Thizy}, {Buil}, {Ribeiro},
  {Andrade}, {Emilio}, {Espinosa Lara}, {Fabregat}, {Janot-Pacheco},
  {Martayan}, {Semaan}, {Suso}, {Baglin}, {Michel}, \& {Samadi}}]{Neiner2009}
{Neiner}, C., {Guti{\'e}rrez-Soto}, J., {Baudin}, F., {et~al.} 2009, \aap, 506,
  143

\bibitem[{{Ouazzani} {et~al.}(2020){Ouazzani}, {Ligni{\`e}res}, {Dupret},
  {Salmon}, {Ballot}, {Christophe}, \& {Takata}}]{Ouazzani2020}
{Ouazzani}, R.~M., {Ligni{\`e}res}, F., {Dupret}, M.~A., {et~al.} 2020, \aap,
  640, A49

\bibitem[{{Palaversa} {et~al.}(2013){Palaversa}, {Ivezi{\'c}}, {Eyer},
  {Ru{\v{z}}djak}, {Sudar}, {Galin}, {Kroflin}, {Mesari{\'c}}, {Munk},
  {Vrbanec}, {Bo{\v{z}}i{\'c}}, {Loebman}, {Sesar}, {Rimoldini}, {Hunt-Walker},
  {VanderPlas}, {Westman}, {Stuart}, {Becker}, {Srdo{\v{c}}}, {Wozniak}, \&
  {Oluseyi}}]{Palaversa2013}
{Palaversa}, L., {Ivezi{\'c}}, {\v{Z}}., {Eyer}, L., {et~al.} 2013, \aj, 146,
  101

\bibitem[{{Pamyatnykh}(1999)}]{Pamyatnykh1999}
{Pamyatnykh}, A.~A. 1999, \actaa, 49, 119

\bibitem[{{P{\'a}pics} {et~al.}(2012){P{\'a}pics}, {Briquet}, {Baglin},
  {Poretti}, {Aerts}, {Degroote}, {Tkachenko}, {Morel}, {Zima}, {Niemczura},
  {Rainer}, {Hareter}, {Baudin}, {Catala}, {Michel}, {Samadi}, \&
  {Auvergne}}]{Papics2012}
{P{\'a}pics}, P.~I., {Briquet}, M., {Baglin}, A., {et~al.} 2012, \aap, 542, A55

\bibitem[{{P{\'a}pics} {et~al.}(2017){P{\'a}pics}, {Tkachenko}, {Van Reeth},
  {Aerts}, {Moravveji}, {Van de Sande}, {De Smedt}, {Bloemen}, {Southworth},
  {Debosscher}, {Niemczura}, \& {Gameiro}}]{Papics2017}
{P{\'a}pics}, P.~I., {Tkachenko}, A., {Van Reeth}, T., {et~al.} 2017, \aap,
  598, A74

\bibitem[{{Pedersen} {et~al.}(2021){Pedersen}, {Aerts}, {P{\'a}pics},
  {Michielsen}, {Gebruers}, {Rogers}, {Molenberghs}, {Burssens}, {Garcia}, \&
  {Bowman}}]{Pedersen2021}
{Pedersen}, M.~G., {Aerts}, C., {P{\'a}pics}, P.~I., {et~al.} 2021, Nature
  Astronomy, 5, 715

\bibitem[{{Pedersen} {et~al.}(2019){Pedersen}, {Chowdhury}, {Johnston},
  {Bowman}, {Aerts}, {Handler}, {De Cat}, {Neiner}, {David-Uraz}, {Buzasi},
  {Tkachenko}, {Sim{\'o}n-D{\'\i}az}, {Moravveji}, {Sikora}, {Mirouh},
  {Lovekin}, {Cantiello}, {Daszy{\'n}ska-Daszkiewicz}, {Pigulski},
  {Vanderspek}, \& {Ricker}}]{Pedersen2019}
{Pedersen}, M.~G., {Chowdhury}, S., {Johnston}, C., {et~al.} 2019, \apjl, 872,
  L9

\bibitem[{{Pedersen} {et~al.}(2020){Pedersen}, {Escorza}, {P{\'a}pics}, \&
  {Aerts}}]{Pedersen2020}
{Pedersen}, M.~G., {Escorza}, A., {P{\'a}pics}, P.~I., \& {Aerts}, C. 2020,
  \mnras, 495, 2738

\bibitem[{P\'erez \& Granger(2007)}]{IPython2007}
P\'erez, F. \& Granger, B.~E. 2007, Computing in Science and Engineering, 9, 21

\bibitem[{{Pigulski} \& {Pojma{\'n}ski}(2009)}]{Pigulski2009}
{Pigulski}, A. \& {Pojma{\'n}ski}, G. 2009, in American Institute of Physics
  Conference Series, Vol. 1170, Stellar Pulsation: Challenges for Theory and
  Observation, ed. J.~A. {Guzik} \& P.~A. {Bradley}, 351--354

\bibitem[{{Poggio} {et~al.}(2021){Poggio}, {Drimmel}, {Cantat-Gaudin}, {Ramos},
  {Ripepi}, {Zari}, {Andrae}, {Blomme}, {Chemin}, {Clementini}, {Figueras},
  {Fouesneau}, {Fr{\'e}mat}, {Lobel}, {Marshall}, {Muraveva}, \&
  {Romero-G{\'o}mez}}]{Poggio2021}
{Poggio}, E., {Drimmel}, R., {Cantat-Gaudin}, T., {et~al.} 2021, \aap, 651,
  A104

\bibitem[{{Poro} {et~al.}(2021){Poro}, {Paki}, {Mazhari}, {Sarabi}, {Kahraman
  Alicavus}, {Ahangarani Farahani}, {Guilani}, {Popov}, {Zubareva}, {Zarei
  Jalalabadi}, {Nourmohammad}, {Davoudi}, {Sabaghpour Arani}, \&
  {Ghalee}}]{Poro2021}
{Poro}, A., {Paki}, E., {Mazhari}, G., {et~al.} 2021, \pasp, 133, 084201

\bibitem[{{Reese} {et~al.}(2021){Reese}, {Mirouh}, {Espinosa Lara}, {Rieutord},
  \& {Putigny}}]{Reese2021}
{Reese}, D.~R., {Mirouh}, G.~M., {Espinosa Lara}, F., {Rieutord}, M., \&
  {Putigny}, B. 2021, \aap, 645, A46

\bibitem[{{Reese} {et~al.}(2013){Reese}, {Prat}, {Barban}, {van 't
  Veer-Menneret}, \& {MacGregor}}]{Reese2013}
{Reese}, D.~R., {Prat}, V., {Barban}, C., {van 't Veer-Menneret}, C., \&
  {MacGregor}, K.~B. 2013, \aap, 550, A77

\bibitem[{{Ricker} {et~al.}(2016){Ricker}, {Vanderspek}, {Winn}, {Seager},
  {Berta-Thompson}, {Levine}, {Villasenor}, {Latham}, {Charbonneau}, {Holman},
  {Johnson}, {Sasselov}, {Szentgyorgyi}, {Torres}, {Bakos}, {Brown},
  {Christensen-Dalsgaard}, {Kjeldsen}, {Clampin}, {Rinehart}, {Deming}, {Doty},
  {Dunham}, {Ida}, {Kawai}, {Sato}, {Jenkins}, {Lissauer}, {Jernigan},
  {Kaltenegger}, {Laughlin}, {Lin}, {McCullough}, {Narita}, {Pepper},
  {Stassun}, \& {Udry}}]{Ricker2016}
{Ricker}, G.~R., {Vanderspek}, R., {Winn}, J., {et~al.} 2016, in Society of
  Photo-Optical Instrumentation Engineers (SPIE) Conference Series, Vol. 9904,
  Space Telescopes and Instrumentation 2016: Optical, Infrared, and Millimeter
  Wave, ed. H.~A. {MacEwen}, G.~G. {Fazio}, M.~{Lystrup}, N.~{Batalha},
  N.~{Siegler}, \& E.~C. {Tong}, 99042B

\bibitem[{{Riess} {et~al.}(2021){Riess}, {Casertano}, {Yuan}, {Bowers},
  {Macri}, {Zinn}, \& {Scolnic}}]{Riess2021}
{Riess}, A.~G., {Casertano}, S., {Yuan}, W., {et~al.} 2021, \apjl, 908, L6

\bibitem[{{Rimoldini et al.}(2022{\natexlab{a}})}]{Rimoldini2022a}
{Rimoldini et al.} 2022{\natexlab{a}}, \aap\ in prep.

\bibitem[{{Rimoldini et al.}(2022{\natexlab{b}})}]{Rimoldini2022b}
{Rimoldini et al.} 2022{\natexlab{b}}, {Gaia DR3 documentation Chapter 10:
  Variability}, Gaia DR3 documentation

\bibitem[{{Ripepi} {et~al.}(2022){Ripepi}, {Catanzaro}, {Clementini}, {De
  Somma}, {Drimmel}, {Leccia}, {Marconi}, {Molinaro}, {Musella}, \&
  {Poggio}}]{Ripepi2022}
{Ripepi}, V., {Catanzaro}, G., {Clementini}, G., {et~al.} 2022, \aap, 659, A167

\bibitem[{{Rivinius} {et~al.}(2003){Rivinius}, {Baade}, \&
  {{\v{S}}tefl}}]{Rivinius2003}
{Rivinius}, T., {Baade}, D., \& {{\v{S}}tefl}, S. 2003, \aap, 411, 229

\bibitem[{{Rivinius} {et~al.}(2013){Rivinius}, {Carciofi}, \&
  {Martayan}}]{Rivinius2013}
{Rivinius}, T., {Carciofi}, A.~C., \& {Martayan}, C. 2013, \aapr, 21, 69

\bibitem[{{Rodr{\'\i}guez} \& {L{\'o}pez-Gonz{\'a}lez}(2000)}]{Rodriguez2000a}
{Rodr{\'\i}guez}, E. \& {L{\'o}pez-Gonz{\'a}lez}, M.~J. 2000, \aap, 359, 597

\bibitem[{{Rodr{\'\i}guez} {et~al.}(2000){Rodr{\'\i}guez},
  {L{\'o}pez-Gonz{\'a}lez}, \& {L{\'o}pez de Coca}}]{Rodriguez2000b}
{Rodr{\'\i}guez}, E., {L{\'o}pez-Gonz{\'a}lez}, M.~J., \& {L{\'o}pez de Coca},
  P. 2000, \aaps, 144, 469

\bibitem[{{Rybizki} {et~al.}(2022){Rybizki}, {Green}, {Rix}, {El-Badry},
  {Demleitner}, {Zari}, {Udalski}, {Smart}, \& {Gould}}]{Rybizki2022}
{Rybizki}, J., {Green}, G.~M., {Rix}, H.-W., {et~al.} 2022, \mnras, 510, 2597

\bibitem[{{Saesen} {et~al.}(2013){Saesen}, {Briquet}, {Aerts}, {Miglio}, \&
  {Carrier}}]{Saesen2013}
{Saesen}, S., {Briquet}, M., {Aerts}, C., {Miglio}, A., \& {Carrier}, F. 2013,
  \aj, 146, 102

\bibitem[{{Saesen} {et~al.}(2010){Saesen}, {Carrier}, {Pigulski}, {Aerts},
  {Handler}, {Narwid}, {Fu}, {Zhang}, {Jiang}, {Vanautgaerden}, {Kopacki},
  {St{\k{e}}{\'s}licki}, {Acke}, {Poretti}, {Uytterhoeven}, {Gielen},
  {{\O}stensen}, {De Meester}, {Reed}, {Ko{\l}aczkowski}, {Michalska},
  {Schmidt}, {Yakut}, {Leitner}, {Kalomeni}, {Cherix}, {Spano}, {Prins}, {van
  Helshoecht}, {Zima}, {Huygen}, {Vandenbussche}, {Lenz}, {Ladjal}, {Puga
  Antol{\'\i}n}, {Verhoelst}, {De Ridder}, {Niarchos}, {Liakos}, {Lorenz},
  {Dehaes}, {Reyniers}, {Davignon}, {Kim}, {Kim}, {Lee}, {Lee}, {Kwon},
  {Broeders}, {van Winckel}, {Vanhollebeke}, {Waelkens}, {Raskin}, {Blom},
  {Eggen}, {Degroote}, {Beck}, {Puschnig}, {Schmitzberger}, {Gelven},
  {Steininger}, {Blommaert}, {Drummond}, {Briquet}, \&
  {Debosscher}}]{Saesen2010}
{Saesen}, S., {Carrier}, F., {Pigulski}, A., {et~al.} 2010, \aap, 515, A16

\bibitem[{{Saio} {et~al.}(2018){Saio}, {Bedding}, {Kurtz}, {Murphy}, {Antoci},
  {Shibahashi}, {Li}, \& {Takata}}]{Saio2018}
{Saio}, H., {Bedding}, T.~R., {Kurtz}, D.~W., {et~al.} 2018, \mnras, 477, 2183

\bibitem[{{Saio} {et~al.}(2017){Saio}, {Ekstr{\"o}m}, {Mowlavi}, {Georgy},
  {Saesen}, {Eggenberger}, {Semaan}, \& {Salmon}}]{Saio2017-PL}
{Saio}, H., {Ekstr{\"o}m}, S., {Mowlavi}, N., {et~al.} 2017, \mnras, 467, 3864

\bibitem[{{Saio} {et~al.}(2015){Saio}, {Kurtz}, {Takata}, {Shibahashi},
  {Murphy}, {Sekii}, \& {Bedding}}]{Saio2015}
{Saio}, H., {Kurtz}, D.~W., {Takata}, M., {et~al.} 2015, \mnras, 447, 3264

\bibitem[{{Salmon} {et~al.}(2014){Salmon}, {Montalb{\'a}n}, {Reese}, {Dupret},
  \& {Eggenberger}}]{Salmon2014}
{Salmon}, S.~J.~A.~J., {Montalb{\'a}n}, J., {Reese}, D.~R., {Dupret}, M.~A., \&
  {Eggenberger}, P. 2014, \aap, 569, A18

\bibitem[{{Schmid} \& {Aerts}(2016)}]{Schmid2016}
{Schmid}, V.~S. \& {Aerts}, C. 2016, \aap, 592, A116

\bibitem[{{Sekaran} {et~al.}(2021){Sekaran}, {Tkachenko}, {Johnston}, \&
  {Aerts}}]{Sekaran2021}
{Sekaran}, S., {Tkachenko}, A., {Johnston}, C., \& {Aerts}, C. 2021, \aap, 648,
  A91

\bibitem[{{Sharma} {et~al.}(2022){Sharma}, {Bedding}, {Saio}, \&
  {White}}]{Sharma2022}
{Sharma}, A.~N., {Bedding}, T.~R., {Saio}, H., \& {White}, T.~R. 2022, \mnras,
  in press, arXiv:2203.02582

\bibitem[{{Smith} \& {Henry}(2021)}]{Smith2021}
{Smith}, M.~A. \& {Henry}, G.~W. 2021, \apj, 915, 13

\bibitem[{{Stankov} \& {Handler}(2005)}]{Stankov2005}
{Stankov}, A. \& {Handler}, G. 2005, \apjs, 158, 193

\bibitem[{{Stello} {et~al.}(2022){Stello}, {Saunders}, {Grunblatt}, {Hon},
  {Reyes}, {Huber}, {Bedding}, {Elsworth}, {Garc{\'\i}a}, {Hekker},
  {Kallinger}, {Mathur}, {Mosser}, \& {Pinsonneault}}]{Stello2021}
{Stello}, D., {Saunders}, N., {Grunblatt}, S., {et~al.} 2022, \mnras, 512, 1677

\bibitem[{{Struve}(1955)}]{Struve1955}
{Struve}, O. 1955, \pasp, 67, 135

\bibitem[{{Szewczuk} \& {Daszy{\'n}ska-Daszkiewicz}(2017)}]{Szewczuk2017}
{Szewczuk}, W. \& {Daszy{\'n}ska-Daszkiewicz}, J. 2017, \mnras, 469, 13

\bibitem[{{Szewczuk} \& {Daszy{\'n}ska-Daszkiewicz}(2018)}]{Szewczuk2018}
{Szewczuk}, W. \& {Daszy{\'n}ska-Daszkiewicz}, J. 2018, \mnras, 478, 2243

\bibitem[{{Szewczuk} {et~al.}(2021){Szewczuk}, {Walczak}, \&
  {Daszy{\'n}ska-Daszkiewicz}}]{Szewczuk2021}
{Szewczuk}, W., {Walczak}, P., \& {Daszy{\'n}ska-Daszkiewicz}, J. 2021, \mnras,
  503, 5894

\bibitem[{{Taylor}(2005)}]{Taylor2005}
{Taylor}, M.~B. 2005, in Astronomical Society of the Pacific Conference Series,
  Vol. 347, Astronomical Data Analysis Software and Systems XIV, ed.
  P.~{Shopbell}, M.~{Britton}, \& R.~{Ebert}, 29

\bibitem[{{Townsend}(2005)}]{Townsend2005}
{Townsend}, R.~H.~D. 2005, \mnras, 364, 573

\bibitem[{{Van Beeck} {et~al.}(2021){Van Beeck}, {Bowman}, {Pedersen}, {Van
  Reeth}, {Van Hoolst}, \& {Aerts}}]{VanBeeck2021}
{Van Beeck}, J., {Bowman}, D.~M., {Pedersen}, M.~G., {et~al.} 2021, \aap, 655,
  A59

\bibitem[{{Van Reeth} {et~al.}(2018){Van Reeth}, {Mombarg}, {Mathis},
  {Tkachenko}, {Fuller}, {Bowman}, {Buysschaert}, {Johnston}, {Garc{\'\i}a
  Hern{\'a}ndez}, {Goldstein}, {Townsend}, \& {Aerts}}]{VanReeth2018}
{Van Reeth}, T., {Mombarg}, J.~S.~G., {Mathis}, S., {et~al.} 2018, \aap, 618,
  A24

\bibitem[{{Van Reeth} {et~al.}(2016){Van Reeth}, {Tkachenko}, \&
  {Aerts}}]{VanReeth2016}
{Van Reeth}, T., {Tkachenko}, A., \& {Aerts}, C. 2016, \aap, 593, A120

\bibitem[{{Van Reeth} {et~al.}(2015){Van Reeth}, {Tkachenko}, {Aerts},
  {P{\'a}pics}, {Triana}, {Zwintz}, {Degroote}, {Debosscher}, {Bloemen},
  {Schmid}, {De Smedt}, {Fremat}, {Fuentes}, {Homan}, {Hrudkova},
  {Karjalainen}, {Lombaert}, {Nemeth}, {{\O}stensen}, {Van De Steene}, {Vos},
  {Raskin}, \& {Van Winckel}}]{VanReeth2015}
{Van Reeth}, T., {Tkachenko}, A., {Aerts}, C., {et~al.} 2015, \apjs, 218, 27

\bibitem[{Virtanen {et~al.}(2020)Virtanen, Gommers, Oliphant, Haberland, Reddy,
  Cournapeau, Burovski, Peterson, Weckesser, Bright, {van der Walt}, Brett,
  Wilson, Millman, Mayorov, Nelson, Jones, Kern, Larson, Carey, Polat, Feng,
  Moore, {VanderPlas}, Laxalde, Perktold, Cimrman, Henriksen, Quintero, Harris,
  Archibald, Ribeiro, Pedregosa, {van Mulbregt}, \& {SciPy 1.0
  Contributors}}]{Virtanen2020}
Virtanen, P., Gommers, R., Oliphant, T.~E., {et~al.} 2020, Nature Methods, 17,
  261

\bibitem[{{White} {et~al.}(2017){White}, {Pope}, {Antoci}, {P{\'a}pics},
  {Aerts}, {Gies}, {Gordon}, {Huber}, {Schaefer}, {Aigrain}, {Albrecht},
  {Barclay}, {Barentsen}, {Beck}, {Bedding}, {Fredslund Andersen}, {Grundahl},
  {Howell}, {Ireland}, {Murphy}, {Nielsen}, {Silva Aguirre}, \&
  {Tuthill}}]{White2017}
{White}, T.~R., {Pope}, B.~J.~S., {Antoci}, V., {et~al.} 2017, \mnras, 471,
  2882

\bibitem[{{Wu} {et~al.}(2020){Wu}, {Li}, {Deng}, {Lin}, {Song}, \&
  {Jiang}}]{Wu2020}
{Wu}, T., {Li}, Y., {Deng}, Z.-m., {et~al.} 2020, \apj, 899, 38

\bibitem[{{Xiong} {et~al.}(2016){Xiong}, {Deng}, {Zhang}, \&
  {Wang}}]{Xiong2016}
{Xiong}, D.~R., {Deng}, L., {Zhang}, C., \& {Wang}, K. 2016, \mnras, 457, 3163

\bibitem[{{Yu} {et~al.}(2020){Yu}, {Bedding}, {Stello}, {Huber}, {Compton},
  {Gizon}, \& {Hekker}}]{Yu2020}
{Yu}, J., {Bedding}, T.~R., {Stello}, D., {et~al.} 2020, \mnras, 493, 1388

\bibitem[{{Yu} {et~al.}(2021){Yu}, {Hekker}, {Bedding}, {Stello}, {Huber},
  {Gizon}, {Khanna}, \& {Bi}}]{Yu2021}
{Yu}, J., {Hekker}, S., {Bedding}, T.~R., {et~al.} 2021, \mnras, 501, 5135

\bibitem[{{Yu} {et~al.}(2018){Yu}, {Huber}, {Bedding}, {Stello}, {Hon},
  {Murphy}, \& {Khanna}}]{Yu2018}
{Yu}, J., {Huber}, D., {Bedding}, T.~R., {et~al.} 2018, \apjs, 236, 42

\bibitem[{{Zechmeister} \& {K{\"u}rster}(2009)}]{Zechmeister2009}
{Zechmeister}, M. \& {K{\"u}rster}, M. 2009, \aap, 496, 577

\bibitem[{{Ziaali} {et~al.}(2019){Ziaali}, {Bedding}, {Murphy}, {Van Reeth}, \&
  {Hey}}]{Ziaali2019}
{Ziaali}, E., {Bedding}, T.~R., {Murphy}, S.~J., {Van Reeth}, T., \& {Hey},
  D.~R. 2019, \mnras, 486, 4348

\end{thebibliography}
